\documentclass[12pt]{elsarticle}
\usepackage[utf8]{inputenc}
\usepackage{setspace}
\usepackage[margin=0.75in]{geometry}
\usepackage{graphicx}
\graphicspath{ {./figures/} }
\usepackage[labelfont=bf, font = normalsize]{caption}
\usepackage{subcaption}
\usepackage{amsmath}
\usepackage{bm}
\usepackage{xcolor}
\usepackage{array}
\usepackage{appendix}
\usepackage{tikz}
\usepackage{floatrow}
\DeclareFloatFont{small}{\footnotesize}
\newcolumntype{P}[1]{>{\centering\arraybackslash}p{#1}}
\newcommand{\myparagraph}[1]{\paragraph{#1}\mbox{}\\}

\providecommand{\keywords}[1]
{
  \small	
  \textbf{\textit{Keywords:}} #1
}
            
\usepackage{natbib}
\title{%
	\LARGE
	A multiphysics modeling approach for in-stent restenosis\\
	\large
	Theoretical aspects and finite element implementation}

\author[1]{Kiran Manjunatha \corref{cor1}}
\ead{manjunatha@ifam.rwth-aachen.de}
\cortext[cor1]{Corresponding author\\kiran.manjunatha@rwth-aachen.de\\ Mies-van-der-Rohe-Str. 1, 52074 Aachen, Germany}

\author[2]{Marek Behr}
\author[3]{Felix Vogt}
\author[1]{Stefanie Reese}

\affiliation[1]{Institute of Applied Mechanics, RWTH Aachen University}
\affiliation[2]{Chair for Computational Analysis of Technical Systems, RWTH Aachen University }
\affiliation[3]{Department of Cardiology, Pulmonology, Intensive Care and Vascular
Medicine, RWTH Aachen University}

\date{}

\begin{document}


\begin{abstract}
Development of \textit{in silico} models are intrinsic in understanding disease progression in soft biological tissues. Within this work, we propose a fully-coupled  Lagrangian finite element framework which replicates the process of in-stent restenosis observed post stent implantation in a coronary artery. Coupled advection-reaction-diffusion reactions are set up that track the evolution of the concentrations of the platelet-derived growth factor, the transforming growth factor-$\beta$, the extracellular matrix, and the density of the smooth muscle cells. A continuum mechanical description of growth incorporating the evolution of arterial wall constituents is developed, and a suitable finite element implementation discussed. Qualitative validation of the computational model are presented by emulating a stented artery. Patient-specific data can be integrated into the model to predict the risk of restenosis and thereby assist in tuning of stent implantation parameters to mitigate the risk.\\
\\
\end{abstract}

\maketitle


\section{Introduction}
Coronary artery disease (CAD) is amongst the largest causes for disease burden, and in fact has been the leading cause of deaths worldwide \citep{nichols2014}. Percutaneous coronary intervention (PCI) is one of the minimally invasive procedures used to overcome CAD by restoring blood flow in clogged coronary arteries wherein a combination of coronary angioplasty and insertion of a supporting structure called stents is utilized. Unfortunately, PCI is associated with several risk factors including in-stent restenosis and stent thrombosis. In-stent restenosis refers to the accumulation of new tissue within the walls of the coronary artery leading to a diminished cross-section of blood flow even after stent implantation, hence defeating the whole purpose of the PCI procedure. Restenosis rates are reported at 15-20\% in ideal coronary lesions, the figures going as high as 30-60\% in case of complex lesions \citep{fattori2003}. \textit{Neointimal hyperplasia} is the underlying mechanism for the restenotic process. It is a collaborative effect of migration and proliferation of smooth muscle cells in the arterial wall, brought about by intricate signalling cascades that are triggered by certain stimuli, either internal or external to the arterial wall. 

Drug-eluting stents have been used effectively in reducing restenosis rates. Antiproliferative agents coated onto polymeric layers of the stents and progressively released into the arterial wall lead to substantial reduction of neointimal hyperplasia \citep{liistro2002,PARK20062432}. But the incidence rate has not yet been reduced significantly \citep{putra2021}. Suspected causes include arterial overstretch,  disturbed flow patterns resulting in low wall shear stresses on the vessel walls, slow reendothelialization, and delayed effects of polymer destabilization. 

An \textit{in silico} model that can successfully capture the mechanisms that bring about neointimal hyperplasia can aid in precisely addressing the risk associated with restenosis after implantation of drug-eluting stents. Additionally, it can help in adapting the PCI parameters that include strut design, artery overstretch and drug release rate. Over the years after the advent of PCI, several computational approaches have been developed that serve as \textit{in silico} models. They are broadly classified into discrete agent-based models (ABM), cellular automata (CA) techniques and continuum models.

\citet{zahedmanesh2012} developed a multiscale ABM unidirectionally coupled to a finite element model and investigated the influence of stent implantation parameters. More recently, \citet{li2019} extended this approach with bidirectional coupling between the agent-based and finite element (FE) models, and examined the lumen-loss rate caused by oscillatory stresses on the vessel wall. They also incorporated reendothelialization studies within their framework. \citet{Keshavarzian2018} included the effects of growth factors, proteases and additional signal molecules within their ABM and studied the responses of arteries to altered blood pressures and varying levels of the vessel wall constituents. \citet{Evans2008} on the other hand proposed the complex autonoma (CxA) approach involving hierarchical coupling of CA and ABM models. Damage-induced cell proliferativity was studied using a coupled ABM-FEA approach in \citet{Nolan2018AnIO} wherein the effects of instantaneous and cyclic loading were studied. The latest work by \citet{Zun2021EffectsOL} involves coupling of a 2D ABM of restenosis to a 1D coronary blood flow model and investigating the effects of blood flow dynamics on the physiology of restenosis. In spite of their capability to reproduce microscale mechanisms with high fidelity, ABMs suffer from the burden of computational cost. In addition, since the ABMs are based on simplistic rules at the cellular level, feeding observable mechanistic data at the macroscopic level into the system to calibrate the large number of parameters represents a challenging and tedious task. 

On the other end of the spectrum, phenomenological continuum models have been developed to model intimal thickening due to restenosis, and in general growth of soft biological tissues.  \citet{rodriguez1994} proposed a kinematic description of growth via a split of the deformation gradient into a volumetric growth part and an elastic part, drawing parallels from the modeling of plasticity. The continuum mechanical treatment of growth dates even further back, dealing with bone remodeling via introduction of mass sources \citep{Cowin1976BoneRI}. \citet{Kuhl2003TheoryAN} on the other hand introduced mass fluxes instead of mass sources in the context of open system thermodynamics and proposed a coupled monolithic approach for bone remodeling. The density preserving aspects outlined in the aforementioned work holds relevant in case of restenosis. \citet{Garikipati2004ACT} developed a similarly coupled framework for modeling biological growth and homeostasis by tracking the nutrients, enzymes and amino acids necessary for the growth process. \citet{Lubarda2002OnTM} proposed a generalized constitutive theory to study the growth of isotropic, transversely isotropic and orthotropic biological tissues, and further suggested the structure of the growth part of the deformation gradient. In addition, specific choices were provided for the  evolution of the growth part of the deformation gradient which are consistent with finite deformation continuum thermodynamics. Models, based on the classical mixture and homogenization theory, that predict mechanically-dependent growth and remodeling in soft tissues by capturing the turnover of constituents in soft tissues \citep{Humphrey2002ACM,Cyron2017GrowthAR} also hold relevance.  \citet{Fereidoonnezhad2016StressSA} formulated a pseudo-elastic damage model to describe discontinuous softening and permanent deformation in soft tissues. Later (see \citet{fereidoo2017}), the model was extended to include  damage-induced growth utilizing the well-established multiplicative split of the deformation gradient. On similar grounds, \citet{he2020} considered damage in plaque and arterial layers caused by stent deployment and developed a damage-dependent growth model incorporating isotropic volumetric growth.

In recent times, multiscale and multiphysics based continuum approaches that take into account the evolution of species of interest, and hence capture active mechanisms in the arterial wall have proven therapeutically insightful. \citet{budu2008} developed a model to track the growth factors and their influence on venous intimal hyperplasia, and proposed an empirical formulation that predicts the luminal radius. \citet{escuer2019} proposed a model wherein the transport of wall constituents and cell proliferative mechanisms were coupled to an isotropic volume growth hypothesis. Combination of fluid-structure interaction (FSI) and multifield scalar transport models have also been proposed. \citet{Yoshihara2014ACF} realized a sequential unidirectionally coupled FSI framework for modeling biological tissue response in multiphysical scenarios including respiratory mechanics. \citet{Thon2018AMA} established the aforementioned framework in the context of modeling early atherosclerosis. 

On a similar rationale, the aim of the current work is to develop a multiphysics continuum model that captures the molecular and cellular mechanisms in an arterial wall at enough resolution to be able to incorporate patient-specific morphological and immunological data and predict the risks associated with in-stent restenosis. A fully-coupled Lagrangian finite element formulation is developed herein based on coupled advection-reaction-diffusion equations and continuum mechanical modeling with the vision of embedding it in a fully-coupled FSI framework. Two continuum theories for finite growth in the restenotic process are hypothesized and evaluated. Key differences to the work of \citet{escuer2019} lie in the capturing of chemotactic and haptotactic movement of smooth muscle cells, the incorporation of anisotropic growth, and the finite element formulation itself. Evolution equations for the wall species and continuum mechanical modeling aspects are discussed in Section \ref{sect_math_model}. Finite element implementation details are elaborated in Section \ref{fe_impl}. Relevant numerical examples are dealt with in Section \ref{sect_num_eval}.

\subsection{In-stent restenosis}\label{section:pathophysiology}

\subsubsection{Structure of the arterial wall}
Before delving into the pathophysiology of in-stent restenosis, it is beneficial to first understand the structure of an arterial wall, that of the coronary arteries in particular. 

The human vascular system consists of two major categories of arteries that include \textit{elastic} and \textit{muscular} arteries, coronary arteries belonging to the latter. An arterial wall, irrespective of the category, consists of three concentric layers: \textit{intima}, \textit{media}, and the \textit{adventitia}. Intima refers to the layer of the wall that lies closest to the blood flow. The intima usually contains a monolayer of endothelial cells over a thin basal lamina. Intima in the muscular arteries contains in addition a few layers of smooth muscle cells embedded in the subendothelial space. Media, the layer beyond the basal lamina, mainly contains smooth muscle cells embedded in an extracellular matrix that includes elastin and collagen. Collagen is typically arranged into helices wrapped around the circumference of the wall. The smooth muscle cells are packed into concentric layers separated by sheets of elastin. The media is bound by the external elastic lamina on the outer end of the arterial wall. Finally, adventitia, the outermost layer of the vessel wall, contains a dense network of collagen fibers, elastin, nerve endings and fibroblasts. The distribution of the wall constituents in each layer, in combination with their orientations, determines the layer-specific mechanical response of the arterial wall.

\subsubsection{Pathophysiological hypothesis}
The current work focuses on four constituents of the arterial wall (referred to as \textit{species} hereinafter) that are crucial in bringing about in-stent restenosis: platelet-derived growth factor (PDGF), transforming growth factor (TGF)-$\beta$, extracellular matrix (ECM) and smooth muscle cells (SMCs).

PDGF refers to a family of disulfide-bonded heterodimeric proteins which has been implicated in vascular remodeling processes, including neointimal hyperplasia, that follow an injury to arterial wall. This can be attributed to its mitogenic and chemoattractant properties. PDGF is secreted by an array of cellular species namely the endothelial cells, SMCs, fibroblasts, macrophages and platelets.

TGF-$\beta$ herein referred to is a family of growth factors, composed of homodimeric or heterodimeric polypeptides, associated with multiple regulatory properties depending on cell type, growth conditions and presence of other polypeptide growth factors. They play a key role in cell proliferation, differentiation and apoptosis.

ECM collectively refers to the noncellular components present within the arterial wall, and is composed mainly of collagen. ECM provides the essential physical scaffolding for cellular constituents and also initiates crucial biochemical and biomechanical cues that are required for tissue morphogenesis, cell differentiation and homeostasis. 

SMCs, also termed \textit{myocytes}, are one of the most significant cellular components in the arterial wall which are primarily responsible for the modulation of vascular resistance and thereby the regulation of blood flow.

\begin{figure}[htbp]
    \centering
    \includegraphics[scale=0.65]{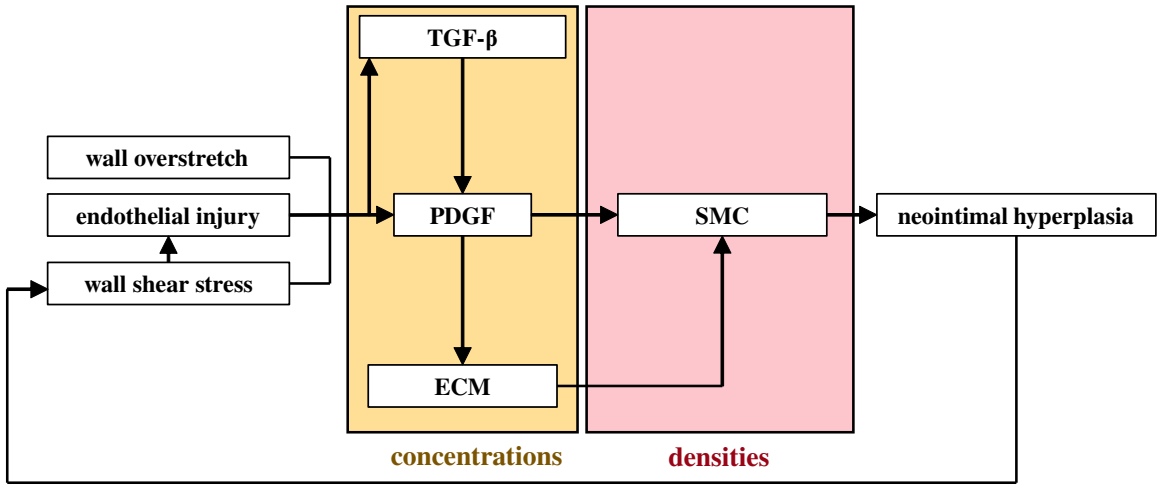}
    \caption{\textbf{schematic of the restenotic process}}
    \label{fig_ISR}
\end{figure}

During the implantation of stents, the endothelial monolayer on the arterial walls gets denuded due to the abrasive action of the stent surface. Additionally, when the stent is underexpanded, stent struts partially obstruct the blood flow creating vortices in their wake regions. This causes oscillatory wall shear stresses and hence further damages the endothelium. Also, depending on the arterial overstretch achieved during the implantation, injuries can occur within the deeper layers of the arterial wall, even reaching the medial layer. Platelets shall aggregate at the sites of the aforementioned injuries as part of the inflammatory response. PDGF and TGF-$\beta$, which are stored in the $\alpha$-granules of the aggregated platelets, are thereby released into the arterial wall. The presence of PDGF upregulates \textit{matrix metalloproteinase} (MMP) production in the arterial wall. ECM, being a network of collagen and glycoproteins surrounding the SMCs, gets degraded due to MMP. SMCs in the media, which are usually held stationary by the ECM, are rendered free for migration within the degraded collagen network. The focal adhesion sites created due to cleaved proteins in the ECM provide directionality to the migration of SMC, the migratory mechanism being termed \textit{haptotaxis}. PDGF also activates a variety of signaling cascades that enhance the motility of SMCs \citep{gerthoffer2007}. Furthermore, the local concentration gradient in PDGF influences the direction of SMC migration, which is termed \textit{chemotaxis}. Both the mechanisms in accordance result in the accumulation of the medial SMCs in the intima of the arterial wall. In addition, a degraded ECM encourages the proliferation of SMCs under the presence of PDGF since they switch their phenotypes from contractile to synthetic under such an ECM environment. A positive feedback loop might occur wherein the migrated SMCs create further obstruction to the blood flow and subsequent upregulation of both growth factors. The uncontrolled growth of vascular tissue that follows can eventually lead to a severe blockage of the lumen.

TGF-$\beta$ is indirectly involved in restenosis through its bimodal regulation of the inflammatory response \citep{battegay1990}. At low concentrations, TGF-$\beta$ upregulates PDGF autosecretion by cellular species in the arterial wall, mainly SMCs. In contrast, at high concentrations of TGF-$\beta$, a scarcity in the receptors for the binding of PDGF on SMCs occurs, thereby reducing the proliferativity of SMCs. 

In summary, a simplified hypothesis for the pathophysiology of in-stent restenosis which aids in the mathematical modeling is presented. A schematic is shown in Fig \ref{fig_ISR} summarizing the entire hypothesis, including the influencing factors, wall constituents and their interactions, and the outcomes.

\section{Mathematical modeling}\label{sect_math_model}
To model the pathophysiological process presented in the previous section, evolution equations are set up for the four species within the arterial wall and coupled to the growth kinematics. The cellular species (SMCs) of the arterial wall are quantified in terms of cell densities. The extracellular species (PDGF, TGF-$\beta$ and ECM) are quantified in terms of their concentrations. The arterial wall is modeled as an open system allowing for transfer of cellular and extracellular species into and out of it. Blood flow within the lumen is not considered within the modeling framework. 

\subsection{Evolution of species in the arterial wall}
The advection-reaction-diffusion equation forms the basis for modeling the transport phenomena governing the evolution of species within the arterial wall. The general form for a scalar field $\phi$ is given below:
\begin{equation}\label{ard_eq_general_form}
    \underset{\lower.5em \hbox{\text{rate}}}{\displaystyle{\left.\frac{\partial \phi}{\partial t}\right|_{\bm{x}}}} + \underbrace{\text{\sf div} \left(\phi\,\boldsymbol{v}\right)}_{\lower.3em \hbox{\text{advection}}} = \underbrace{\text{\sf div} \left(k\,\text{\sf grad} \phi\right)}_{\lower.35em \hbox{\text{diffusion}}} + \underbrace{\overset{\text{source}}{R_{so}} - \overset{\text{sink}}{R_{si}}}_{\lower.4em \hbox{\text{reaction}}}.
\end{equation}
\vspace{0.1in}\\
Here, $\bm{v}$ denotes the velocity of the medium of transport and $k$, the diffusivity of
$\phi$ in the medium. The above general form is valid for arbitrary points within a continuum body in its current configuration represented by the domain $\Omega$. The terms on the right hand side of Equation \ref{ard_eq_general_form} shall now be particularized for the individual species of the arterial wall. Table \ref{transp_vars} lists the variables associated with each species and their respective units. 

\begin{table}[hbt!]\
\centering
\caption{\textbf{Transport variables}}
\label{transp_vars} 
\begin{tabular}{p{3cm}p{1.5cm}p{3cm}p{2cm}}
\noalign{\hrule height 0.05cm}\noalign{\smallskip}
\\
variable type & variable & associated species & units  \\
\\
\noalign{\hrule height 0.05cm}\noalign{\smallskip}
\\
 & $c_{{}_P}$ & PDGF & [mol/mm$^3$]\\
 \\
 concentration & $c_{{}_T}$ & TGF-$\beta$ & [mol/mm$^3$]\\ 
 \\
 & $c_{{}_E}$ & ECM & [mol/mm$^3$]\\
 \\
 \hline\noalign{\smallskip}
 \\
cell density & $\rho_{{}_S}$ & SMC & [cells/mm$^3$]\\
\\
\noalign{\hrule height 0.05cm}\noalign{\smallskip}
\end{tabular}
\end{table}

\subsubsection{Prelude}
\begin{figure}[!htb]
    \centering
    \begin{minipage}{.5\textwidth}
        \centering
        \includegraphics[scale = 0.65]{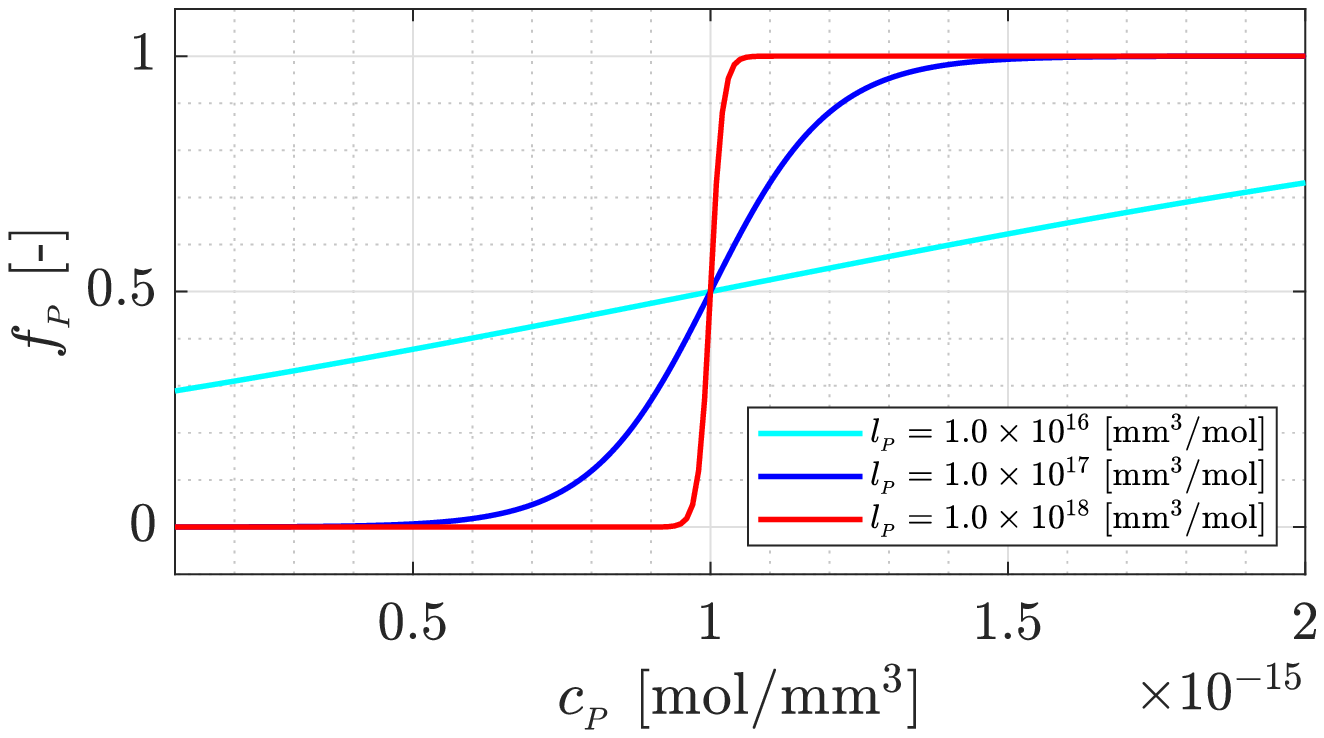}
        \subcaption{PDGF dependent scaling}
    \end{minipage}%
    \begin{minipage}{0.5\textwidth}
        \centering
        \includegraphics[scale = 0.65]{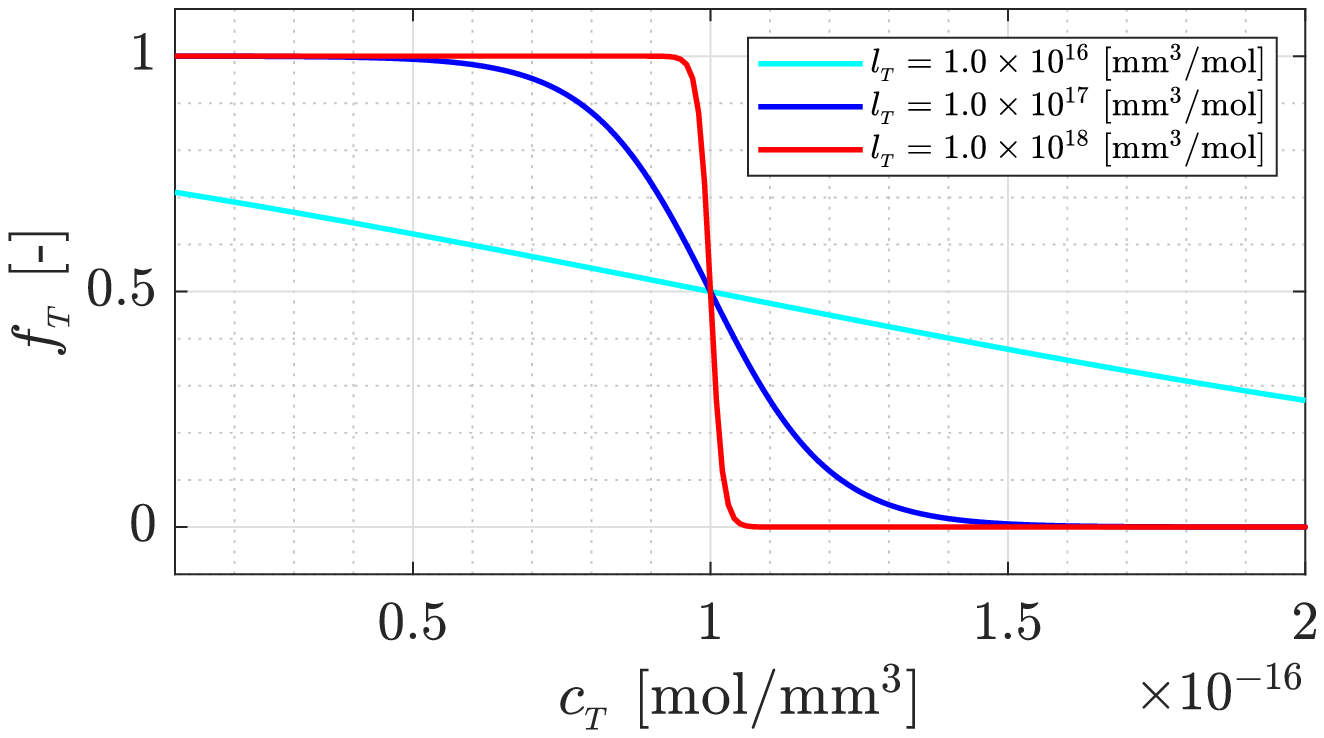}
        \subcaption{TGF$-\beta$ dependent scaling}
    \end{minipage}
    \caption{\textbf{Scaling functions:} They possess values between 0 and 1. One can control the smoothness of the switching on/off of biochemical phenomena by adjusting the steepness coefficients $l_{{}_P}$ and $l_{{}_T}$ respectively.}
    \label{figs:scaling}
\end{figure}

It is benefecial at this stage of the mathematical modeling process to introduce the following scaling functions that shall often be utilized in the particularization of Eq. \ref{ard_eq_general_form} to individual species. They are based on the general logistic function, and assist in smooth switching of certain biochemical phenomena between on and off states.
\begin{enumerate}
    \item[(a)] PDGF dependent scaling function: 
    \begin{equation}
        f_{{}_P} = \frac{1}{1 + e^{-l_{{}_P}\left(c_{{}_P} - c_{{}_{P,th}}\right)}}.
        \label{pdgf_scal}
    \end{equation}
    \item[(b)] TGF-$\beta$ dependent scaling function: 
    \begin{equation}
        f_{{}_T} = \frac{1}{1 + e^{l_{{}_T}\left(c_{{}_T} - c_{{}_{T,th}}\right)}}.
        \label{tgf_scal}
    \end{equation}
\end{enumerate}
In the above equations, $l_{{}_P}$  and $l_{{}_T}$ are termed the respective steepness coefficients, while $c_{{}_{P,th}}$ and $c_{{}_{T,th}}$ are predefined PDGF and TGF$-\beta$ thresholds at which the switching is intended. Fig \ref{figs:scaling} illustrates the behavior of the above functions for varying exemplary steepness coefficients $l_{{}_P}$ and $l_{{}_T}$. $c_{{}_{P,th}}$ and $c_{{}_{T,th}}$ are prescribed to be $10^{-15}$ [mol/mm$^3$] and $10^{-16}$ [mol/mm$^3$] respectively for illustratory purposes.\\
\vspace{0.5in}
\vspace{-0.2in}
\subsubsection{Growth factors}
Typically, growth factors exhibit short-range diffusivity within the interstitium of soft tissues. The different modes of diffusion-based transport of growth factors include (a) free diffusion, (b) hindered diffusion, and (c) facilitated diffusion \citep{fan2014}. We restrict ourselves to the free mode of diffusion wherein the molecules disperse freely from the source to the target cells. Furthermore, the action of growth factors is significantly localized, courtesy of their short half-lives. They are hence modeled with significantly low diffusivities.   

\myparagraph{Platelet-derived growth factor}
PDGF enters the arterial wall from within the $\alpha$-granules of the aggregated platelets at sites of arterial and/or endothelial injury. It is assumed to freely diffuse throughout the arterial wall. As mentioned in Section \ref{section:pathophysiology}, TGF-$\beta$ brings about autocrine secretion of PDGF by SMCs. This is reflected in a source term proportional to the local TGF-$\beta$ concentration introduced into the governing equation below. Finally, the migration and proliferation of SMCs occur at the cost of internalization of PDGF receptors post activation, which is modeled via a sink term. At high concentrations of TGF$-\beta$, fewer PDGF receptors are expressed by SMCs. This results in lower rates of PDGF consumption. This phenomenon is taken care of by introducing the scaling function $f_{{}_T}$ into the sink term (See Eq. \ref{tgf_scal}). The level of TGF$-\beta$ beyond which there is a drop in PDGF receptor expression is controlled by the threshold value $c_{{}_{T,th}}$. The particularized advection-reaction-diffusion equation hence reads
\begin{equation}\label{pdgf_bal}
    \displaystyle{\left.\frac{\partial c_{{}_P}}{\partial t}\right|_{\bm{x}}} + \text{\sf div} \left(c_{{}_P}\,\boldsymbol{v}\right) = \underbrace{\text{\sf div} \left(D_{{}_{P}}\,\text{\sf grad}\,c_{{}_{P}}\right)}_{\text{diffusion}} +  \underbrace{\eta_{{}_P} \,\rho_{{}_{S}}\, c_{{}_{T}}}_{\substack{\text{autocrine secretion}\\\text{by SMCs}}} - \underbrace{\varepsilon_{{}_P} \,f_{{}_T}\,\rho_{{}_{S}}\, c_{{}_{P}}}_{\substack{\text{receptor}\\\text{internalization}}},
\end{equation}
where $D_{{}_P}$ refers to the diffusivity of PDGF in the arterial wall. Additionally, $\eta_{{}_P}$ is termed the autocrine PDGF secretion coefficient, and $\varepsilon_P$ the PDGF receptor internalization coefficient.

\myparagraph{Transforming growth factor - $\bm{\beta}$}
Similar to PDGF, TGF-$\beta$ is also assumed to freely diffuse through the arterial wall. In contrast to PDGF, TGF-$\beta$ is not secreted by SMCs but rather by cells infiltrating the arterial wall, namely lymphocytes, monocytes and platelets. In the context of our simplified pathophysiological hypothesis, TGF-$\beta$ enters the system only via boundary conditions mimicking platelet aggregation and subsequent degranulation. The governing equation is hence particularized as

\begin{equation}\label{tgf_bal}
    \displaystyle{\left.\frac{\partial c_{{}_T}}{\partial t}\right|_{\bm{x}}} + \text{\sf div} \left(c_{{}_T}\,\boldsymbol{v}\right) = \underbrace{\text{\sf div} \left(D_{{}_{T}}\,\text{\sf grad}\,c_{{}_{T}}\right)}_{\text{diffusion}} - \underbrace{\varepsilon_{{}_T} \,\rho_{{}_{S}}\, c_{{}_{T}}}_{\substack{\text{receptor}\\\text{internalization}}},
\end{equation}
where $D_{{}_T}$ refers to the diffusivity of TGF-$\beta$ within the arterial wall, and $\varepsilon_{{}_T}$ is termed the TGF-$\beta$ receptor internalization coefficient.

\subsubsection{Extracellular matrix}
The medial layer of the arterial wall mainly contains SMCs that are densely packed into the ECM, consisting of glycoproteins including collagen, fibronectin, and elastin. Additionally, the interstitial matrix contains proteoglycans that regulate movement of molecules through the network as well as modulate the bioactivity of inflammatory mediators, growth factors and cytokines \citep{korpetinou2014}. Amongst those listed, collagen is the major constituent within the ECM which regulates cell behavior in inflammatory processes. In our modeling framework, collagen is hence considered to be the sole ingredient of ECM in the arterial wall.

Presence of PDGF within the arterial wall induces MMP production, specifically MMP-2. The signaling pathways involved in MMP production under the influence of PDGF are elucidated in \citep{cui2014}. Interstitial collagen is cleaved by MMPs via collagenolysis \citep{fields2013}, which is modeled via a sink term in the evolution equation. Collagen catabolism results in switching of SMC phenotype from quiescent to synthetic due to the loss of structural scaffolding within which the SMCs are tethered. Synthesis of collagen is exacerbated by the aforementioned phenotype switch \citep{thyberg1996}. An ECM source term, which results in a logistic evolution of collagen concentration, is introduced in this regard, and an asymptotic threshold for collagen concentration $c_{{}_{E,th}}$ prescribed. Collagen is a non motile species and hence the diffusion term is absent in the governing equation. The evolution of ECM concentration therefore reads as follows:
\begin{equation}\label{ecm_bal}
    \displaystyle{\left.\frac{\partial c_{{}_E}}{\partial t}\right|_{\bm{x}}} + \text{\sf div} \left(c_{{}_E}\,\boldsymbol{v}\right) = \underbrace{\eta_{{}_E} \rho_{{}_{S}} \left(1 - \displaystyle{\frac{c_{{}_{E}}}{c_{{}_{E,th}}}}\right)}_{\substack{\text{secretion by}\\\text{synthetic SMCs}}} - \underbrace{\varepsilon_{{}_E} \, c_{{}_{P}}\,c_{{}_{E}}}_{\substack{\text{MMP-induced}\\\text{degradation}}},
\end{equation}
where $\eta_{{}_E}$ is termed the collagen secretion coefficient, and $\varepsilon_{{}_E}$ the collagen degradation coefficient.

\begin{figure}[!htb]
    \centering
    \includegraphics[scale=0.7]{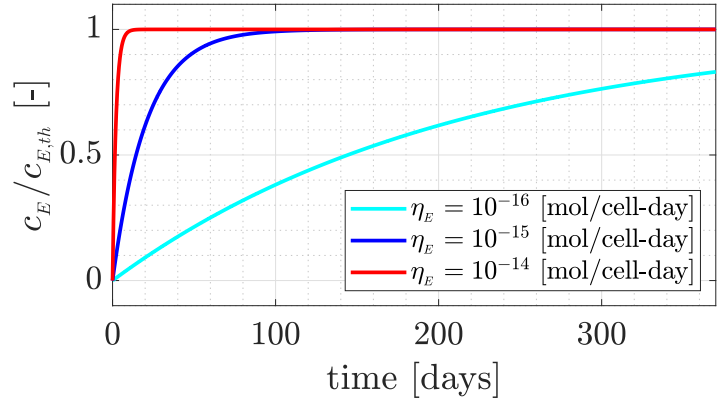}
    \caption{\textbf{Asymptotic behavior of the ECM secretion term:} The concentration of the extracellular matrix is normalized using the threshold concentration $c_{{}_{E,th}}$. The rate of collagen synthesis can be controlled via the collagen secretion coefficient $\eta_{{}_E}$. }
    \label{ecm_source}
\end{figure}

The asymptotic behavior of the source term can be realized by solving the reduced ordinary differential equation
\begin{equation}
    \displaystyle{\left.\frac{\partial c_{{}_E}}{\partial t}\right|_{\bm{x}}} = \eta_{{}_E} \rho_{{}_{S}} \left(1 - \displaystyle{\frac{c_{{}_{E}}}{c_{{}_{E,th}}}}\right)
\end{equation}
at a fixed SMC density value. The analytical solution to the above ODE is
\begin{equation}
    c_{{}_E}(t) = c_{{}_{E,th}}\,\displaystyle{\left[1 - \displaystyle{e^{-\left(\eta_{{}_E}\,\rho_{{}_S}\,t/c_{{}_{E,th}}\right)}} \right]},
\end{equation}
assuming an initially fully degraded ECM, i.e. $c_{{}_E}(t = 0) = 0$ [mol/mm$^3$]. Fig \ref{ecm_source} depicts the evolution of ECM concentration through a period of one year for varying values of the collagen secretion coefficient $\eta_{{}_E}$.

\subsubsection{Smooth muscle cells}
In a healthy homeostatic artery, the SMCs adhere to the ECM acquiring the quiescent phenotype. But they retain the ability to migrate and proliferate in response to vascular injuries \citep{clowes1983}. Injries to the vessel wall engendedr a degraded collagen environment. This results in the phenotypic modulation of medial SMCs, further leading to their migration and proliferation, thereby inducing neointimal hyperplasia. Growth factors, mainly PDGF, assist in remodeling the extracellular matrix and making it conducive for migratory and proliferative mechanisms. For details regarding the cellular signaling cascades that stem from PDGF activation of SMCs, readers are directed to \citet{gerthoffer2007} and \citet{newby2000}. The phenotypic modulation is not explicitly modeled within the current work in contrast to \citep{escuer2019}. 

\myparagraph{SMC migration}\\
Within the current modeling framework, two migratory mechanisms are to be captured, namely \textit{chemotaxis} and \textit{haptotaxis}. Both are modeled via the chemotaxis term suggested in the seminal work of \citet{keller1971}.\\
\vspace{0.02in}\\
\noindent\textit{Chemotaxis} refers to the directed migration of motile species in response to chemical stimuli. Within the medial layer of the arterial wall, SMCs experience polarized chemotactic forces due to PDGF gradients in the interstitial matrix. Also, migration of SMCs under chemotactic forces require focal adhesion sites for the extended lamellipodia to bind on to, which are supplied by a degradation in the ECM. Hence the motile sensitivity appearing in the chemotaxis term is scaled according to the local ECM concentration.\\
\vspace{0.02in}\\
\noindent\textit{Haptotaxis} is the directional migration of motile species up the gradient of focal adhesion sites. This gradient in the focal adhesion sites is indirectly captured by the gradient of degradation in the ECM. Also, PDGF is necessary to activate signaling cascades that result in extension of the lamellipodia. The mechanism is dominant only beyond a certain threshold of PDGF concentration $c_{{}_{P,th}}$ since enough lamellipodia are required to sense the disparity in focal adhesion sites and determine the direction of motility. But the lamellipodia extension quickly reaches its saturation level. Hence the motile sensitivity in the haptotaxis term is scaled according to local PDGF concentration via the scaling function $f_{{}_P}$ (See Eq. \ref{pdgf_scal}).\\

\begin{figure}[htbp]
    \centering
    \includegraphics[scale=0.24]{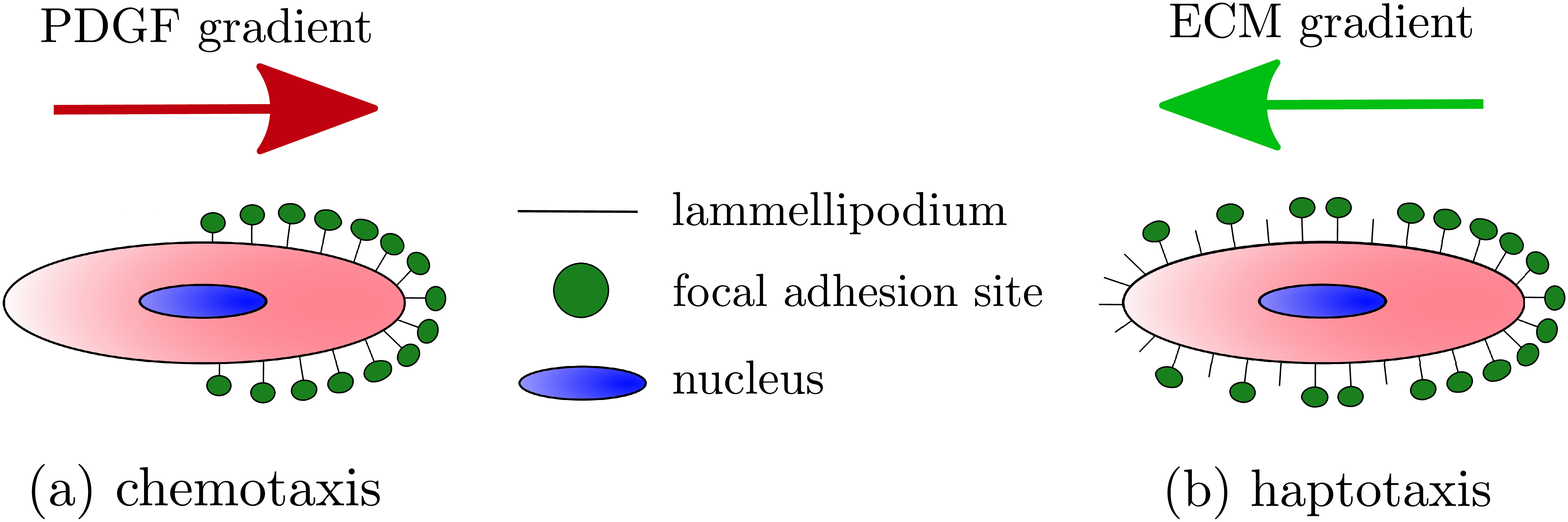}
    \caption{\textbf{SMC migration mechanisms - directional motility achieved by the induced polarity in motile forces.} \\(a) Chemotaxis is the directed migration of SMCs towards higher PDGF presence \\(b) Haptotaxis is the directed migration of SMCs down the degraded ECM pathway }
    \label{fig_motility}
\end{figure}

\myparagraph{SMC proliferation}\\
Within a degraded ECM, SMCs acquire a synthetic phenotype and hence multiply. Although the effect of PDGF on the proliferation of SMCs is nonlinear \citep{koyama1994}, a linear dependence is assumed in the current work. Hence the proliferation term in the governing equation is linearly scaled according to the local ECM and PDGF concentration. Additionally, at high concentrations of TGF-$\beta$, PDGF receptors that activate the proliferative mechanisms become scarce and hence the proliferativity of SMCs is reduced \citep{battegay1990}. This effect is captured by the introduction of the  scaling function $f_{{}_T}$ (See Eq. \ref{tgf_scal}) into the proliferation term which is dependent on the local TGF-$\beta$ concentration. This results in switching of the proliferativity of SMCs to the off state when TGF$-\beta$ concentration exceeds the threshold level $c_{{}_{T,th}}$. \\

\noindent The particularized governing equation for the SMC density is therefore formulated as
\begin{equation}\label{smc_bal}
\begin{aligned}
    \displaystyle{\left.\frac{\partial \rho_{{}_S}}{\partial t}\right|_{\bm{x}}} + \text{\sf div} \left(\rho_{{}_S}\,\boldsymbol{v}\right) = 
    & - \underbrace{\text{\sf div}\left(\chi_{{}_C} \left(1 - \displaystyle{\frac{c_{{}_{E}}}{c_{{}_{E,th}}}}\right) \,\rho_{{}_{S}} \,\text{\sf grad}\,c_{{}_{P}}\right)}_{\text{chemotaxis}} \\
    & + \underbrace{ \text{\sf div}\left(\chi_{{}_H} \,f_{{}_P} \,\rho_{{}_{S}} \,\text{\sf grad}\,c_{{}_{E}}\right)}_{\text{haptotaxis}}\\
    & + \underbrace{\eta_{{}_S}\,f_{{}_T}\,c_{{}_{P}}\, \rho_{{}_{S}} \left(1 - \displaystyle{\frac{c_{{}_{E}}}{c_{{}_{E,th}}}}\right)}_{\text{proliferation}},
\end{aligned}
\end{equation}
\noindent where $\chi_{{}_C}$ is the chemotactic sensitivity, $\chi_{{}_H}$ is the haptotactic sensitivity, and $\eta_{{}_S}$ is the SMC proliferation coefficient.\\

\subsection{Continuum mechanical modeling}
The structural behavior of the arterial wall is predominantly influenced by the medial and adventitial layers and hence only these are considered for modeling. Each layer is assumed to be composed of two families of collagen fibres embedded in an isotropic ground matrix. SMCs are assumed to be the drivers of the growth process within the isotropic ground matrix. Collagen, and hence the extracellular matrix, is assumed to strongly influence the compliance of the arterial wall.   

\subsubsection{Kinematics}\label{kinematics_section}
If $\boldsymbol{\varphi}$ is the deformation map between the reference configuration $\Omega_0$ at time time $t_0$ and the current configuration $\Omega$ at time $t$ of a continuum body, a particle at position $\boldsymbol{X}$ in the reference configuration is mapped to that at $\boldsymbol{x}$ in the current configuration via the deformation gradient $\boldsymbol{F} = \partial\,\boldsymbol{\varphi}(\boldsymbol{X},t)/\partial \bm{X}$. The right Cauchy-Green tensor is further defined as $\boldsymbol{C} = \boldsymbol{F}^T \,\boldsymbol{F} $.

For the description of growth, the well established multiplicative decomposition of the deformation gradient \citep{rodriguez1994} is adopted, i.e.
\begin{equation}
    \boldsymbol{F} = \boldsymbol{F}_e\,\boldsymbol{F}_g,
\end{equation}
wherein an intermediate incompatible configuration which achieves a locally stress-free state is assumed. Upon the polar decomposition of the growth part of the deformation gradient, i.e., $\bm{F}_g = \bm{R}_g\,\bm{U}_g$, one can write
\begin{equation}\label{F_split}
\bm{F} = \underbrace{\bm{F}_e\,\bm{R}_g}_{:=\bm{F}_{*}}\,\bm{U}_g = \bm{F}_{*}\,\bm{U}_g,
\end{equation}
where the elastic deformation gradient $\bm{F}_e$ ensures the compatibility of the total deformation in the continuum, $\bm{R}_g$ is an orthogonal tensor representing the rotational part of $\bm{F}_g$, and $\bm{U}_g$ is the right stretch tensor associated with growth. It is benefecial to define at this point the tensor residing in the reference configuration,
\begin{equation}\label{rcg_*}
\boldsymbol{C}_* = \bm{F}^T_*\,\bm{F}_* = \boldsymbol{U}_g^{-1}\, \boldsymbol{C}\, \boldsymbol{U}_g^{-1}.
\end{equation}
Based on Eq. \ref{F_split}, the volumetric change associated with the deformation gradient $\bm{F}$ is deduced to be
\begin{equation}\label{detFsplit}
    J = \text{\sf det}\,\bm{F} = J_*\,J_g, \quad J_* = \sqrt{\det \bm{C}_*}, \quad J_g = \det \bm{U}_g.
\end{equation}

\subsubsection{Helmholtz free energy}
The Helmholtz free energy per unit volume in the reference configuration $\Omega_0$ is split into an isotropic part associated with the isotropic ground matrix, and an anisotropic part corresponding to the collagen fibers, i.e.,
\begin{equation}
\psi := \psi_{iso}(\boldsymbol{C}_*,\bm{U}_g) + \psi_{ani}(\boldsymbol{C}, \boldsymbol{H}_1, \boldsymbol{H}_2, c^0_{{}_E}).  \label{hfe}
\end{equation}
Of course, we assume a simplified form in the above equation wherein the isotropic and anisotropic parts are assigned equal weightage. One can choose a more general form where the terms in the Helmholtz free energies are weighted according to the volume fractions of the associated constituents. This aspect has been extensively evaluated in the context of tissue engineered biohybrid aortic heart valves in \citet{stapleton2015}. 

In Eq. \ref{hfe}, the right stretch tensor associated with growth, i.e. $\bm{U}_g$, acts as an internal variable for only the isotropic part of the Helmholtz free energy which is dependent on $\bm{C}_*$. This is based on the assumption that SMCs are the main drivers for growth, and they are considered a part of the isotropic ground matrix. On the other hand, the anisotropic part of the Helmholtz free energy is assumed to be dependent on the full $\bm{C}$ since any stretch associated with growth can still stretch the collagen fibers.

The specific choice for the isotropic part is assumed to be of Neo-Hookean form, given by
\begin{equation}
\psi_{iso} (\bm{C}_*,\bm{U}_g) = \displaystyle{\frac{\mu}{2}}\left(\text{tr}\,\boldsymbol{C}_* - 3\right) - \mu\,\text{ln}\,J_* + \displaystyle{\frac{\Lambda}{4}}\left(J_*^2 - 1 - 2\,\text{ln}\,J_* \right),
\label{iso_hfe}
\end{equation} 
where the definition of $\bm{C}_*$ from Eq. \ref{rcg_*}, and that of $J_*$ from Eq. \ref{detFsplit} are utilized.
The anisotropic part is particularized to be of exponential form \citep{holzapfel2000} as
\begin{equation}
\psi_{ani} (\boldsymbol{C}, \boldsymbol{H}_1, \boldsymbol{H}_2, c^0_{{}_E}) = \displaystyle{\frac{k_1}{2k_2}}\sum_{i=1,2} \left(\text{\sf exp}\left[k_2\langle E_i\rangle ^2 \right]-1\right).
\label{aniso_hfe}
\end{equation}
The stress-like material parameter $k_1$, introduced above, is here designed to be a linear function of the local ECM concentration in the reference configuration $c^{0}_{{}_E}$, i.e.,
\begin{equation}
    k_1 := \bar{k}_1 \,\displaystyle{\left(\frac{c^{0}_{{}_E}}{c_{{}_{E,eq}}}\right)},
\end{equation}
$\bar{k}_1$ being the stress-like material parameter for healthy collagen, and $c_{{}_{E,eq}}$ referring to the homeostatic ECM concentration in a healthy artery.

In Eq. \ref{aniso_hfe},  $\boldsymbol{H}_i$ ($i = 1,2$) are the generalized structural tensors constructed from the local collagen orientations $\boldsymbol{a}_{0i}$ in the reference configuration using the following relation,
\begin{eqnarray}
\boldsymbol{H}_i &:=& \kappa\,\boldsymbol{I} + \left(1 - 3\,\kappa \right)\,\boldsymbol{a}_{0i} \otimes \boldsymbol{a}_{0i}\label{struct_tens2}\,\,,
\end{eqnarray}
where $\kappa$ is a dispersion parameter \citep{gasser2006} accounting for a von Mises distribution of collagen orientations. The Green-Lagrange strain $E_i$ is calculated from the right Cauchy-Green tensor $\bm{C}$ utilizing the relationship 
\begin{equation}\label{GL_strain}
    E_i := \boldsymbol{H}_i : \boldsymbol{C} - 1,
\end{equation}
wherein the definition of the scalar product of second order tensors $\boldsymbol{A}:\boldsymbol{B} = A_{ij}\,B_{ij}$ (Einstein summation convention) is applied. The Macaulay brackets around $E_i$ in Eq. \ref{aniso_hfe} ensure that the fibers are activated only in tension and hence only the positive part of the strain is considered within the free energy.

\subsubsection{Growth theories}\label{section_growth_theories}
Two separate growth theories are proposed in this section based on two histological cases. 
\vspace{-0.1in}
\begin{enumerate}
\item[(a)] stress-free anisotropic growth\\ \textbf{histological case:}
two \textit{distinct collagen orientations} with \textit{negligible dispersion}  
\item[(b)] isotropic matrix growth\\ \textbf{histological case:} two \textit{diffuse collagen orientations} with \textit{high dispersion}
\end{enumerate}
\pagebreak

\myparagraph{(a) Stress-free anisotropic growth}

\noindent A stress-free incompatible grown state can be formulated if we assume that the orientations of collagen fibers lack any dispersion. Mathematically, if $\kappa = 0$, Eq. \ref{struct_tens2} boils down to the simple form 
\begin{equation}\label{struct_tens_zero_dispersion}
    {\bm{H}}_i = \bm{a}_{0i}\otimes\bm{a}_{0i}.
\end{equation}
\citet{Lubarda2002OnTM} suggest a form of $\bm{F}_g$ for transversely isotropic mass growth, given by
\begin{equation}\label{F_g-lubarda}
\bm{F}_g = \vartheta_2\,\bm{I} + (\vartheta_1 - \vartheta_2)\,\bm{\gamma} \otimes \bm{\gamma},
\end{equation}
wherein $\vartheta_1$ is the stretch in the direction of the fibers ($\bm{\gamma}$), and $\vartheta_2$ is the stretch associated with any direction orthogonal to $\bm{\gamma}$. In our case, it is intuitive to assume that the growth takes place in a direction orthogonal to the plane containing $\bm{a}_{01}$ and $\bm{a}_{02}$ to achieve a stress-free state. Based on Eq. \ref{F_g-lubarda}, $\bm{U}_g$ is now suggested to be
\begin{equation}\label{aniso_growth}
    \bm{U}_g := \bm{I} + (\vartheta - 1)\, \bm{\gamma}\otimes \bm{\gamma} ,
\end{equation}
 where $\vartheta$ is the growth stretch, and $\bm{\gamma}$ is the unit vector in the direction of presumed growth given by
\begin{equation}\label{aniso_growth_orientation}
    \bm{\gamma} := \frac{\bm{a}_{01} \times \bm{a}_{02}}{||\bm{a}_{01} \times \bm{a}_{02}||}.
\end{equation}
In the above equation, $|| (\bullet) ||$ refers to the $L^2$ norm of the vector $(\bullet)$. The growth stretch, formulated under the assumption of preservation of SMC density, is given by
\begin{equation}\label{growth_stretch_aniso}
    \vartheta := \frac{\rho^0_{{}_S}}{\rho_{{}_S,eq}},
\end{equation}
where $\rho^0_{{}_S}$ is the SMC density in the reference configuration and $\rho_{{}_{S,eq}}$ is the homeostatic SMC density of a healthy artery. 

\myparagraph{(b) Isotropic matrix growth}

\noindent In the presence of dispersed collagen fibres, a stress-free grown state is unobtainable since at least some of the collagen fibres are inevitably stretched under any kind of anisotropic growth assumption. We then resort to the simplest isotropic form of growth of the matrix, i.e., 
\begin{equation}\label{iso_growth}
   \bm{U}_g := \vartheta\,\bm{I} ,
\end{equation}
where $\vartheta$ is the growth stretch and $\bm{I}$ the second order identity tensor. The growth stretch can again be formulated under the assumption of preservation of SMC density as in Eq. \ref{growth_stretch_aniso} to be
\begin{equation}\label{growth_stretch_iso}
   \vartheta := \left(\displaystyle{\frac{\rho^{0}_{{}_S}}{\rho_{{}_{S,eq}}}}\right)^{1/3}.
\end{equation}
Clearly, the grown tissue is not stress-free in this case.

\paragraph{Remark:} Within this work, we restrict ourselves to growth formulations which have the form of $\bm{U}_g$ directly prescribed. This renders the evolution of $\bm{U}_g$ directly dependent on the governing PDE for the evolution of SMC density. More general and elaborate continuum mechanical models for growth and remodeling of soft biological tissues can be derived utilizing the framework for modeling anisotropic inelasticity via structural tensors, introduced in \citet{reese2021}. The anisotropic growth formulation developed in \citet{lamm2022} is also relevant in this regard wherein the growth is  stress-driven. 

\subsection{Boundary and initial conditions}\label{bc}

\begin{table}[htbp!]\
\centering
\caption{\textbf{Boundary conditions}}
\label{table_bc} 
\begin{tabular}{p{1.cm}p{3.5cm}p{9.5cm}}
\noalign{\hrule height 0.05cm}\noalign{\smallskip}
variable & Dirichlet & Neumann  \\
\noalign{\hrule height 0.05cm}\noalign{\smallskip}
 $c_{{}_P}$ & \(c_{{}_P} = \hat{c}_{{}_P}\quad \text{on} \quad \Gamma^D_P\) &  \(\bm{q}_{{}_P}\cdot\bm{n} = - D_{{}_P}\,\text{\sf grad}\,(c_{{}_P})\cdot\bm{n} = \bar{q}_{{}_P}(c_{{}_P}) =  p_{{}_{en}} \,(\bar{c}_{{}_P} - c_{{}_P})\quad \text{on} \quad \Gamma^N_{{}_{GF}}\)\\
 \\
$c_{{}_T}$ & \(c_{{}_T} = \hat{c}_{{}_T}\quad \text{on} \quad \Gamma^D_T\) &  \(\bm{q}_{{}_T}\cdot\bm{n} =- D_{{}_T}\,\text{\sf grad}\,(c_{{}_T})\cdot\bm{n} = \bar{q}_{{}_T}(c_{{}_T}) =  p_{{}_{en}} \,(\bar{c}_{{}_T} - c_{{}_T})\quad \text{on} \quad \Gamma^N_{{}_{GF}}\)\\
\\
$c_{{}_E}$ & $-$ &  \(\text{\sf grad}\,(c_{{}_E})\cdot\bm{n} = 0 \quad \text{on} \quad \Gamma\)\\
\\
$\rho_{{}_S}$ & $-$ &  \(\text{\sf grad}\,(\rho_{{}_S})\cdot\bm{n} = 0 \quad \text{on} \quad \Gamma\)\\
\\
$\bm{u}$ & $\bm{u} = \hat{\bm{u}} \quad \text{on}\quad \Gamma_{0,u}$ & \(\bm{T} = \bm{P}\cdot\bm{N} = \hat{\bm{T}}\quad \text{on}\quad \Gamma_{0,T}\)\\
\noalign{\hrule height 0.05cm}\noalign{\smallskip}
\end{tabular}
\end{table}
\vspace{-0.1in}
All the relevant boundary conditions are summarized in Table \ref{table_bc}. PDGF and TGF-$\beta$ enter the arterial wall as a consequence of platelet aggregation. This effect can be modeled by prescribing influxes along the normal $\bm{n}$ at the injury sites on the vessel wall $\Gamma^N_{{}_{GF}}$. These influxes can directly be prescribed as time-varying profiles or as functions of the wall shear stresses $\bm{\tau}$ observed at the endothelium \citep{hossain2012}. $p_{{}_{en}} = p_{{}_{en}}(\bm{\tau})$ refers to the wall shear stress dependent permeability of the injured regions of the vessel wall. Concentration profiles can also be directly prescribed on the Dirichlet boundaries $\Gamma^D_P$ and $\Gamma^D_T$ for PDGF and TGF-$\beta$ respectively. The boundary in the current configuration is therefore $\Gamma = \Gamma^N_{{}_{GF}} \cup \Gamma^D_{{}_{GF}}$.

The ECM and the SMCs are considered to be restrained within the arterial wall and hence zero flux boundary conditions are prescribed on the entire boundary of the system $\Gamma$.

Displacements are prescribed on the boundary $\Gamma_{{0,u}}$ in the reference configuration, and tractions on the boundary $\Gamma_{{0,T}}$ in the reference configuration. Also, the total boundary in the reference configuration $\Gamma_{0} = \Gamma_{{0,u}} \cup \Gamma_{{0,T}}$.

The initial ECM concentrations and SMC densities are prescribed to be those of a healthy homeostatic artery in equilibrium. PDGF and TGF$-\beta$ are considered initially absent in the vessel wall. Table \ref{table_ic} summarizes the relevant initial conditions.

\begin{table}[htbp!]\
\centering
\caption{\textbf{Initial conditions}}
\label{table_ic} 
\begin{tabular}{p{2.cm}p{2.5cm}}
\noalign{\hrule height 0.05cm}\noalign{\smallskip}
variable &  initial condition\\
 & (\(\forall \bm{x} \in \Omega\))\\
\noalign{\hrule height 0.05cm}\noalign{\smallskip}
$c_{{}_P}$ & $0$\\
\\
$c_{{}_T}$ & $0$ \\
\\
$c_{{}_E}$ & $c_{{}_{E,eq}}$\\
\\
$\rho_{{}_S}$ & $\rho_{{}_{S,eq}}$\\
\\
\noalign{\hrule height 0.05cm}\noalign{\smallskip}
\end{tabular}
\end{table}
\vspace{-0.1in}

\section{Finite element implementation}\label{fe_impl}
Eqs. \ref{pdgf_bal}, \ref{tgf_bal}, \ref{ecm_bal} and \ref{smc_bal} describe the transport of species in the arterial wall in an Eulerian setting. It is fairly common in the fluid mechanics community to adopt the Eulerian description since the flow velocity $\bm{v}$ is one of the primary variables in the governing PDEs for fluid flow (e.g., Navier-Stokes equations). In contrast, displacements serve as the primary variable in structural mechanical balance equations (balance of linear momentum in the current case). Terms involving the velocity $\bm{v}$ therefore have to be deduced by approximating the time derivatives of either the displacements or deformation gradients. Errors in such approximations can propagate through the solutions, and can in some cases lead to instabilities. Additionally, a concrete fluid carrier that transports the wall constituents is absent in the current framework. The bulk of the soft tissue is itself the transport medium, and hence lacks flow complexities like flow reversals and vortices where the Eulerian description has proven itself to be most beneficial. It is hence favorable to convert all the aforementioned equations to the Lagrangian description, which has been shown to be accurate in the presence of moving boundaries and complex geometries.

\subsection{Strong forms}
The equations which are transformed from the Eulerian to the Lagrangian setting read 

\begin{eqnarray}
    \displaystyle{\left.\frac{\partial c^0_{{}_P}}{\partial t}\right|_{\bm{X}}} &=& \text{\sf Div} \left(D_{{}_{P}}\,\bm{C}^{-1}\,\text{\sf Grad}\,c^0_{{}_{P}}\right)  - \text{\sf Div} \left(D_{{}_{P}}\,\frac{c^0_{{}_P}}{J}\,\bm{C}^{-1}\,\text{\sf Grad}\,J\right) \nonumber\\
    \nonumber\\
    &+&  \frac{\eta_{{}_P}}{J} \,\rho^0_{{}_{S}}\, c^0_{{}_{T}} - \frac{\varepsilon_{{}_P}}{J} \,\left( \frac{1}{1 + e^{l_{{}_T}\left(c^0_{{}_T}\,J^{-1} - c_{{}_{T,th}}\right)}}\right)\, \rho^0_{{}_{S}}\, c^0_{{}_{P}},
\end{eqnarray} 
\begin{eqnarray}    
    \displaystyle{\left.\frac{\partial c^0_{{}_T}}{\partial t}\right|_{\bm{X}}} &=& \text{\sf Div} \left(D_{{}_{T}}\,\bm{C}^{-1}\,\text{\sf Grad}\,c^0_{{}_{T}}\right) - \text{\sf Div} \left(D_{{}_{T}}\,\frac{c^0_{{}_T}}{J}\,\bm{C}^{-1}\,\text{\sf Grad}\,J\right) - \frac{\varepsilon_{{}_T}}{J} \,\rho^0_{{}_{S}}\, c^0_{{}_{T}},    \\ 
    \nonumber\\
    \nonumber\\
    \displaystyle{\left.\frac{\partial c^0_{{}_E}}{\partial t}\right|_{\bm{X}}} &=& \eta_{{}_E} \rho^0_{{}_{S}} \left(1 - \displaystyle{\frac{c^0_{{}_{E}}}{J\,c_{{}_{E,th}}}}\right) - \frac{\varepsilon_{{}_E}}{J} \, c^0_{{}_{P}}\,c^0_{{}_{E}},\\ 
    \nonumber\\
    \nonumber\\
    \displaystyle{\left.\frac{\partial \rho^0_{{}_S}}{\partial t}\right|_{\bm{X}}} = 
     &-& \text{\sf Div}\left(\frac{\chi_{{}_C}}{J} \left(1 - \displaystyle{\frac{c^0_{{}_{E}}}{J\,c_{{}_{E,th}}}}\right) \,\rho^0_{{}_{S}} \,\bm{C}^{-1}\,\text{\sf Grad}\,c^0_{{}_{P}}\right) \nonumber\\
     \nonumber\\
     &+& \text{\sf Div} \left(\frac{\chi_{{}_C}}{J} \left(1 - \displaystyle{\frac{c^0_{{}_{E}}}{J\,c_{{}_{E,th}}}}\right)\,\rho^0_{{}_{S}}\,\frac{c^0_{{}_P}}{J}\,\bm{C}^{-1}\,\text{\sf Grad}\,J\right) \nonumber\\
    \nonumber\\
     &+& \text{\sf Div}\left(\frac{\chi_{{}_H}}{J} \left( \frac{1}{1 + e^{-l_{{}_P}\left(c^0_{{}_P}\,J^{-1} - c_{{}_{P,th}}\right)}}\right) \,\rho^0_{{}_{S}} \,\bm{C}^{-1}\,\text{\sf Grad}\,c^0_{{}_{E}}\right) \nonumber\\
    \nonumber\\
    &-& \text{\sf Div}\left(\frac{\chi_{{}_H}}{J} \left( \frac{1}{1 + e^{-l_{{}_P}\left(c^0_{{}_P}\,J^{-1} - c_{{}_{P,th}}\right)}}\right) \,\rho^0_{{}_{S}} \,\frac{c^0_{{}_E}}{J}\,\bm{C}^{-1}\,\text{\sf Grad}\,J\right) \nonumber\\
    \nonumber\\
     &+& \frac{\eta_{{}_S}}{J^2}\,c^0_{{}_{P}} \rho^0_{{}_{S}} \left(1 - \displaystyle{\frac{c^0_{{}_{E}}}{J\,c_{{}_{E,th}}}}\right)\,\left(\frac{1}{1 + e^{l_{{}_{T}}(c^0_{{}_T}\,J^{-1} - c_{{}_{T,th}})}}\right).
\end{eqnarray}
Here, $(\bullet)^0 = J\,(\bullet)$ refer to the species variables in the reference configuration. The interested reader is referred to \ref{appendix:etl} for details regarding the transfer of quantities from the Eulerian to the Lagrangian description. Finally, the balance of linear momentum governing the quasi-static equilibrium of the arterial wall structure reads
\begin{equation}\label{mom_bal}
{\sf Div}\, \boldsymbol{P} + \boldsymbol{B} = \boldsymbol{0},
\end{equation}
where $\bm{B}$ is the body force vector. The first Piola-Kirchhoff stress tensor $\bm{P}$ is deduced from the Helmholtz free energy function by imposing the fulfilment of the second law of thermodynamics and subsequently applying the Coleman-Noll procedure \citep{coleman_noll1963}, leading to
\begin{equation}
\boldsymbol{P} = \displaystyle{\frac{\partial \psi}{\partial \boldsymbol{F}}}.
\end{equation}

\subsection{Weak forms}

Further, the aforementioned strong forms along with the balance of linear momentum in Eq. \ref{mom_bal} are converted to their respective weak forms by multiplying the terms with the test functions $\delta c^0_{{}_P}$,  $\delta c^0_{{}_T}$, $\delta c^E_{{}_0}$, $\delta \rho^0_{{}_S}$, and $\delta \bm{u}$ and integrating over the continuum domain in the reference configuration. Evaluating the integrals by parts and utilizing the Gauss divergence theorem for the terms involving the divergence operators, one arrives at the residual equations which read 

\begin{eqnarray}
    g_{{}_P} &:=& 
    \left\{\begin{array}{c}
        \displaystyle{\int_{\Omega_0}} \left[\dot{c^0_{{}_P}}  - \frac{\eta_{{}_P}}{J} \,\rho^0_{{}_{S}}\,c^0_{{}_{T}} + \frac{\varepsilon_{{}_P}}{J} \,\left( \frac{1}{1 + e^{l_{{}_T}\left(c^0_{{}_T}\,J^{-1} - c_{{}_{T,th}}\right)}}\right)\,\rho^0_{{}_{S}}\, c^0_{{}_{P}}\right]     \delta c^0_{{}_P}\, dV\\
        \vspace{0.05in}\\
        + \displaystyle{\int_{\Omega_0}} D_{{}_P}\,\text{\sf Grad}^T\,(c^0_{{}_P}) \, \bm{C}^{-1}\,\text{\sf Grad}\,(\delta c^0_{{}_P})\, dV\\
        \vspace{0.05in}\\
        - \displaystyle{\int_{\Omega_0}} D_{{}_P}\,\frac{c^0_{{}_P}}{J}\,\text{\sf Grad}^T\,(J) \, \bm{C}^{-1}\,\text{\sf Grad}\,(\delta c^0_{{}_P})\, dV\\
        \vspace{0.05in}\\
        - \displaystyle{\int_{\Gamma^N_{0,P}}} \left(D_{{}_P}\,\bm{C}^{-1}\,\text{\sf Grad} (c^0_{{}_P})\cdot\bm{N}\right)\,\delta c^0_{{}_P} dA\\
    \end{array}\right\}  = 0, \nonumber\\\label{wf_p}
\end{eqnarray}

\begin{eqnarray}
    g_{{}_T} &:=& 
    \left\{\begin{array}{c}
        \displaystyle{\int_{\Omega_0}} \left[\dot{c^0_{{}_T}} + \frac{\varepsilon_{{}_P}}{J} \,\rho^0_{{}_{S}}\, c^0_{{}_{T}}\right]\delta c^0_{{}_T}\, dV\\
        \vspace{0.05in}\\
        + \displaystyle{\int_{\Omega_0}} D_{{}_T}\,\text{\sf Grad}^T\,(c^0_{{}T}) \, \bm{C}^{-1}\,\text{\sf Grad}\,(\delta c^0_{{}_T})\, dV\\
        \vspace{0.05in}\\
        - \displaystyle{\int_{\Omega_0}} D_{{}_T}\,\frac{c^0_{{}_T}}{J}\,\text{\sf Grad}^T\,(J) \, \bm{C}^{-1}\,\text{\sf Grad}\,(\delta c^0_{{}_T})\, dV\\
        \vspace{0.05in}\\
        - \displaystyle{\int_{\Gamma^N_{0,T}}} \left(D_{{}_T}\,\bm{C}^{-1}\,\text{\sf Grad} (c^0_{{}_T})\cdot\bm{N}\right)\,\delta c^0_{{}_T} dA\\
    \end{array}\right\}  = 0, \label{wf_t}
\end{eqnarray}
\vspace{0.5in}
\begin{eqnarray}
    g_{{}_E} &:=& 
        \displaystyle{\int_{\Omega_0}} \left[\dot{c^0_{{}_E}}  - \eta_{{}_E} \rho^0_{{}_{S}} \left(1 - \displaystyle{\frac{c^0_{{}_{E}}}{J\,c_{{}_{E,th}}}}\right) +  \frac{\varepsilon_{{}_E}}{J} \, c^0_{{}_{P}}\,c^0_{{}_{E}} \right]\delta c^0_{{}_E} dV = 0, \label{wf_e} 
\end{eqnarray}
\\
\vspace{1in}\\
\\
\begin{eqnarray}
    g_{{}_S} &:=& 
    \left\{\begin{array}{c}
        \displaystyle{\int_{\Omega_0}} \left[\dot{\rho^0_{{}_S}} - \frac{\eta_{{}_S}}{J^2}\,c^0_{{}_{P}} \rho^0_{{}_{S}} \left(1 - \displaystyle{\frac{c^0_{{}_{E}}}{J\,c_{{}_{E,th}}}}\right)\,\left(\frac{1}{1 + e^{l_{{}_{T}}(c^0_{{}_T}\,J^{-1} - c_{{}_{T,th}})}}\right)\right]\delta \rho^0_{{}_S} \,dV\\
        \vspace{0.1in}\\
        + \displaystyle{\int_{\Omega_0}} \frac{\chi_{{}_H}}{J} \left( \frac{1}{1 + e^{-l_{{}_P}\left(c^0_{{}_P}\,J^{-1} - c_{{}_{P,th}}\right)}}\right) \,\rho^0_{{}_{S}} \,\text{\sf Grad}^T (c^0_{{}_{E}})\, \bm{C}^{-1}\,\text{\sf Grad}\,(\delta \rho^0_{{}_S}) \,dV\\
        \vspace{0.1in}\\
        - \displaystyle{\int_{\Omega_0}} \frac{\chi_{{}_H}}{J} \left( \frac{1}{1 + e^{-l_{{}_P}\left(c^0_{{}_P}\,J^{-1} - c_{{}_{P,th}}\right)}}\right) \,\rho^0_{{}_{S}}\,\frac{c^0_{{}_E}}{J} \,\text{\sf Grad}^T (J)\, \bm{C}^{-1}\,\text{\sf Grad}\,(\delta \rho^0_{{}_S}) \,dV\\
        \vspace{0.1in}\\
        - \displaystyle{\int_{\Omega_0}} \frac{\chi_{{}_C}}{J}\,\left(1 - \displaystyle{\frac{c^0_{{}_{E}}}{J\,c_{{}_{E,th}}}}\right) \,\rho^0_{{}_{S}} \,\text{\sf Grad}^T (c^0_{{}_{P}})\, \bm{C}^{-1}\,\text{\sf Grad}\,(\delta \rho^0_{{}_S})\, dV\\
        \vspace{0.1in}\\
        + \displaystyle{\int_{\Omega_0}} \frac{\chi_{{}_C}}{J}\,\left(1 - \displaystyle{\frac{c^0_{{}_{E}}}{J\,c_{{}_{E,th}}}}\right) \,\rho^0_{{}_{S}}\,\frac{c^0_{{}_P}}{J} \,\text{\sf Grad}^T (J)\, \bm{C}^{-1}\,\text{\sf Grad}\,(\delta \rho^0_{{}_S})\, dV\\
    \end{array}\right\}  = 0,  \nonumber\\
    \label{wf_s}
\end{eqnarray}
\begin{eqnarray}
    g_{{}_u} &:=& \displaystyle{\int_{\Omega_0}} \bm{P}:\delta \bm{F}\, dV - \displaystyle{\int_{\Omega_0}} (\bm{B}\cdot\delta\bm{u})\, dV - \displaystyle{\int_{\Gamma_{0,T}}} \bm{T}\cdot\delta\bm{u}\,dA = 0.\label{wf_u}
\end{eqnarray}
The material time derivatives of the species are referred to using the notation $\dot{(\bullet)}$ in the above equations. Additionally, $\Gamma_0$ refers to the boundary surfaces of the domain, $\Gamma^N_{0,(\bullet)}$ refers to the Neumann boundaries for the respective wall species $(\bullet)$, and $\bm{N}$ is the normal to the respective Neumann boundaries in the reference configuration. Flux terms are absent in the equations for ECM and SMCs since zero flux boundary conditions are assumed (See Section \ref{bc}).  
\subsection{Temporal discretization}\label{section_temporal_discretization}
The material time derivatives appearing in the evolution equations for the species in the arterial wall are obtained using the backward Euler method. Two variations shall be implemented in this regard. All the terms on the right side of the evolution equations are grouped and denoted as the functions $f_{{}_{(\bullet)}}$. Variables with subscripts $n$ and $n+1$ indicate those at times step $t_{n}$ and time step $t_{n+1}$ respectively.
\subsubsection{Fully-implicit backward Euler method}
Here, all the field variables are modeled with implicit dependence i.e., all the $f_{{}_{(\bullet)}}$ are implicit functions of the field variables. Hence the temporally discretized weak forms attain the format
\begin{eqnarray}
    \dot{c^0_{{}_P}} &=& \frac{(c^0_{{}_P})_{n+1} - (c^0_{{}_P})_{n}}{\Delta t} = f_{{}_P}\left((c^0_{{}_P})_{n+1},(c^0_{{}_T})_{n+1},(c^0_{{}_E})_{n+1},(\rho^0_{{}_S})_{n+1}\right) \nonumber\\
    \dot{c^0_{{}_T}} &=& \frac{(c^0_{{}_T})_{n+1} - (c^0_{{}_T})_{n}}{\Delta t} = f_{{}_T}\left((c^0_{{}_P})_{n+1},(c^0_{{}_T})_{n+1},(c^0_{{}_E})_{n+1},(\rho^0_{{}_S})_{n+1}\right) \nonumber\\
    \dot{c^0_{{}_E}} &=& \frac{(c^0_{{}_E})_{n+1} - (c^0_{{}_E})_{n}}{\Delta t} = f_{{}_E}\left((c^0_{{}_P})_{n+1},(c^0_{{}_T})_{n+1},(c^0_{{}_E})_{n+1},(\rho^0_{{}_S})_{n+1}\right) \nonumber\\
    \dot{\rho^0_{{}_S}} &=& \frac{(\rho^0_{{}_S})_{n+1} - (\rho^0_{{}_S})_{n}}{\Delta t} = f_{{}_S}\left((c^0_{{}_P})_{n+1},(c^0_{{}_T})_{n+1},(c^0_{{}_E})_{n+1},(\rho^0_{{}_S})_{n+1}\right).\label{fi_be}
\end{eqnarray}
\subsubsection{Semi-implicit backward Euler method}
Here only the variables that are temporally discretized in the respective weak form equations are modeled with implicit dependence. The $f_{{}_{(\bullet)}}$ are therefore explicit functions of the rest of the field variables. Hence the temporally discretized weak forms attain the format
\begin{eqnarray}
    \dot{c^0_{{}_P}} &=& \frac{(c^0_{{}_P})_{n+1} - (c^0_{{}_P})_{n}}{\Delta t} = f_{{}_P}\left((c^0_{{}_P})_{n+1},(c^0_{{}_T})_{n},(c^0_{{}_E})_{n},(\rho^0_{{}_S})_{n}\right)\nonumber\\
    \dot{c^0_{{}_T}} &=& \frac{(c^0_{{}_T})_{n+1} - (c^0_{{}_T})_{n}}{\Delta t} = f_{{}_T}\left((c^0_{{}_P})_{n},(c^0_{{}_T})_{n+1},(c^0_{{}_E})_{n},(\rho^0_{{}_S})_{n}\right)\nonumber\\
    \dot{c^0_{{}_E}} &=& \frac{(c^0_{{}_E})_{n+1} - (c^0_{{}_E})_{n}}{\Delta t} = f_{{}_E}\left((c^0_{{}_P})_{n},(c^0_{{}_T})_{n},(c^0_{{}_E})_{n+1},(\rho^0_{{}_S})_{n}\right)\nonumber\\
    \dot{\rho^0_{{}_S}} &=& \frac{(\rho^0_{{}_S})_{n+1} - (\rho^0_{{}_S})_{n}}{\Delta t} = f_{{}_S}\left((c^0_{{}_P})_{n},(c^0_{{}_T})_{n},(c^0_{{}_E})_{n},(\rho^0_{{}_S})_{n+1}\right).\label{si_be}
\end{eqnarray}

\subsection{Spatial discretization}
Eqs. \ref{wf_p}, \ref{wf_t}, \ref{wf_e}, \ref{wf_s}, and  \ref{wf_u} are linearized about the states at $t_{n+1}$ (See \ref{appendix:lwf}). The computational domain in the reference configuration is spatially approximated via finite elements, i.e.,
\begin{equation}
    \Omega_0 \approx \displaystyle{\bigcup^{n_{e}}_{i = 1}}\, \Omega^e_0.
\end{equation}
The solution variables $\langle\bullet\rangle$ and their variations $\delta \langle\bullet\rangle$ are discretized using the isoparametric concept via tri-linear Lagrange shape functions as follows:
\begin{eqnarray}\label{interp}
    \langle\bullet\rangle(\bm{X}) &\approx& \langle\bullet\rangle^{h}(\bm{X}) = \bm{N}^L(\xi,\eta,\zeta)\cdot \langle\bullet\rangle^e\nonumber\\
    \delta \langle\bullet\rangle(\bm{X}) &\approx& \delta \langle\bullet\rangle^h(\bm{X}) = \bm{N}^L(\xi,\eta,\zeta))\cdot\delta \langle\bullet\rangle^e, \quad \forall \bm{X} \in \Omega_0^e, 
\end{eqnarray}
where $\bm{N}^L$ are Lagrange shape function vectors expressed in terms of the isoparametric coordinates $\xi$, $\eta$, and $\zeta$, and $\langle\bullet\rangle^e$ are the vectors containing the nodal values of the element. The gradients of the species variables and their variations are evaluated using the derivatives of the shape functions accumulated in the matrix $\bm{B}$ via the relations
\begin{eqnarray}
    \text{\sf Grad}\, \langle\bullet\rangle (\bm{X}) &\approx& \text{\sf Grad}^h\, \langle\bullet\rangle (\bm{X}) = \bm{B}(\xi,\eta,\zeta)\, \langle\bullet\rangle^e\nonumber\\
    \text{\sf Grad}\, \delta \langle\bullet\rangle (\bm{X}) &\approx& \text{\sf Grad}^h\, \delta \langle\bullet\rangle (\bm{X}) = \bm{B}(\xi,\eta,\zeta)\, \delta\langle\bullet\rangle^e,\quad \forall \bm{X} \in \Omega_0^e. \label{grad_s}
\end{eqnarray}
The gradient of the displacement field is calculated using the matrix $\bm{B}_{{}_U}$ wherein the derivatives of the shape functions are assembled in a different form and according to the arrangement of nodal values in the element displacement vector $\bm{U}^e$. Therefore
\begin{eqnarray}
    \text{\sf Grad}\, \bm{u} (\bm{X}) &\approx& \text{\sf Grad}^h\, \bm{u} (\bm{X}) = \bm{B}_{{}_u}(\xi,\eta,\zeta)\, \bm{U}^e\nonumber\\
    \text{\sf Grad}\, \delta\bm{u} (\bm{X}) &\approx& \text{\sf Grad}^h\, \delta\bm{u} (\bm{X}) = \bm{B}_{{}_u}(\xi,\eta,\zeta)\, \delta\bm{U}^e, \quad \forall \bm{X} \in \Omega_0^e. \label{grad_u}
\end{eqnarray}
Substituting Eqs. \ref{interp}, \ref{grad_s} and \ref{grad_u} into the linearized weak form (See \ref{appendix:lwf}), two forms of system systiffness matrices are obtained for the two types of temporal discretizations elucidated in Eqs. \ref{fi_be} and \ref{si_be}.

\paragraph{Fully-implicit backward Euler method} results in a fully coupled system of linear equations at the element level, the stiffness matrix for which reads
\begin{equation}\label{eq_mono}
\begin{bmatrix}
    \bm{K}^e_{{}_{PP}} & \bm{K}^e_{{}_{PT}} & \bm{0} & \bm{K}^e_{{}_{PS}} &\bm{K}^e_{{}_{Pu}}\\
    \bm{0} & \bm{K}^e_{{}_{TT}} & \bm{0} & \bm{K}^e_{{}_{TS}} & \bm{K}^e_{{}_{Tu}} \\
    \bm{K}^e_{{}_{EP}} & \bm{0} & \bm{K}^e_{{}_{EE}} & \bm{K}^e_{{}_{ES}} & \bm{K}^e_{{}_{Eu}} \\
    \bm{K}^e_{{}_{SP}} & \bm{K}^e_{{}_{ST}} & \bm{K}^e_{{}_{SE}} & \bm{K}^e_{{}_{SS}} & \bm{K}^e_{{}_{Su}} \\
    \bm{0} & \bm{0} & \bm{K}^e_{{}_{uE}} & \bm{K}^e_{{}_{uS}} & \bm{K}^e_{{}_{uu}}
\end{bmatrix}.
\end{equation}
The resulting assembled global system of equations is hence unsymmetric, and forms the monolithic construct. 

\paragraph{Semi-implicit backward Euler method} results in a decoupled systems of linear equations at the element level. The stiffness matrix for the subsystem of equations for the species in the arterial wall is hence a block diagonal matrix and reads
\begin{equation}\label{eq_staggered}
\begin{bmatrix}
    \bm{K}^e_{{}_{PP}} & \bm{0} & \bm{0} & \bm{0}  \\
    \bm{0} & \bm{K}^e_{{}_{TT}} & \bm{0} & \bm{0}  \\
    \bm{0} & \bm{0} & \bm{K}^e_{{}_{EE}} & \bm{0}  \\
    \bm{0} & \bm{0} & \bm{0} & \bm{K}^e_{{}_{SS}}  
\end{bmatrix},
\end{equation}
and that for the displacement field is $\bm{K}^e_{{}_{uu}}$.
The resulting assembled global system of equations for the wall species is symmetric. Additionally, due to the semi-implicitness of the temporal discretization of the species variables and the linearity of the terms involved in the decoupled equations, the associated subsystem is devoid of nonlinearities and hence can be solved in a single iteration of the Newton-Raphson method. Hence a staggered construct is preferred wherein the updates for the wall species are first calculated and handed over to the structural subsystem for calculation of displacements within every time step of the computation. 

\subsection{Flux interface}
To incorporate the flux boundary conditions described in Section \ref{bc}, an interface element is desirable since the fluxes are dependent on the PDGF and TGF$-\beta$ concentrations in the current configuration, resulting in additional contributions to the global residual vector as well as the global tangent matrix throughout the solution process. In addition, in line with the final goal of developing an FSI framework for modeling in-stent restenosis, the interface element shall aid in transferring quantities across the fluid-structure interfaces.

\begin{figure}[htbp!]
    \centering
    \includegraphics[scale=0.6]{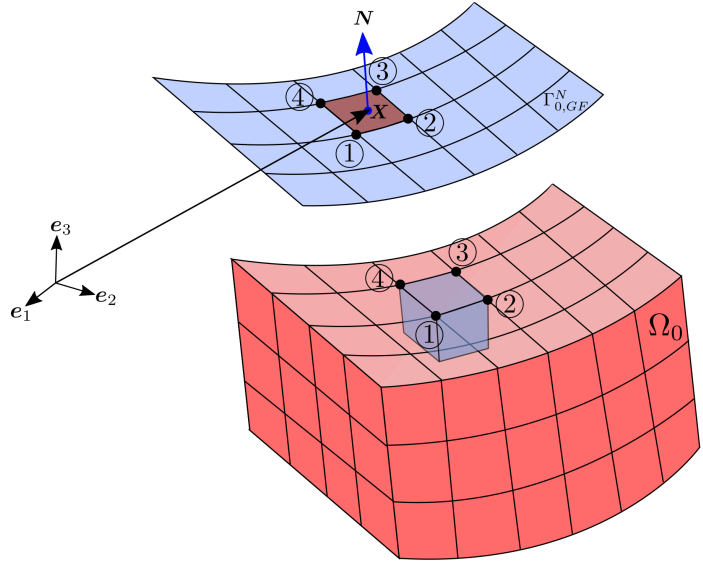}
    \caption{\textbf{Bulk mesh projected to the Neumann boundary}}
    \label{fig_flux_inter}
\end{figure}

From the weak forms presented in Eqs. \ref{wf_p} and \ref{wf_t}, the general form of the residual contributions to be evaluated on the respective Neumann boundary surfaces $\Gamma^N_{0,GF}$ in the reference configuration are of the form
\begin{equation}\label{flux_integral}
    g^N_{{}_{0,GF}} := - \displaystyle{\int_{\Gamma^N_{0,GF}}} \left(\bm{Q}_{{}_{GF}}\cdot\bm{N}\right)\,\delta c^0_{{}_{GF}}\, dA,
\end{equation}
where $\bm{Q}_{{}_{GF}}$ are the fluxes on the Neumann boundaries, subscripts $GF = P \text{ or } T$ referring to the growth factors PDGF and TGF-$\beta$ respectively. The normal flux can be reformulated as 
\begin{equation}
\bm{q}_{{}_{GF}}\cdot\bm{n} = (\bar{q}(c_{{}_{GF}})\bm{n})\,\,\cdot \bm{n}
\end{equation}
since $\bm{n}$ is a unit vector. Transforming the growth factor flux from current to the reference configuration using the Piola identity, we obtain
\begin{equation}\label{flux_transformation}
    \bm{Q}_{{}_{GF}} = J\,\bar{q}\,\bm{F}^{-1}\,\bm{n}.
\end{equation}
Using the above equation in Eq. \ref{flux_integral}, we get
\begin{equation}\label{flux_integral_transformed}
     g^N_{{}_{0,GF}} := - \displaystyle{\int_{\Gamma^N_{0,GF}}} \,J\,\bar{q}\,\left(\bm{n}^T\,\bm{F}^{-T}\,\bm{N}\right)\,\delta c^0_{{}_{GF}}\, dA,
\end{equation}
where
\begin{equation}
    \bar{q}(c^0_{{}_{GF}}) = p_{{}_{en}} \,\left(\bar{c}_{{}_{GF}} - \frac{c^0_{{}_{GF}}}{J}\right).
\end{equation}
To evaluate the integral in Eq. \ref{flux_integral_transformed} in the finite element setting, a discretized Neumann boundary is obtained in the reference configuration by projecting the bulk 3-D mesh onto the Neumann boundary surface as shown in Fig \ref{fig_flux_inter}. For example, Nodes \textcircled{1} through \textcircled{4} are shared between the elements in the bulk mesh and its projected surface mesh. The position vectors in the reference and current configurations are interpolated within the surface using
\begin{eqnarray}
    \bm{X} &\approx& \bm{X}^h = \bar{\bm{N}}^L(\xi, \eta)\,\bm{X}^e + \zeta\,\bm{N}, \quad \forall \bm{X} \in \Omega_0^e\nonumber\\
    \bm{x} &\approx& \bm{x}^h = \bar{\bm{N}}^L(\xi, \eta)\,\bm{x}^e + \zeta\,\bm{n}, \quad \forall \bm{x} \in \Omega_e
\end{eqnarray}
where $\bar{\bm{N}}^L(\xi, \eta)$ are the bilinear Lagrange shape functions. As observed in the equations above, the position vector interpolation along the $\zeta$ direction is accomplished using the surface normals $\bm{N}$ and $\bm{n}$ in the reference and current configurations respectively, given by
\begin{eqnarray}
    \bm{N} = \displaystyle{\frac{\frac{\partial \bm{X}^h}{\partial \xi}\,\times\,\frac{\partial \bm{X}^h}{\partial \eta}}{\left|\left| \frac{\partial \bm{X}^h}{\partial \xi}\,\times\,\frac{\partial \bm{X}^h}{\partial \eta}\right|\right|}}, \quad \bm{n} = \displaystyle{\frac{\frac{\partial \bm{x}^h}{\partial \xi}\,\times\,\frac{\partial \bm{x}^h}{\partial \eta}}{\left|\left| \frac{\partial \bm{x}^h}{\partial \xi}\,\times\,\frac{\partial \bm{x}^h}{\partial \eta}\right|\right|}} = \displaystyle{\frac{\bm{F}\,\frac{\partial \bm{X}^h}{\partial \xi}\,\times\,\bm{F}\,\frac{\partial \bm{X}^h}{\partial \eta}}{\left|\left| \bm{F}\,\frac{\partial \bm{X}^h}{\partial \xi}\,\times\,\bm{F}\,\frac{\partial \bm{X}^h}{\partial \eta}\right|\right|}}.
\end{eqnarray}

The solution variables and their variations are interpolated using the bilinear Lagrange shape functions, i.e.,
\begin{eqnarray}\label{interp_surf}
    \langle\bullet\rangle(\bm{X}) &\approx& \langle\bullet\rangle^{h}(\bm{X}) = \bar{\bm{N}}^L(\xi,\eta)\cdot \langle\bullet\rangle^e\nonumber\\
    \delta \langle\bullet\rangle(\bm{X}) &\approx& \delta \langle\bullet\rangle^h(\bm{X}) = \bar{\bm{N}}^L(\xi,\eta)\cdot\delta \langle\bullet\rangle^e, \quad \forall \bm{X} \in \Omega_0^e. 
\end{eqnarray}
Finally, the deformation gradient necessary for the evaluation of the surface integral in Eq. \ref{flux_integral_transformed} is evaluated using
\begin{equation}
    \bm{F} = \bm{j}\cdot\bm{J}^{-1}, 
\end{equation}
where
\begin{eqnarray}
    \bm{J} &=& \left[\frac{\partial \bm{X}^h}{\partial \xi},\,\frac{\partial \bm{X}^h}{\partial \eta},\bm{N} \right]\nonumber\\
    \nonumber\\
    \bm{j} &=& \left[\frac{\partial \bm{x}^h}{\partial \xi},\,\frac{\partial \bm{x}^h}{\partial \eta},\bm{n} \right].
\end{eqnarray}
Due to the dependence of the flux integrals (Eq. \ref{flux_integral_transformed}) on the deformation gradient $\bm{F}$, additional contributions appear in the global stiffness matrix at the nodes shared between the bulk mesh and the elements on the Neumann boundary surface.

\section{Numerical evaluation}\label{sect_num_eval}
The finite element formulation presented in this work is incorporated into the software package $FEAP$ by means of user-defined elements \citep{taylor_feap_2020}. To evaluate the efficacy of the developed finite element framework in predicting in-stent restenosis, several examples are computed in this section. To determine the set of the model parameters that macroscopically reflect the physics of restenosis, an unrestrained block model is first setup and the growth theories presented in Section \ref{section_growth_theories} are evaluated. Additionally, the computational efficiencies of the monolithic and staggered solution strategies obtained as a consequence of differences in the temporal discretization (Eqs. \ref{fi_be} and \ref{si_be}) are evaluated using the block model. Further, simplified models representing an artery post balloon angioplasty as well as a stented artery are setup, evaluated, and comparisons to the macroscopic growth behavior during in-stent restenosis presented.

\subsection{Unrestrained block}

\begin{figure}[htbp!]
    \centering
    \includegraphics[scale=0.23]{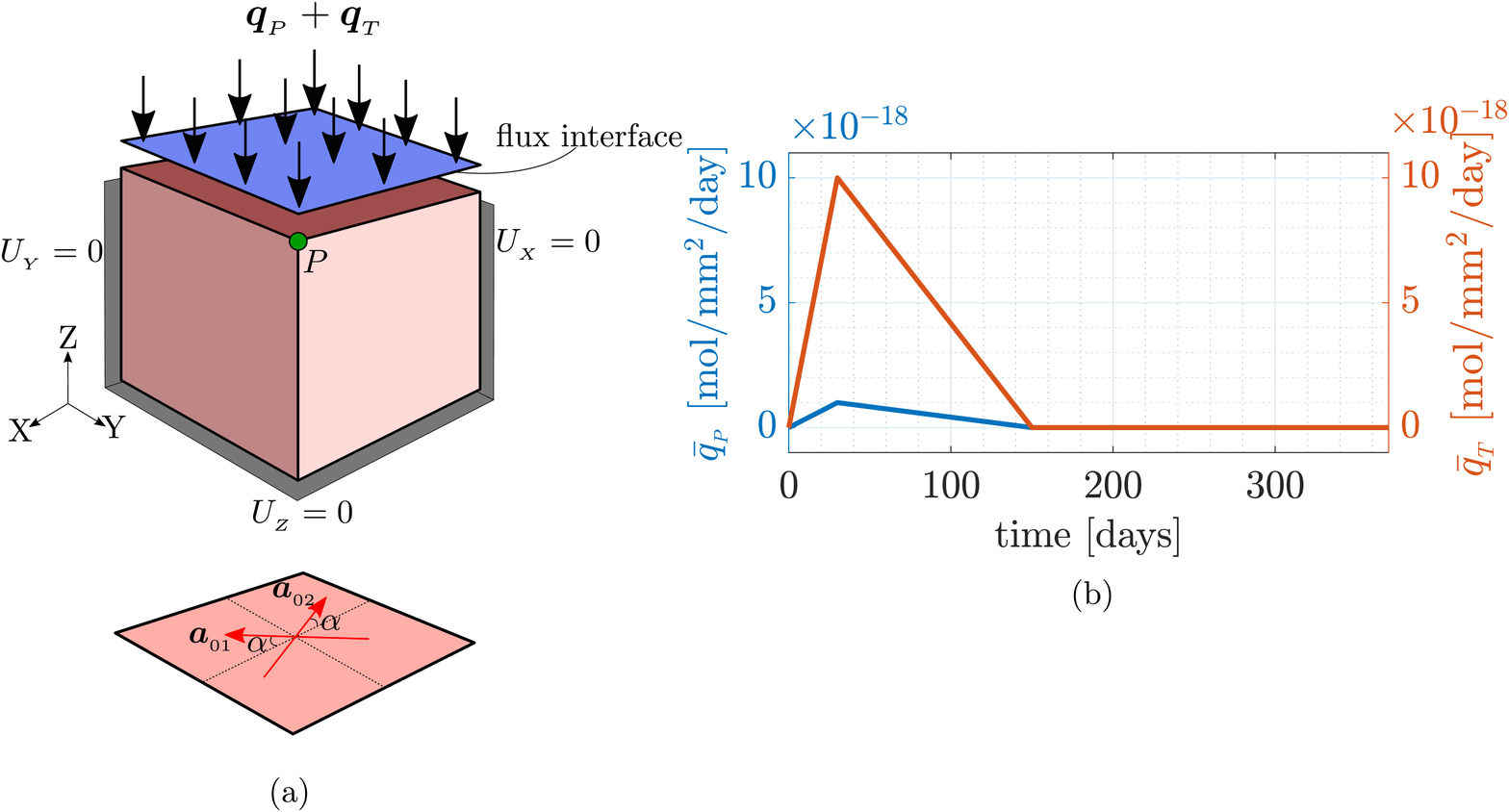}
    \caption{\textbf{Unrestrained block test.} \\(a) Problem setup: collagen orientation vectors are assumed to be on the $X-Y$ plane. The block is fixed against normal displacements on the faces marked in grey. PDGF and TGF$-\beta$ influxes along the normal to the top surface are prescribed.\\(b) Influx profiles: The normal fluxes $\bar{q}_{{}_P} = \bm{q}_{{}_P}\cdot \bm{n}$ and $\bar{q}_{{}_T} = \bm{q}_{{}_T}\cdot \bm{n}$ mimic the process of endothelium damage and recovery.}
    \label{fig_cube_problem}
\end{figure}
A cubic block of side length $1$ [mm] is generated as shown in Fig \ref{fig_cube_problem}(a). The collagen orientations are chosen to be embedded primarily within the $X-Y$ plane.

\subsubsection*{Discretization} The block is meshed with $4 \times 4 \times 4$ trilinear hexahedral elements. The problem is temporally discretized using a time step size of $\Delta t = 1$ [days].

\subsubsection*{Boundary conditions} Fixation is provided along the  normal directions on sides marked in grey so that rigid body motions are arrested. PDGF and TGF-$\beta$ influxes are prescribed for a period of 370 days (approximately a year) on the flux interface since crucial restenotic mechanisms are observed on this time span. The influx profiles on the current configuration, shown in Fig \ref{fig_cube_problem}(b), mimic the process of endothelium damage and recovery. The ratio between the PDGF and TGF$-\beta$ influxes reflect the ratio between serum levels of the respective growth factors \citep{czarkowska2006}.  

\subsubsection*{Parameters} The \textit{in vivo} cellular and molecular mechanisms in the arterial wall are difficult to replicate and quantify \textit{in vitro}. For mechanisms that are replicated, the model parameters are carried over from literature, and for those that are not, the parameters are chosen in such a way that they qualitatively reflect the macroscopic phenomena. They are listed in Table \ref{model_params}.\\
\\
Both the growth models described in Section \ref{section_growth_theories} are evaluated and compared within the fully-coupled monolithic solution framework.

\begin{table}[htbp!]\
\centering
\caption{\textbf{Unrestrained block - Model parameters}}
\label{model_params} 
\begin{tabular}{p{1.5cm}p{4.5cm}p{2cm}p{2.5cm}p{3cm}}
\noalign{\hrule height 0.05cm}\noalign{\smallskip}
parameter & description & value & units & reference  \\
\noalign{\hrule height 0.05cm}\noalign{\smallskip}
 \textbf{PDGF} & & & \\
 $D_{{}_P}$ & diffusivity & $0.1$ & [mm$^2$/day] & \citep{budu2008}\\
 $\eta_{{}_P}$ & autocrine secretion coefficient & $1.0 \times 10^{-6}$ & [mm$^3$/cell/day] & choice\\ 
 $\varepsilon_{{}_P}$ & receptor internalization coefficient & $1.0 \times 10^{-7}$ & [mm$^3$/cell/day] & choice\\
  $c_{{}_{P,th}}$ & PDGF threshold for the scaling function & $1.0 \times 10^{-15}$ & [mol/mm$^3$] & choice\\
   $l_{{}_{P}}$ & steepness coefficient for PDGF dependent scaling & $1.0 \times 10^{16}$ & [mm$^3$/mol] & choice\\ 
 \\
 \textbf{TGF-}$\bm{\beta}$ & & & \\
 $D_{{}_T}$ & diffusivity& $0.1$ & [mm$^2$/day] & \citep{budu2008}\\
 $\varepsilon_{{}_T}$ & receptor internalization coefficient & $1.0 \times 10^{-7}$ & [mm$^3$/cell/day] & choice\\
  $c_{{}_{T,th}}$ & TGF-$\beta$ threshold for the scaling function & $1.0 \times 10^{-16}$ & [mol/mm$^3$] & choice\\ 
   $l_{{}_{T}}$ & steepness coefficient TGF$-\beta$ dependent scaling & $1.0 \times 10^{16}$ & [mm$^3$/mol] & choice\\
 \\
 \textbf{ECM} & & & \\
 $\eta_{{}_E}$ & collagen secretion coefficient & $1.0 \times 10^{-7}$ & [mol/cell/day] & \citep{cilla2014}\\
 $\varepsilon_{{}_E}$ & collagen degradation coefficient & $1.0 \times 10^{21}$ & [mm$^3$/mol/day] & choice\\ 
 $c_{{}_{E,eq}}$ & collagen concentration in a healthy artery & $7.0 \times 10^{-9}$ & [mol/mm$^3$] & \citep{sae2013} \\
 $c_{{}_{E,th}}$ & collagen secretion threshold & $7.0007 \times 10^{-9}$ & [mol/mm$^3$] & choice \\
 \\
 \textbf{SMC} & & & \\
 $\chi_{{}_C}$ & chemotactic sensitivity & $1.0 \times 10^{11}$ & [mm$^5$/mol/day] & \citep{budu2008}\\
 $\chi_{{}_H}$ & haptotactic sensitivity & $1.0 \times 10^{6}$ & [mm$^5$/mol/day] & choice\\
 $\eta_{{}_S}$ & proliferation coefficient & $1.0 \times 10^{14}$ & [mm$^3$/cell/day] & choice\\
  $\rho_{{}_{S,eq}}$ & SMC density of a healthy artery & $3.7 \times 10^{5}$ & [cells/mm$^3$] & \citep{OConnell2008TheTM}\\
 \\
 \textbf{structural} & & & \\
 $\mu$ & shear modulus for the matrix & $0.02$ & [M Pa] & \citep{he2020}\\
 $\Lambda$ & Lam\'{e} parameter for the matrix & $10$ & [M Pa] & \citep{he2020}\\
 $\bar{k}_1$ & stress like parameter for collagen fibres & $0.112$ & [M Pa] & \citep{he2020}\\
 $k_2$ & exponential coefficient for collagen fibres & $20.61$ & [-] & \citep{he2020}\\
 $\kappa$ & dispersion parameter & $0.1$ & [-] & \citep{he2020}\\
 $\alpha$ & collagen orientation angle w.r.t X-axis & $41$ & [${}^{\circ}$] & \citep{he2020}\\
\noalign{\hrule height 0.05cm}\noalign{\smallskip}
\end{tabular}
\end{table}

\begin{figure}[htb!] 
  \begin{subfigure}[b]{0.5\linewidth}
    \centering
    \includegraphics[width=0.5\linewidth]{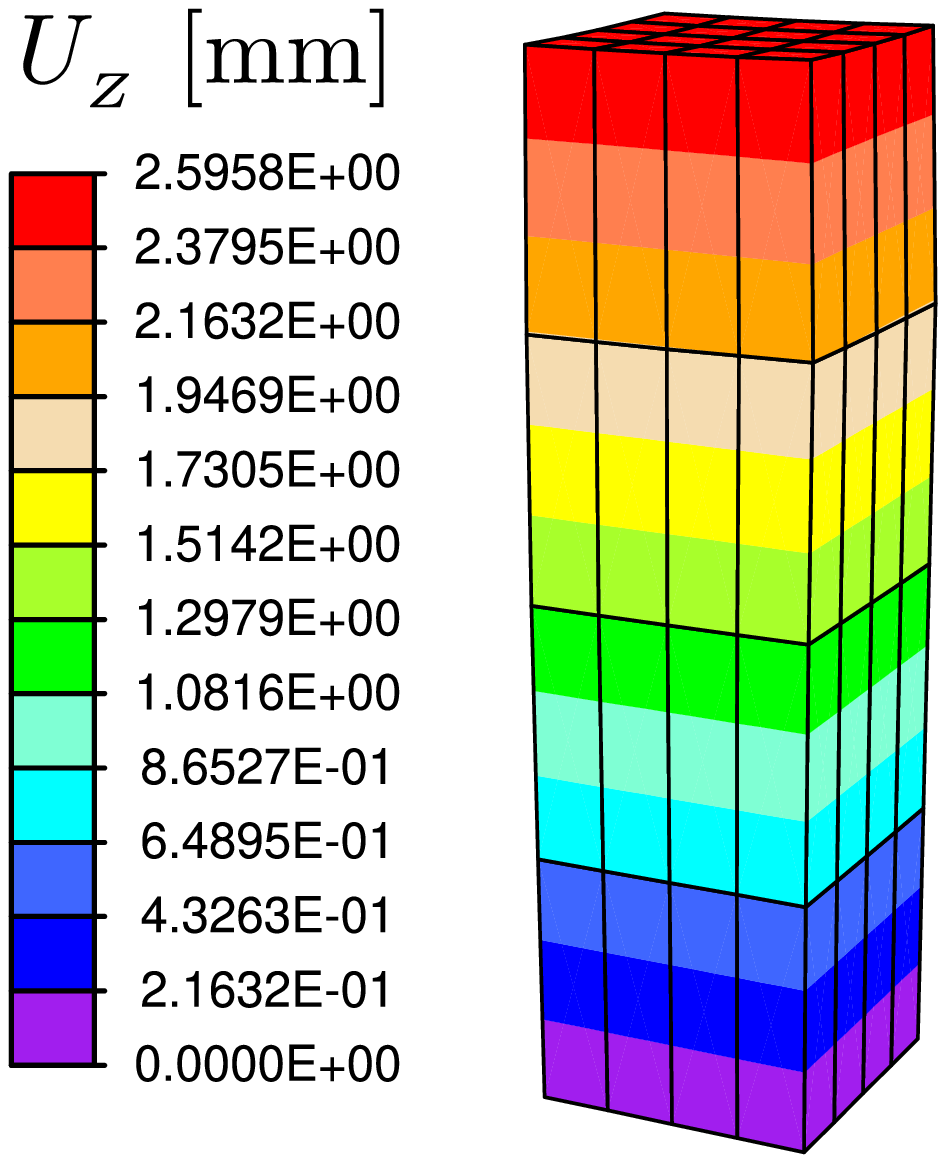} 
    \caption{stress-free anisotropic growth} 
    \label{uz:b} 
    \vspace{4ex}
  \end{subfigure}
  \begin{subfigure}[b]{0.5\linewidth}
    \centering
    \includegraphics[width=0.7\linewidth]{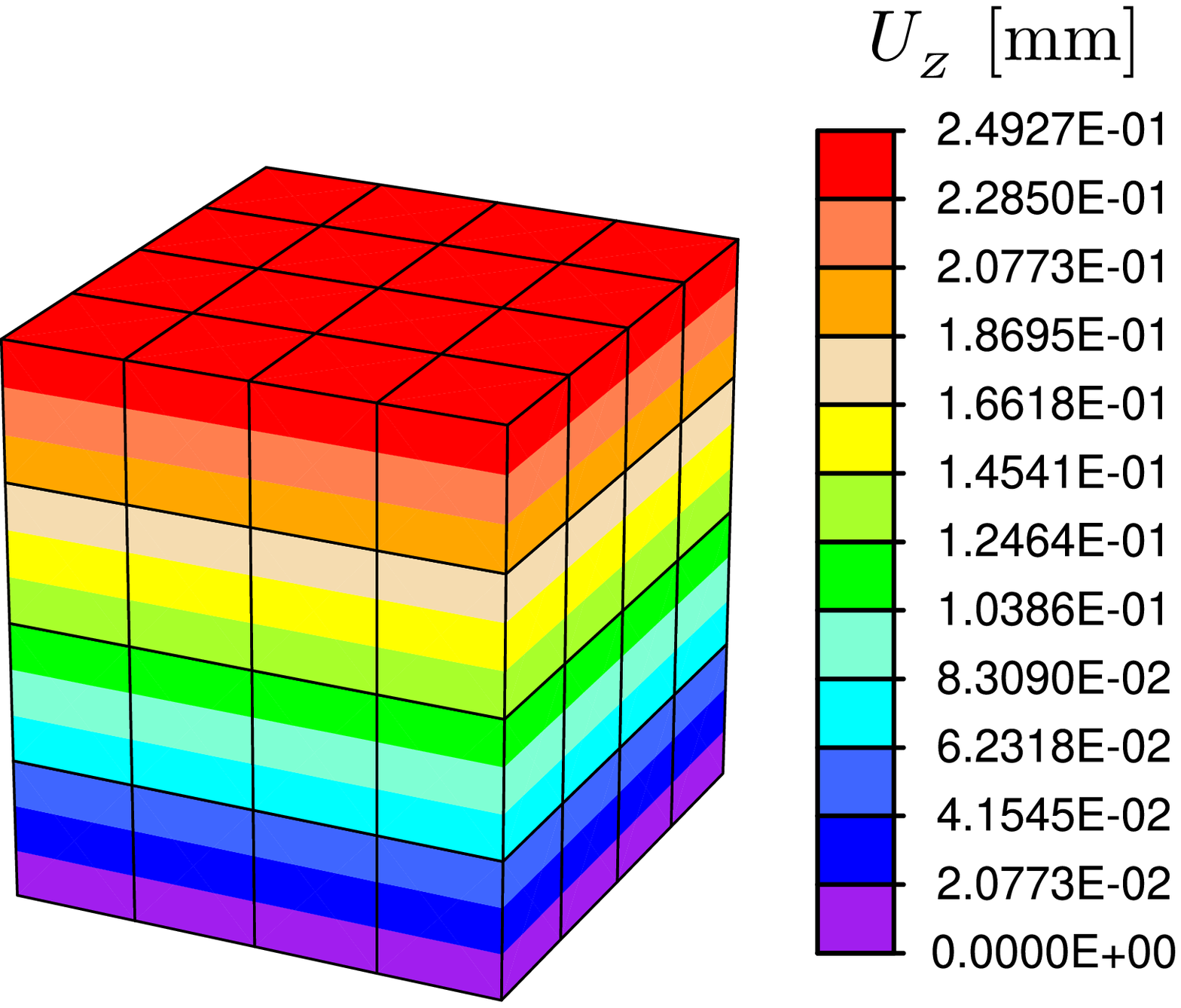} 
    \caption{isotropic matrix growth ($\kappa = 0.3$)}
    \label{uz:c} 
    \vspace{4ex}
  \end{subfigure}
    \vspace{0.7in}\\ 
  \begin{subfigure}[b]{\linewidth}
    \centering
    \includegraphics[width=0.7\linewidth]{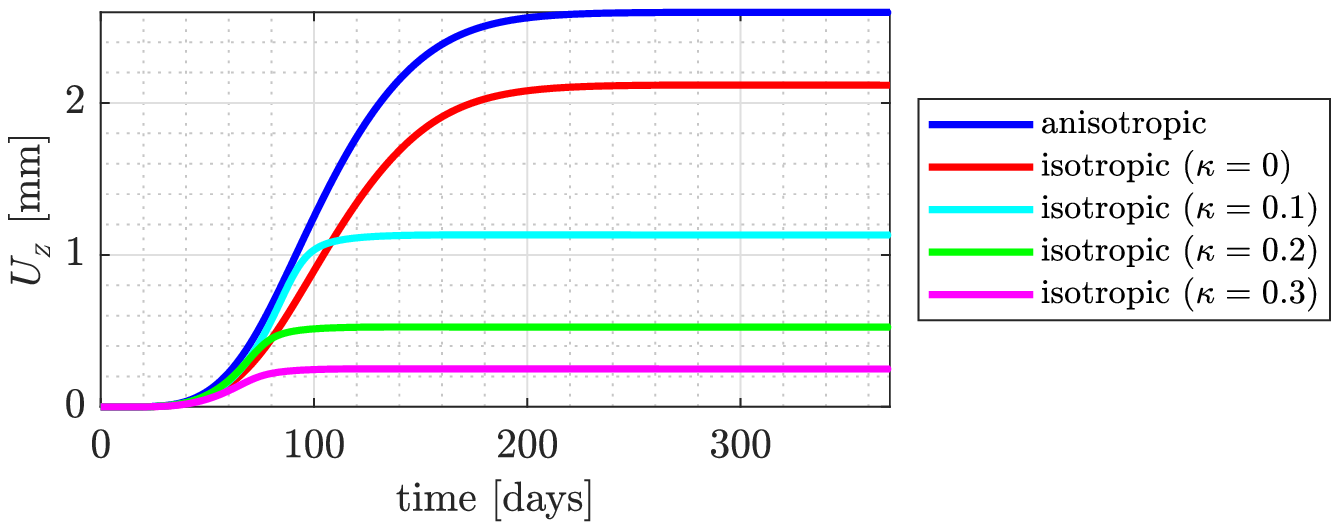} 
    \caption{{Evolution of vertical displacement at point $P$ (see Fig \ref{fig_cube_problem}(a))}} 
    \label{uz:a} 
    \vspace{4ex}
  \end{subfigure}
  \caption{\textbf{Comparison of the two growth theories:} The value of the macroscopic quantity $U_{{}_Z}$ i.e., the vertical displacement, at point P is used for the comparison. \\(a) Stress-free anisotropic growth: Maximum growth is observed along the direction perpendicular to the plane consisting the collagen orientations. \\(b) Isotropic matrix growth: At $\kappa = 0.3$, the collagen orientations are distributed more or less in an isotropic manner. Hence the growth response is also isotropic. \\(c) As $\kappa$ approaches zero, the response of the isotropic matrix growth theory converges towards that of the stress-free anisotropic growth theory.}
  \label{uz} 
\end{figure}

\begin{figure}[htb!] 
  \begin{subfigure}[b]{0.5\linewidth}
    \centering
    \includegraphics[width=0.9\linewidth]{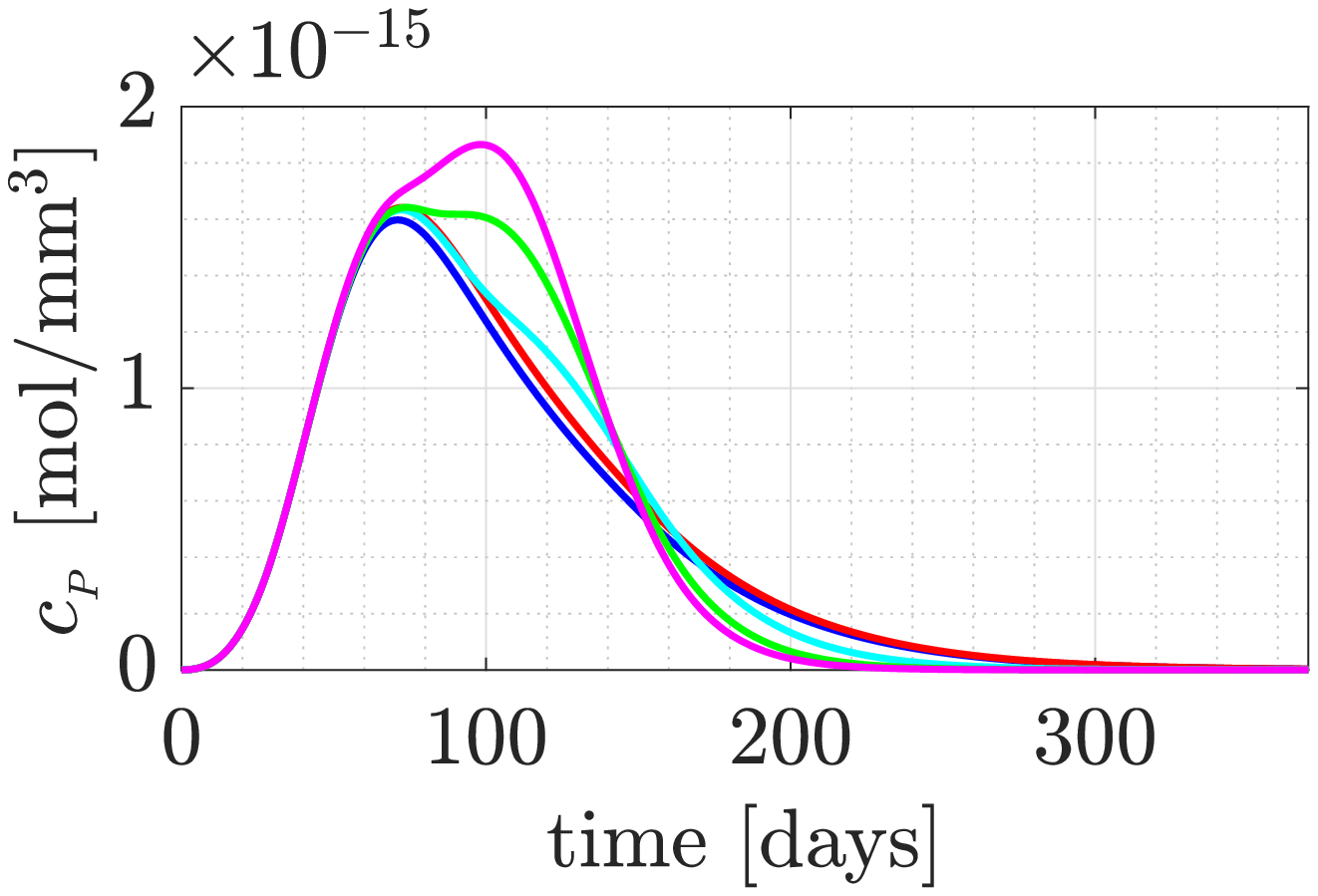} 
    \caption{{Evolution of PDGF concentration}}
    \label{evols:a} 
    \vspace{4ex}
  \end{subfigure}
  \begin{subfigure}[b]{0.5\linewidth}
    \centering
    \includegraphics[width=0.9\linewidth]{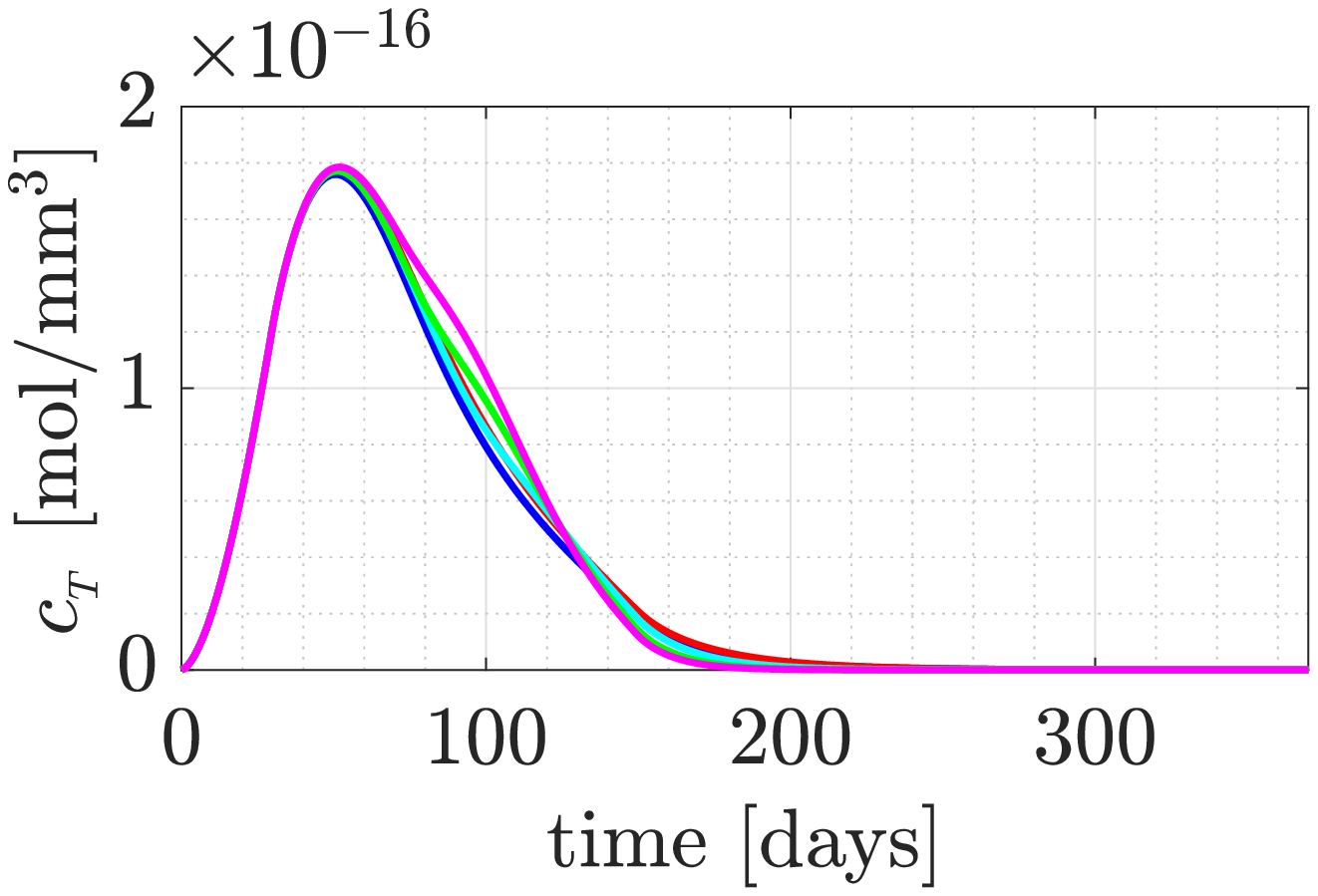} 
    \caption{{Evolution of TGF$-\beta$ concentration}}
    \label{evols:a} 
    \vspace{4ex}
  \end{subfigure}
  \begin{subfigure}[b]{\linewidth}
    \centering
    \includegraphics[width=0.2\linewidth]{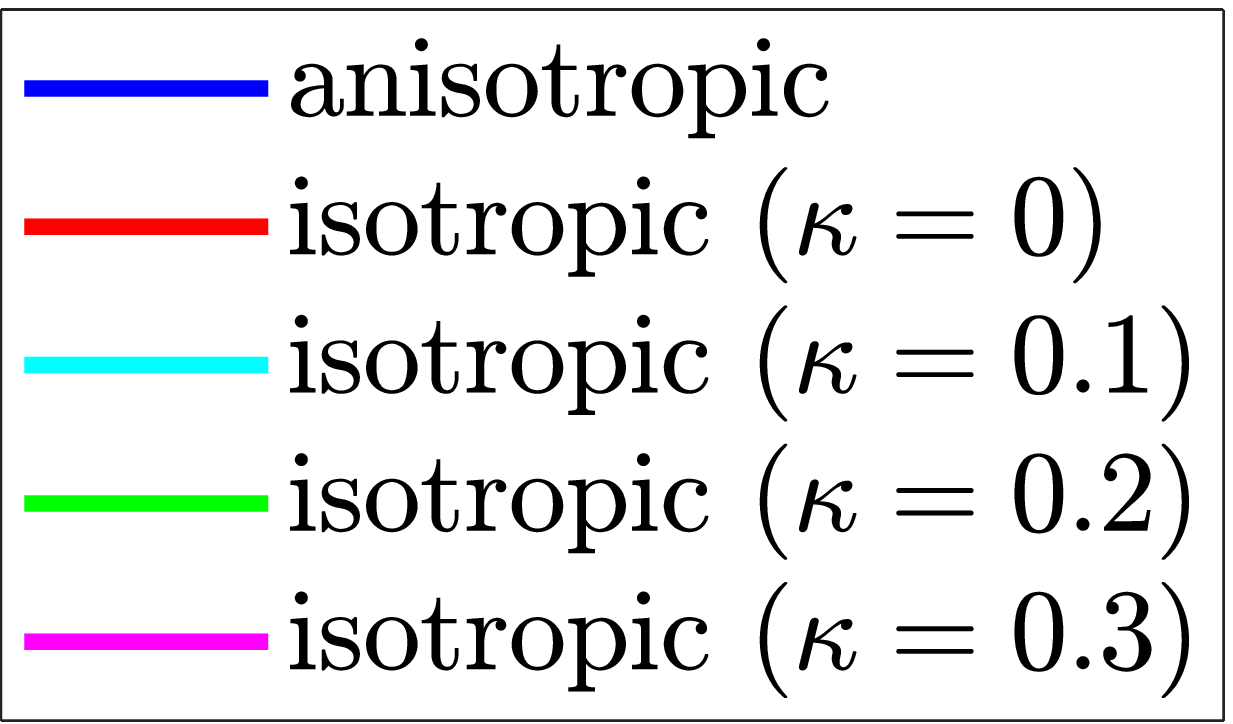} 
    \label{evols:a} 
    \vspace{4ex}
  \end{subfigure}  
  \begin{subfigure}[b]{0.5\linewidth}
    \centering
    \includegraphics[width=0.9\linewidth]{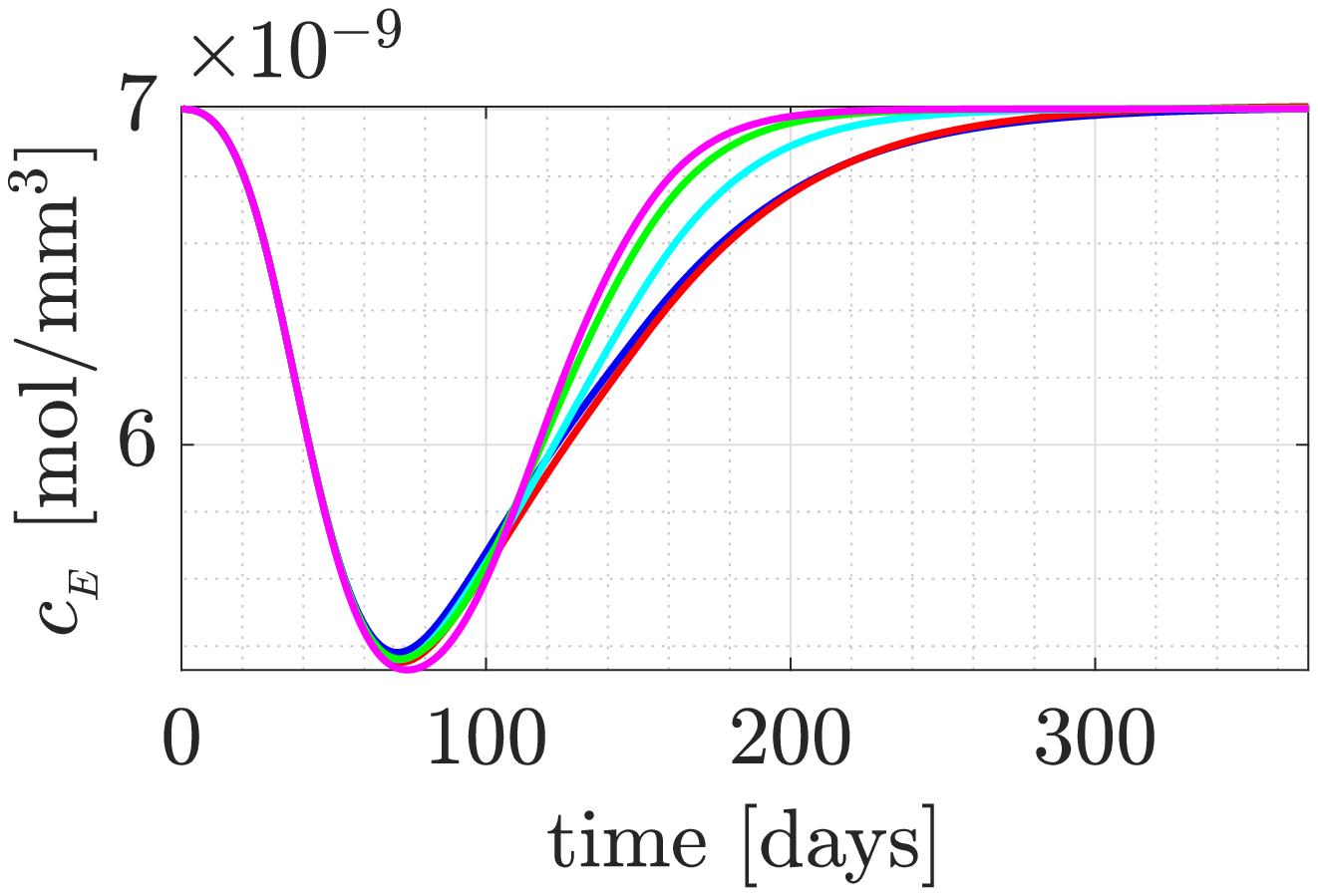} 
    \caption{{Evolution of ECM concentration}} 
    \label{evols:a} 
  \end{subfigure}
  \begin{subfigure}[b]{0.5\linewidth}
    \centering
    \includegraphics[width=0.9\linewidth]{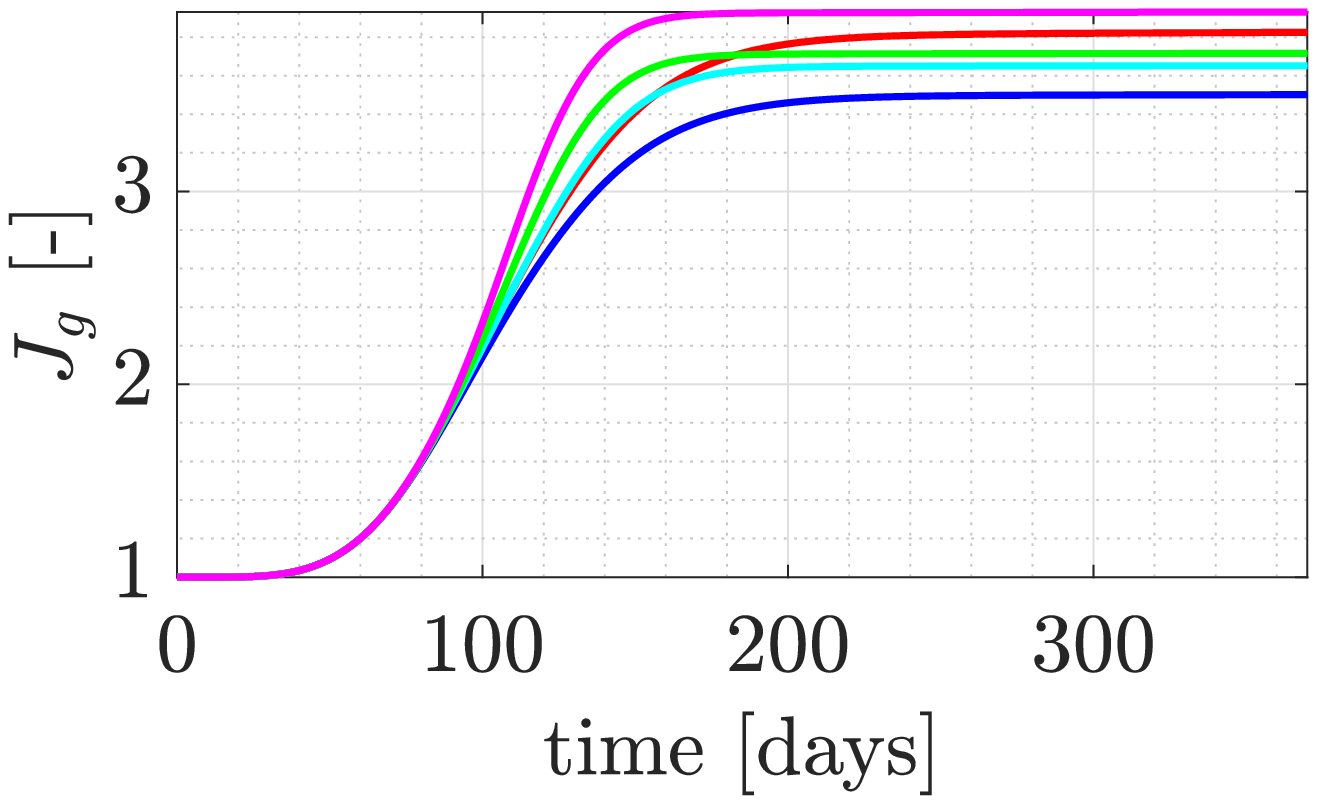} 
    \caption{{Evolution of growth volumetric change}} 
    \label{evols:a} 
  \end{subfigure} 
  \caption{\textbf{Evolution of quantities of interest.}\\(a) PDGF attains its peak further down the timeline due to secretion mechanisms.\\(b) Evolution of TGF$-\beta$ conforms to the influx profile.\\(c) ECM degrades due to PDGF presence, and heals due to collagen secretion.\\(d) Growth exhibits a sigmoidal pattern.}
  \label{evols} 
\end{figure}

\subsubsection*{Results and discussion} The evolutions of the wall species in the deformed configurations and the volume change due to growth for both the growth theories, at the point $P$ annotated in Fig \ref{fig_cube_problem}(a), are plotted in Figs \ref{evols} (a)-(d). As expected, the evolution profiles of TGF-$\beta$ closely follow the influx profile since there are no secretion processes to enhance its presence in the wall (Fig \ref{evols}(b)). On the other hand, PDGF is secreted by SMCs in presence of TGF-$\beta$ and in a degraded ECM environment which is reflected in the fact that the peak in PDGF concentration is achieved later than that of TGF-$\beta$ (Fig \ref{evols}(a)). As observed in Fig \ref{evols}(c), the ECM is initially degraded by PDGF. Once the SMCs proliferate and secrete collagen, the healing process begins and ECM reaches its equilibrium value when all of PDGF is consumed. The behavior of the ECM is in line with the physiology of the matrix formation phase of the wound healing process described in \citet{forrester1991}. The macroscopic description of growth volumetric change $J_g$ also conforms to those presented in \citet{fereidoo2017} and \citet{schwartz1996}(Fig \ref{evols}(d)). Since the model is evidently sensitive to patient specific data, it is sufficient at this point that the results qualitatively reflect the pathophysiology.

From Figs \ref{evols} (a)-(d), the effect of incorporation of dispersion in collagen orientations is clearly understood. Interesting is the fact that the evolution of wall species in the isotropic matrix growth model converge to those of the anisotropic growth model as $\kappa$ approaches zero, but do not exactly coincide. The discrepancy at $\kappa = 0$ can clearly be explained by the differences in the hypotheses for the two growth models. In the stress-free anisotropic growth hypothesis, the stress-free grown configuration is defined independently of the local ECM concentration. On the other hand, isotropic matrix growth leads to residual stresses in the intermediate grown configuration which are dependent on the local ECM concentration.  At some point in the evolution of growth, if low ECM concentrations are encountered, the isotropic matrix growth model experiences low residual stresses, thereby conforming to a more isotropic form of growth. One additional observation is that prescribing $\kappa$ very close to $1/3$ results in an isotropic dispersion of collagen fiber orientations, leading to an isotropic growth response as seen in Fig \ref{uz}(b). A parameter sensitivity, study for those parameters that can be deemed patient-specific, is provided in \ref{appendix:pst}.

\subsubsection*{Comparison of coupling constructs} 
Using the isotropic matrix growth model, the monolithic and staggered coupling strategies, that are a result of the fully-implicit and semi-implicit temporal discretizations respectively (Section \ref{section_temporal_discretization}), are compared using the evolution of the volmetric change due to growth. 
\begin{figure}[htb!] 
  \begin{subfigure}[b]{0.5\linewidth}
    \centering
    \includegraphics[width=0.85\linewidth]{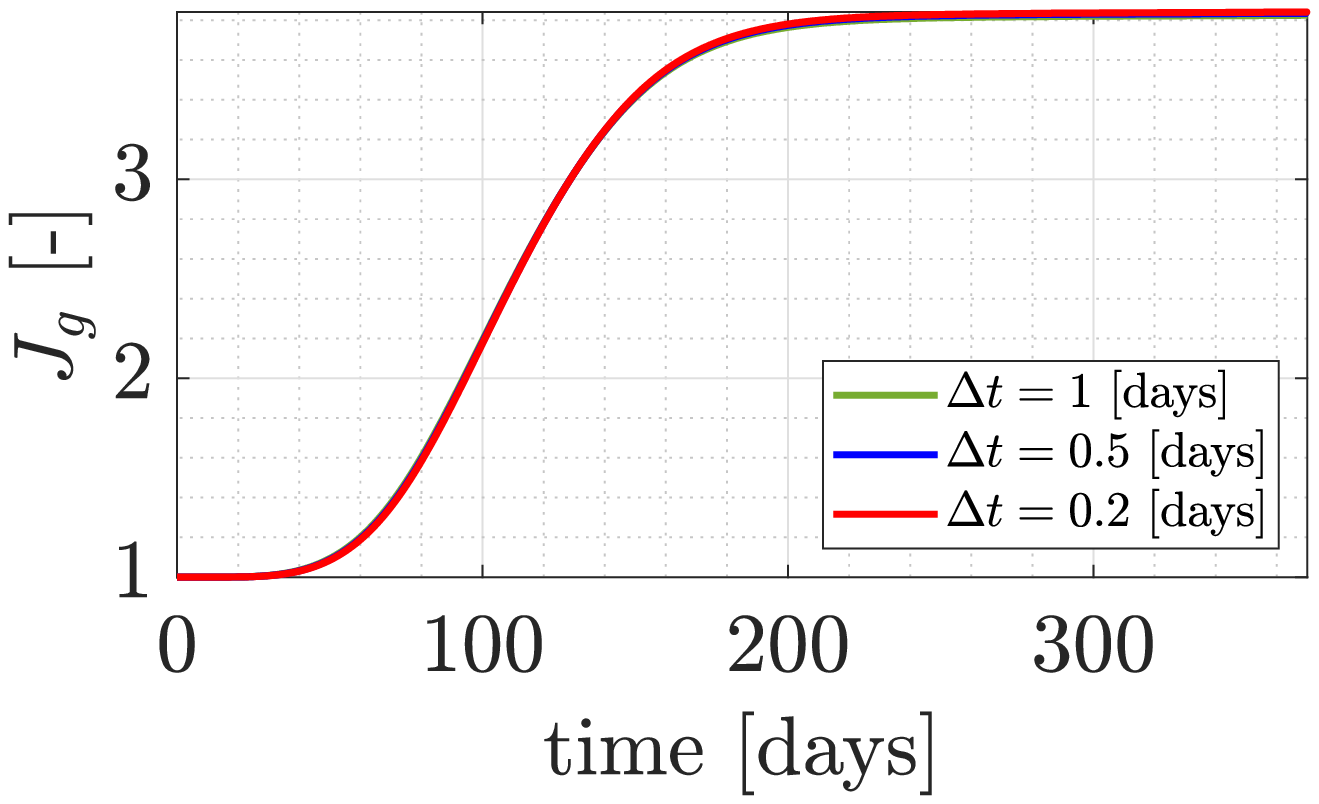} 
    \caption{{Time step convergence - monolithic}} 
    \label{convergence:a} 
  \end{subfigure}
  \begin{subfigure}[b]{0.5\linewidth}
    \centering
    \includegraphics[width=0.85\linewidth]{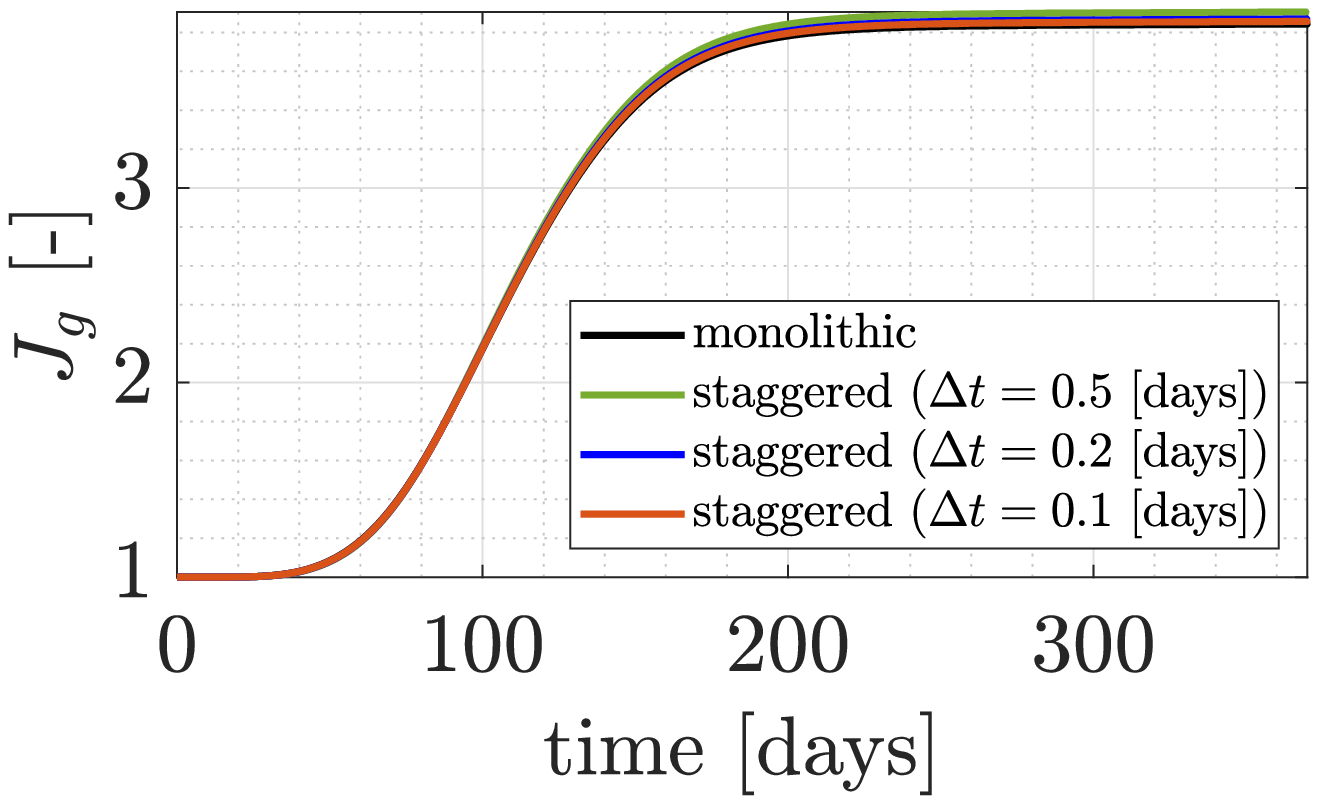} 
    \caption{{Time step convergence - staggered}} 
    \label{convergence:a} 
  \end{subfigure} 
  \begin{subfigure}[b]{0.5\linewidth}
    \centering
    \includegraphics[width=0.85\linewidth]{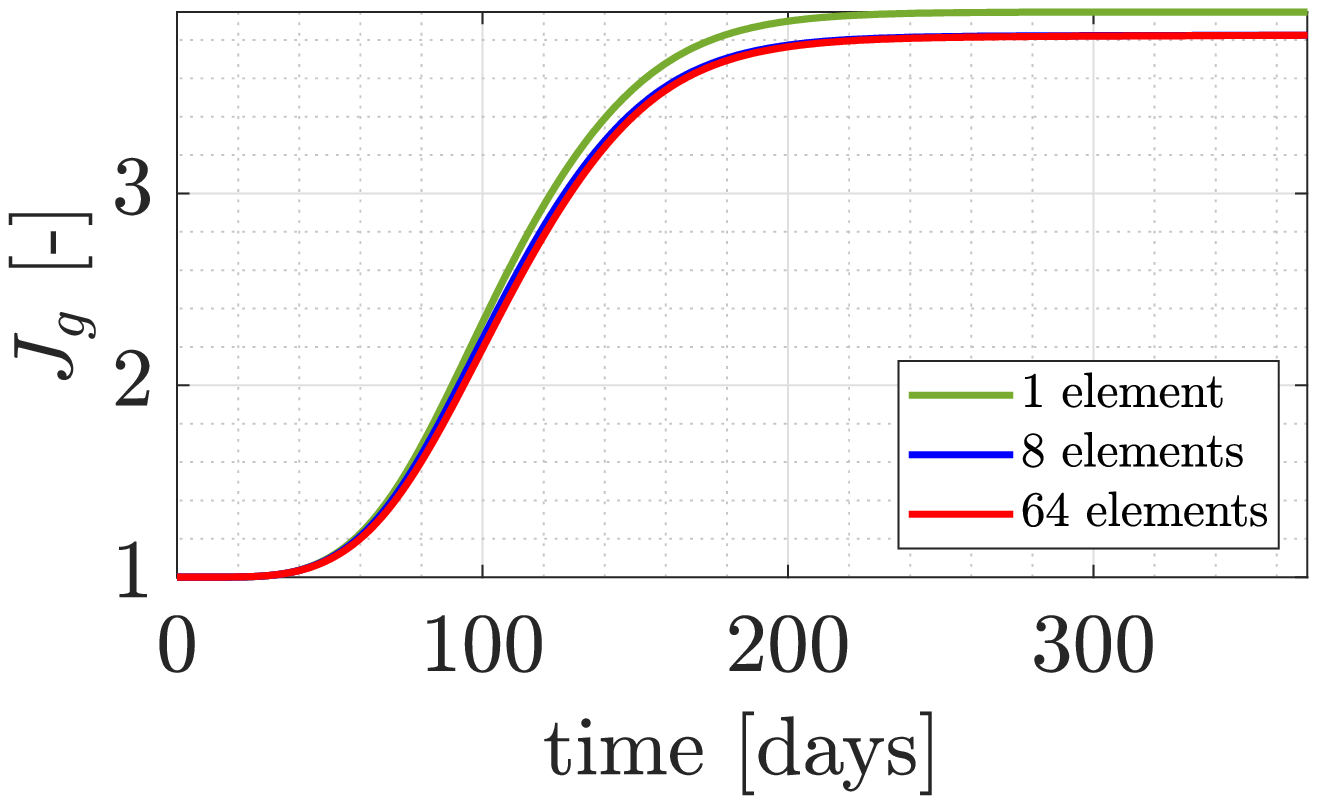} 
    \caption{{Mesh convergence - monolithic}} 
    \label{convergence:a} 
  \end{subfigure}
  \begin{subfigure}[b]{0.5\linewidth}
    \centering
    \includegraphics[width=0.85\linewidth]{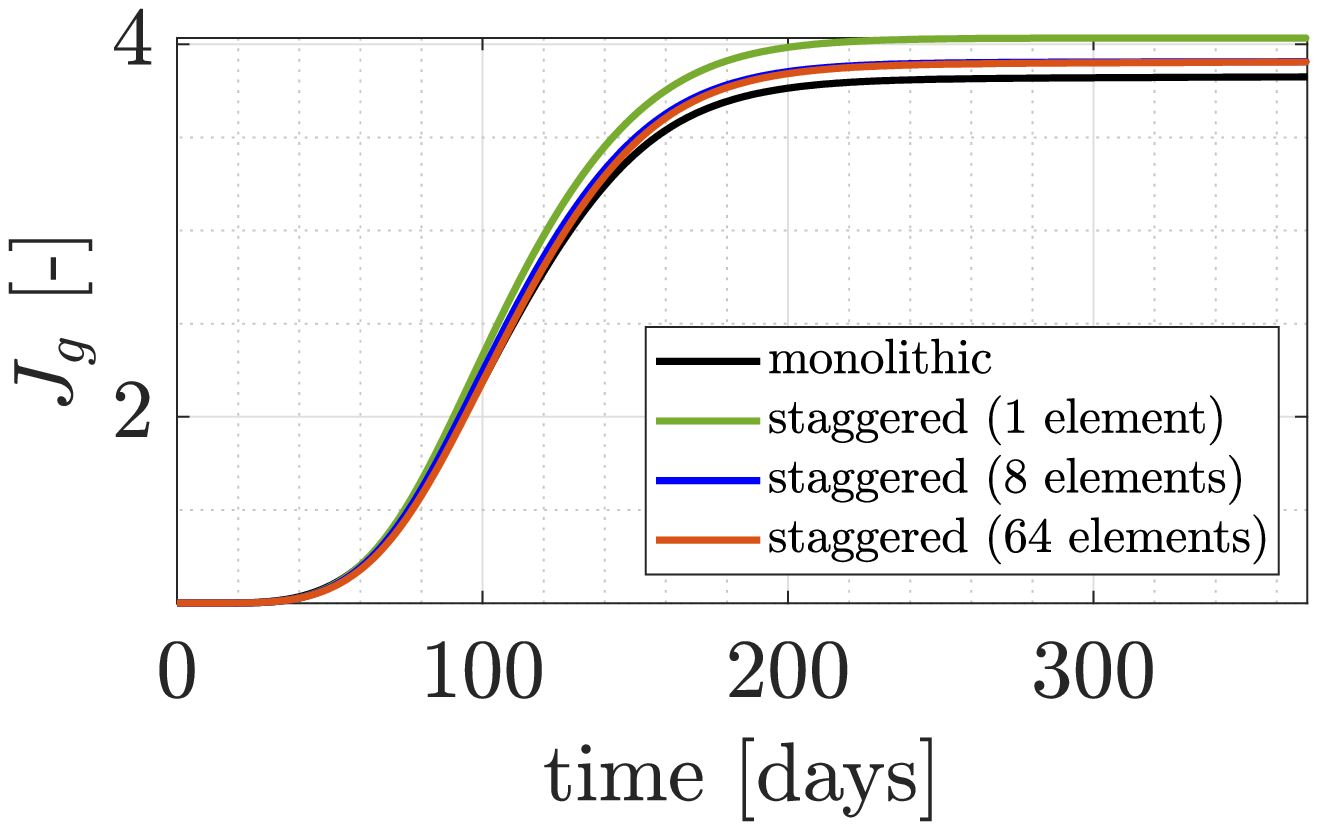} 
    \caption{{Mesh convergence - staggered}} 
    \label{convergence:a} 
  \end{subfigure} 
  \caption{\textbf{Time step and mesh convergence analyses.}\\(a) The monolithic coupling strategy provides sufficient accuracy even at large time steps.\\(b) The staggered coupling strategy does not lead to convergence for time step sizes larger than $0.5$ [days], but sufficiently accurate solutions are obtained for time steps smaller than $0.5$ [days].\\(c) Sufficiently accurate solutions are obtained for coarse mesh sizes via the monolithic construct.\\(d) The solutions are  relatively inaccurate for coarse mesh sizes when compared to those of the monolthic construct. Also, the solutions do not coincide with the monolithic solution even for fine meshes.}
  \label{convergence} 
\end{figure}

The isotropic matrix growth model implemented in a monolithic framework already achieves time step convergence at $\Delta t = 1$ [days] as observed in Fig \ref{convergence}(a). The staggered framework also demonstrates sufficient accuracy when $\Delta t$ is below $0.5$ [days] (See Fig \ref{convergence}(b)). But the simulation breaks down for time step sizes greater than that due to accumulation of errors.

The monolithic coupling strategy demonstrates great accuracy for coarse spatial discretizations as seen in Fig \ref{convergence}(c). Staggered strategy on the other hand achieves mesh convergence for coarse spatial discretizations, but the solutions do not coincide with those of the monolithic one. This is attributed to the semi-implicitness in the temporal discretization.

\begin{figure}[htb!] 
  \begin{subfigure}[b]{0.5\linewidth}
    \centering
    \includegraphics[width=0.9\linewidth]{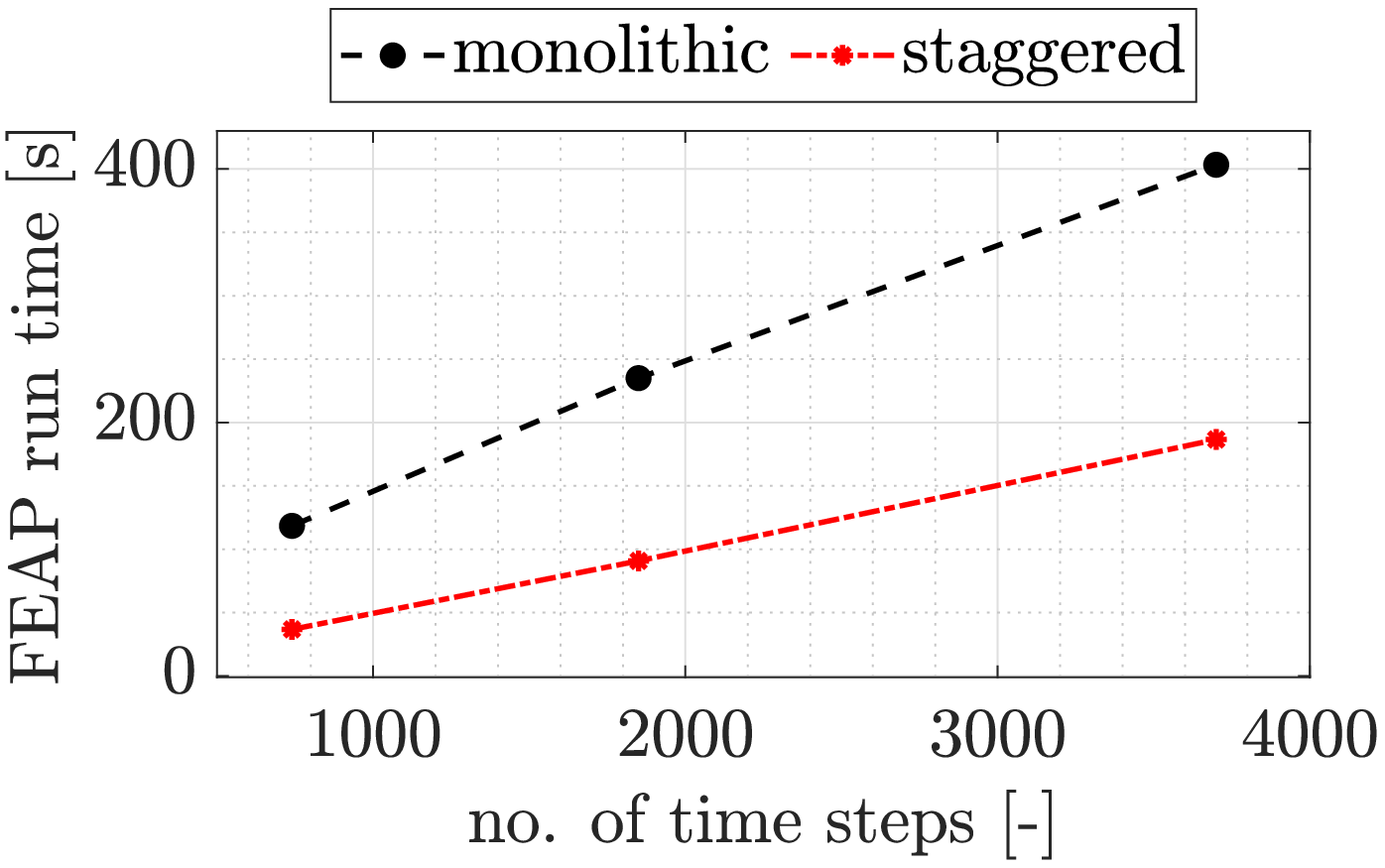} 
    \caption{{FEAP run times - varying time step sizes}} 
    \label{runtime:a} 
  \end{subfigure}
  \begin{subfigure}[b]{0.5\linewidth}
    \centering
    \includegraphics[width=0.85\linewidth]{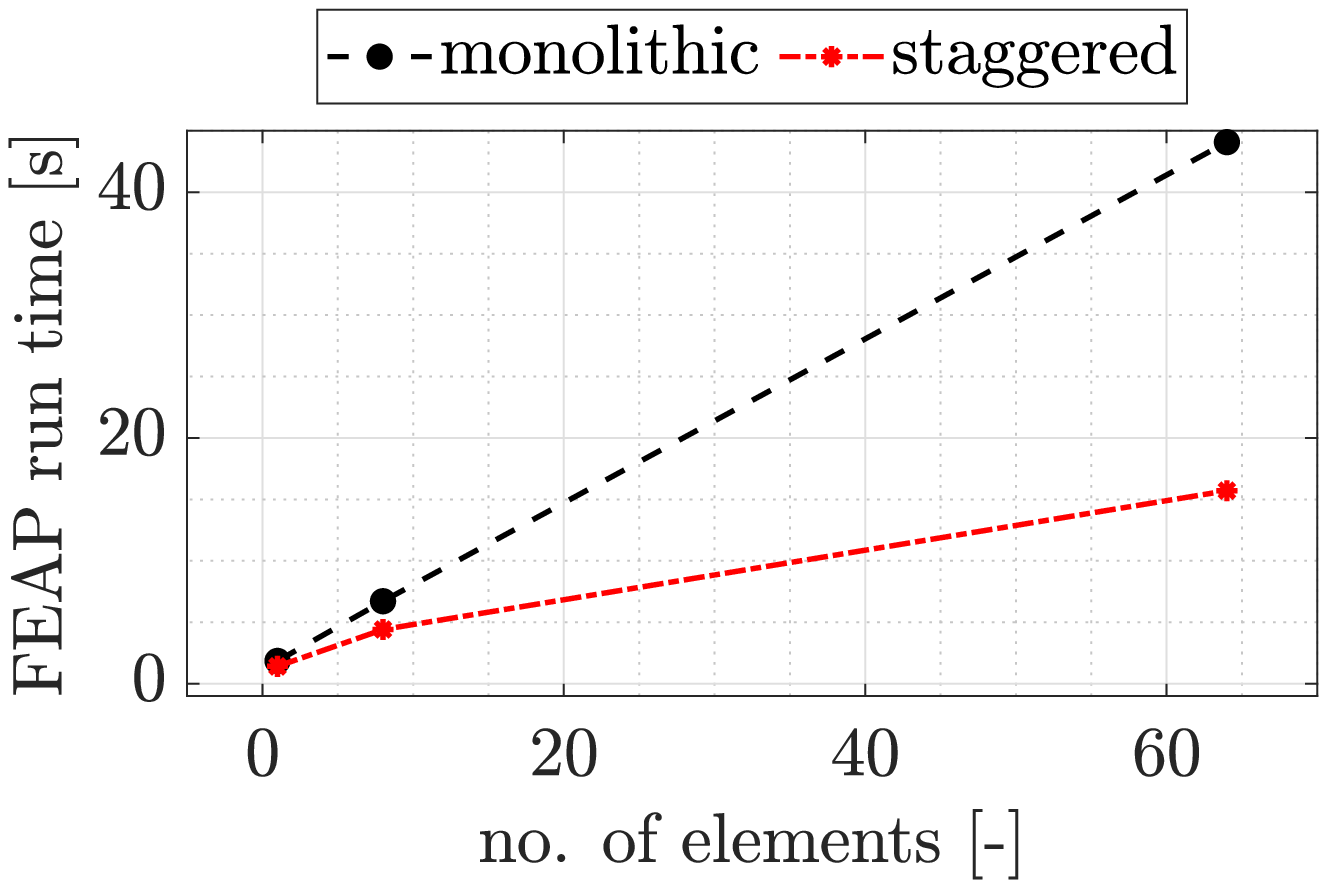} 
    \caption{{FEAP run times - varying mesh sizes}} 
    \label{runtime:a} 
  \end{subfigure} 
  \caption{\textbf{FEAP run time comparison.} \\(a) Varying time step sizes: substantially high run times are observed for smaller time step sizes in case of the monolithic construct. \\(b) Varying mesh sizes: when the spatial discretization gets finer, the run times are evidently higher for the monolithic construct.}
  \label{runtime} 
\end{figure}

As the time step size decreases, the FEAP run time associated with the monolithic coupling strategy increases drastically, which can be seen in Fig \ref{runtime}(a). This marked increase in computational effort is attributed to the dense structure of the system matrix associated with the monolithic approach (Eq. \ref{eq_mono}). In contrast, the staggered approach leads to symmetric sparse system matrices for the wall species (Eq. \ref{eq_staggered}), the inversion of which is inexpensive. In addition, the species variables can be updated with just one single inversion of the associated system matrices due to the semi-implicit temporal discretization and lack of nonlinearity. The displacement system on the other hand requires several iterations to achieve convergence via Netwon-Raphson iterations. Overall, the monolithic solution strategy hence results in a relatively large number of matrix inversions compared to the staggered approach, and therefore higher run times.

A similar trend was observed in the FEAP run times when the mesh density was increased (Fig \ref{runtime}(b)). The difference in computational effort is here attributed to the size of the system matrices rather than the number of inversions necessary in case of varying time step sizes. 

It is therefore concluded that when the complexity of the finite element system necessitates small time step sizes (e.g., contact problems), the inexpensiveness of the staggered approach can be taken advantage of. Meanwhile, if the accuracy of the solution is of high importance, and there are no restrictions on the time step size, the monolithic coupling can be utilized. 

\subsection{Restenosis after balloon angioplasty}\label{subsection_bal_ang}
\begin{figure}[htb!]
    \centering
    \includegraphics[scale=0.45]{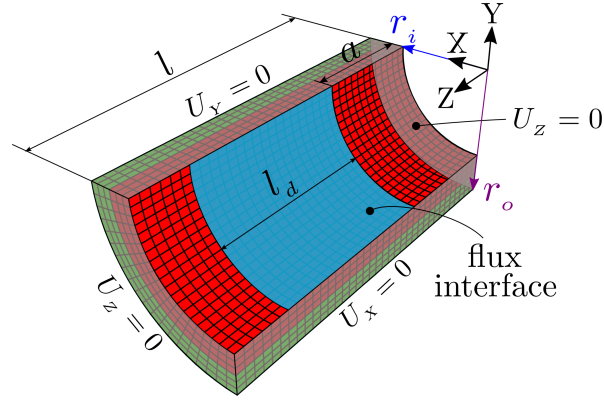}
    \caption{\textbf{Balloon angioplasty - Problem setup}}
    \label{fig_bal_ang}
\end{figure}

Owing to symmetry, a quadrant of an arterial wall is generated in $FEAP$ as shown in Fig \ref{fig_bal_ang} with $l = 6$ [mm], $r_{{}_i} = 1.55$ [mm], and $r_{{}_o} = 2.21$ [mm]. The medial layer of the artery is modeled to be $0.34$ [mm] thick, and is marked in red. The adventitial layer is considered to be $0.32$ [mm] thick, and is marked in green. These dimensions resemble those of a rat aorta. A region of length $l_{{}_d} = 3$ [mm], beginning at a distance of $a = 2$ [mm] along the longitudinal direction, is considered damaged due to endothelial denudation as a result of balloon angioplasty. The monolithic construct in combination with the isotropic matrix growth model is utilized for this example.
\vspace{-0.05in}

\subsubsection*{Discretization} The geometry is meshed using trilinear hexahedral elements. Each layer of the arterial wall is meshed with 3 elements across their thicknesses, 20 elements along the circumferential direction, and 36 elements along the longitudinal direction. The region where the endothelium is denuded is meshed with bilinear quadrilateral elements which are projected from the bulk mesh. Time step size of $\Delta t = 1$ [days] is used for the simulation.

\subsubsection*{Boundary conditions} The sides marked in translucent gray are fixed against displacements in their normal directions as depicted. The influx profile on the deformed configuration is prescribed as shown in Fig \ref{fig_cube_problem}(b), the peak value of $\bar{q}_{{}_P}$ being $1 \times 10^{-19}$ [mol/mm$^2$/day] and that of $\bar{q}_{{}_T}$ being $1 \times 10^{-18}$ [mol/mm$^2$/day].  

\subsubsection*{Parameters} Most of the model parameters are taken over from Table \ref{model_params}. The ones that differ from those listed in Table \ref{model_params} are tabulated below in Table \ref{model_params_artery}. The collagen orientation angle $\alpha$ is now prescribed with respect to the longitudinal direction $Z$ within the circumferential-longitudinal plane.

\begin{table}[hbt!]\
\centering
\caption{\textbf{Balloon angioplasty - Model parameters}}
\label{model_params_artery} 
\begin{tabular}{p{2cm}p{2cm}p{2.5cm}p{2.5cm}p{3cm}}
\noalign{\hrule height 0.05cm}\noalign{\smallskip}
parameter & media & adventitia & units & reference\\
\noalign{\hrule height 0.05cm}\noalign{\smallskip}
 $D_{{}_P}$ & $0.01$ & $0.005$ & [mm$^2$/day] & choice\\
 $\varepsilon_{{}_E}$ & $3.0 \times 10^{23}$ & $3.0 \times 10^{23}$ & [mm$^3$/mol/day]& choice\\
 $\chi_{{}_C}$ & $1.0 \times 10^{11}$ & $0.0$ & [mm$^5$/mol/day]& choice\\
$\chi_{{}_H}$ & $1.0 \times 10^{6}$ & $0.0$ & [mm$^5$/mol/day]& choice\\
$\eta_{{}_S}$ & $1.0 \times 10^{13}$ & $0.0$ & [mm$^3$/cell/day]& choice\\
$\mu$ & $0.02$ & $0.008$ & [M Pa]& \citep{he2020} \\
$\Lambda$ & $10$ & $10$ & [M Pa]& \citep{he2020} \\
$\bar{k}_1$ & $0.112$ & $0.362$ & [M Pa]& \citep{he2020} \\
$k_2$ & $20.61$ & $7.089$ & [-]& \citep{he2020}\\
$\kappa$ & $0.24$ & $0.17$ & [-]& \citep{he2020} \\
$\alpha$ & $41$ & $50.1$ & [${}^{\circ}$]& \citep{he2020} \\
\noalign{\hrule height 0.05cm}\noalign{\smallskip}
\end{tabular}
\end{table}
\vspace{-0.2in}
\subsubsection*{Results and discussion} 
\begin{figure}[htb!]
    \centering
    \includegraphics[scale=0.46]{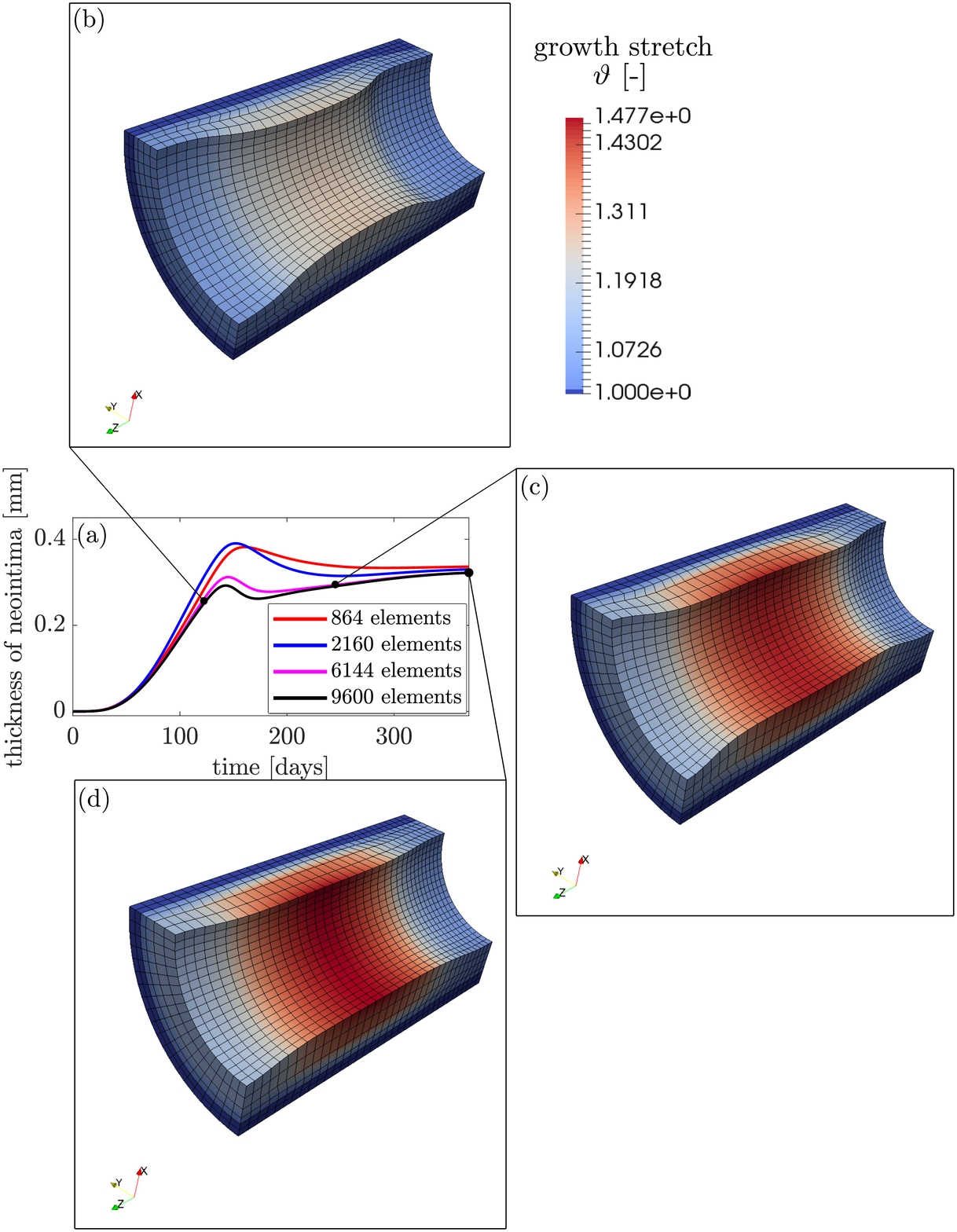}
    \caption{\textbf{Restenosis after balloon angioplasty.} \\
    (a) Evolution of the thickness of neointima along $Z = 3.5$ [mm] line. There is a slight reduction in neointima observed after achieving a peak at about 150 days. This can be attributed to Poisson's effect due to stretching of the adjacent tissue.\\
    (b) growth stretch contour at $t = 125$ [days]\\
    (c) growth stretch contour at $t = 250$ [days]\\
    (d) growth stretch contour at $t = 370$ [days]}
    \label{fig_bal_ang_res}
\end{figure}

Fig \ref{fig_bal_ang_res}(a) shows the evolution of neointimal thickness over a period of 370 days, along the line $Z = 3.5$ [mm] on the lumen surface. Figs \ref{fig_bal_ang_res}(b)-(c) show the evolution of the growth stretch $\vartheta$ over time. Due to the low diffusivities of the growth factors in the arterial wall, the neointimal growth is localized initially at the injury site. As the growth factors diffuse along the longitudinal direction, the growth can be seen also outside of the injury sites. Since the adventitia contains negligible amount of SMCs, the cell movement and proliferation is switched off within the adventitia by the prescription of zero values for the chemotactic and haptotactic sensitivities as well the SMC proliferation coefficient. 

\pagebreak
Additionally, a peak in the neointimal thickness is observed at around $t = 150$ [days], and beyond that a slight reduction is observed. The diffusing growth factors and ensuing growth stretches the tissue adjacent to the $Z = 3.5$ [mm] line on the lumen surface, which explains the compression at this region as a result of the Poisson effect. This effect can be validated by the experimental results presented in \citet{Zun2017}. Beyond 180 days, no significant change in the neointimal thickness is observed.

\subsection{In-stent restenosis}

Finally, to evaluate the capability of the developed formulation to model in-stent restenosis, a quadrant of an artery is modeled similar to that in Section \ref{subsection_bal_ang}. $l = 3$ [mm] here and all the other dimensions are the same as that in Fig \ref{fig_bal_ang}. The monolithic approach incorporating the stress-free anisotropic growth model ($\kappa = 0$) is utilized for this example. 

\begin{figure}[htbp!]
    \centering
    \includegraphics[scale=0.45]{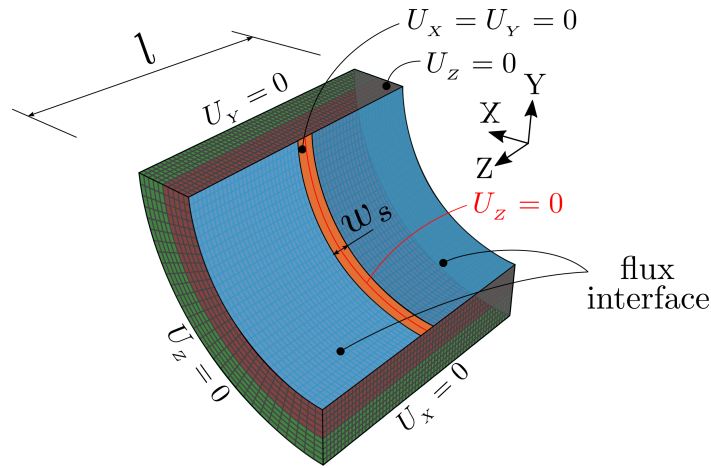}
    \caption{\textbf{Stented artery - Problem setup}}
    \label{fig_stented_artery}
\end{figure}

\subsubsection*{Discretization} The geometry is again meshed using trilinear hexahedral elements. Each layer of the arterial wall is meshed with 5 elements across their thicknesses, 30 elements along the circumferential direction, and 60 elements along the longitudinal direction. 

\subsubsection*{Boundary conditions} The sides marked in translucent gray are again fixed against displacements along their normals. Additionally, a small region of width $w_s = 0.1$ [mm] across the $Z = l/2$ line is fixed as shown in Fig \ref{fig_stented_artery}. This mimics a simplified stent strut held against the arterial wall. The flux interface is defined across the entire lumen surface of the artery except for the region where the stent strut is assumed to be present. To avoid the movement of stent strut surface, the nodes that lie on the lumen along $Z = l/2$ line are fixed against longitudinal displacements as shown. Self contact is prescribed on the lumen surface.

\subsubsection*{Parameters} The model parameters are the same as those listed for the balloon angioplasty case.

\subsubsection*{Results and discussion} 
\begin{figure}[htb!]
    \centering
    \includegraphics[scale=0.25]{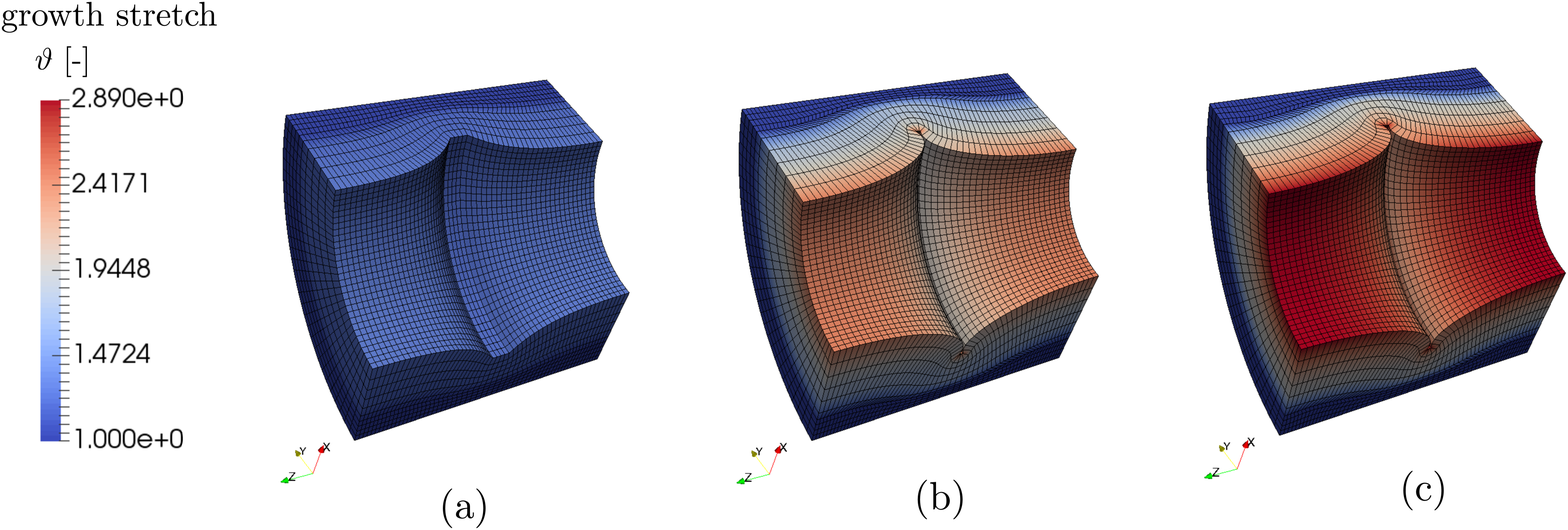}
    \caption{\textbf{Restenosis post implantation of an idealized stent.}\\
The stent gets completely engulfed by the soft tissue 180 days after implantation. \\(a) growth stretch contour at $t = 60$ [days]\\ (b) growth stretch contour at $t = 120$ [days]\\(c) growth stretch contour at $t = 180$ [days] }
    \label{fig_stented_artery_res}
\end{figure}

The contours of the growth stretch $\vartheta$ at 60, 120 and 180 days are plotted in Figs \ref{fig_stented_artery_res}(a)-(c). It is clearly seen that the stented area is completely engulfed by the neointima as expected. There was no neointimal growth observed in this model beyond 180 days.

\section{Conclusion and outlook}
A finite element framework that can successfully model in-stent restenosis based on the damage sustained by the endothelial monolayer was developed in this work. The framework considers the significant molecular and cellular mechanisms that result in neointimal hyperplasia and couples it to the two theories of finite growth developed herein. Although the multiphysics framework has been exploited by several authors for modeling atherosclerosis as well as neointimal hyperplasia, a fully coupled 3-dimensional Lagrangian formulation has not yet been explored, and hence is considered a novelty of this work. Additionally, the flux interface element developed as part of this work enables coupling the formulation to fluid mechanics simulations within an FSI framework.

The wide array of parameters associated with the developed model provides enough flexibility to factor in patient-specific aspects. Due to lack of experimental data pertaining to isolated effects of the considered species of the arterial wall, the model could unfortunately not be validated at the molecular and cellular levels. Only the macroscopic effects could be replicated and qualitatively compared. Experimental validation remains part of the future work that will follow.

Quantification of endothelium damage and subsequent prescription of wall shear stress and endothelium permeability dependent influx of growth factors also falls within the scope of further developments of the formulation.

One key aspect that affects neointimal hyperplasia is the deep injuries sustained during balloon angioplasty and stent implantation. Quantification of the damage sustained in the deep layers of the arterial wall, and addition of damage-dependent growth factor sources shall enhance the fidelity of the formulation.

Furthermore, collagen secretion and close packing of SMCs are all considered to reduce the entropy of the system. Introducing entropy sinks into the balance of entropy of the system can provide thermodynamic restrictions to the evolution as well as direction of growth, and shall therefore be a key part of future work on the formulation. Also, stress/stretch driven growth as well as collagen remodeling effects are ignored in the current framework, and shall therefore be another significant aspect to be considered in future developments.

Finally, the usage of trilinear elements for modeling the balance equations is known to induce locking effects. Finite element formulations incorporating reduced integration and hourglass stabilization shall be beneficial in this context. They are also associated with significant reduction in computational effort. The solid beam formulation (Q1STb \citep{frischkorn2013}) is relevant in modeling filigree structures like stents. Implementing it as part of the current framework shall aid in modeling stent expansion and endothelium damage efficiently.   

\appendix
\appendixpage
\addappheadtotoc
\section{Model related transparencies}
\subsection{Transfer of evolution equations for arterial wall species from the Eulerian to the Lagrangian description}
\label{appendix:etl}
If $\phi$ is a scalar variable that represents a species in the arterial wall, consider the following general form of evolution equation (Eq. \ref{ard_eq_general_form})
\begin{equation}\label{ard_genereal_form_simple}
   {\displaystyle{\left.\frac{\partial \phi}{\partial t}\right|_{\bm{x}}}} + {\text{\sf div} \left(\phi\,\boldsymbol{v}\right)}   = {\text{\sf div} \left(k\,\text{\sf grad} \phi\right)} + R_{so} - R_{si}.
\end{equation}
Consider now the material time derivative of the scalar field $\phi^0$, where $\phi^0 = J\,\phi$, $J = \text{\sf det} \bm{F}$. We here use the short hand notation of $\dot{(\bullet)}$ to represent the material time derivative of the quantity ${(\bullet)}$ By using chain rule of differentiation,
\begin{equation} \label{material_time_derivative}
    \dot{\phi^0} = J\,\dot{\phi} + \phi\,\dot{J}.
\end{equation}
It is known that for any second order tensor $\bm{A}$,
\begin{equation}\label{det_deriv_wrt_own}
    \frac{\partial (\text{\sf det}\, \bm{A})}{\partial \bm{A}} = (\text{\sf det}\, \bm{A}) \,\, \bm{A}^{-T}.
\end{equation}
By using chain rule of differentiation and the above relation,
\begin{eqnarray}
    \dot{J} &=& \frac{\partial J}{\partial \bm{F}}: \dot{\bm{F}}\nonumber\\
    &=& J\,\bm{F}^{-T}:\dot{\bm{F}}.
\end{eqnarray}
Using the definition of the spatial velocity gradient $\bm{l} = \dot{\bm{F}}\cdot\bm{F}^{-1}$, we get
\begin{eqnarray}
    \dot{J} &=& J\,\bm{F}^{-T}:\bm{l}\cdot\bm{F}\nonumber\\
    &=& J\,\bm{F}^{-T}\cdot\bm{F}^{T}: \bm{l}\nonumber\\
    &=& J\,\text{\sf tr} (\bm{l})\nonumber\\
    &=& J\, \text{\sf div } (\bm{v}) \label{j_dot}
\end{eqnarray}
Also,
\begin{equation}\label{phi_dot}
    \dot{\phi} = {\displaystyle{\left.\frac{\partial \phi}{\partial t}\right|_{\bm{x}}}} + \bm{v}\cdot \text{\sf grad}\,(\phi),
\end{equation}
and
\begin{equation}\label{divergence}
     {\text{\sf div} \left(\phi\,\boldsymbol{v}\right)} = \phi\,\text{\sf div} (\bm{v}) + \bm{v}\cdot\text{\sf grad} (\phi). 
\end{equation}
Substituting Eqs. \ref{j_dot}, \ref{divergence}, and \ref{phi_dot} in the left hand side of Eq. \ref{ard_genereal_form_simple}, we get
\begin{equation}
    {\displaystyle{\left.\frac{\partial \phi}{\partial t}\right|_{\bm{x}}}} + {\text{\sf div} \left(\phi\,\boldsymbol{v}\right)} = \left(\frac{1}{J}\right)\,\dot{\phi^0}.
\end{equation}
To convert the right hand side terms in Eq. \ref{ard_genereal_form_simple} to the Lagrangian form, we use the following identity:
\begin{eqnarray}
    \text{\sf div} (k\,\text{\sf grad}\,\phi) &=& \frac{1}{J}\,\text{\sf Div} \left(J\,\bm{C}^{-1}\,k\,\text{\sf Grad}\,\phi \right)\nonumber\\
    &=& \frac{1}{J}\,\text{\sf Div} \left[\bm{C}^{-1}\,k\,\text{\sf Grad}\,\phi^0 - k\,\frac{\phi^0}{J}\,\text{\sf Grad}\,J \right].
\end{eqnarray}
Further, all the source and sink terms are expressed in terms of $\phi^0 = J\,\phi$.

\subsection{Linearized weak forms}
\label{appendix:lwf}
The weak forms linearized about the variables at time $t_{n+1}$ are derived from Eqs. \ref{wf_p} - \ref{wf_u} and are listed below.
\begin{eqnarray}
    \Delta g_{{}_P} =& \left.\displaystyle{\left(\frac{\partial g_{{}_P}}{\partial c^0_{{}_P}}\right)}\right|_{t_{n+1}}\,\Delta c^0_{{}_P}\nonumber\\
    &+ \left.\displaystyle{\left(\frac{\partial g_{{}_P}}{\partial c^0_{{}_T}}\right)}\right|_{t_{n+1}}\,\Delta c^0_{{}_T}\nonumber\\
    &+ \left.\displaystyle{\left(\frac{\partial g_{{}_P}}{\partial \rho^0_{{}_S}}\right)}\right|_{t_{n+1}}\,\Delta \rho^0_{{}_S}\nonumber\\
    &+ \left.\displaystyle{\left(\frac{\partial g_{{}_P}}{\partial \bm{F}}\right)}\right|_{t_{n+1}}\,\Delta \bm{F}\label{lwf_p}
\end{eqnarray}

\begin{eqnarray}
    \Delta g_{{}_T} =& \left.\displaystyle{\left(\frac{\partial g_{{}_T}}{\partial c^0_{{}_T}}\right)}\right|_{t_{n+1}}\,\Delta c^0_{{}_T}\nonumber\\
    &+ \left.\displaystyle{\left(\frac{\partial g_{{}_T}}{\partial \rho^0_{{}_S}}\right)}\right|_{t_{n+1}}\,\Delta \rho^0_{{}_S}\nonumber\\
    &+ \left.\displaystyle{\left(\frac{\partial g_{{}_T}}{\partial \bm{F}}\right)}\right|_{t_{n+1}}\,\Delta \bm{F}\label{lwf_t}
\end{eqnarray}

\begin{eqnarray}
    \Delta g_{{}_E} =& \left.\displaystyle{\left(\frac{\partial g_{{}_E}}{\partial c^0_{{}_P}}\right)}\right|_{t_{n+1}}\,\Delta c^0_{{}_P}\nonumber\\
    &+ \left.\displaystyle{\left(\frac{\partial g_{{}_E}}{\partial \rho^0_{{}_S}}\right)}\right|_{t_{n+1}}\,\Delta \rho^0_{{}_S}\nonumber\\
    &+ \left.\displaystyle{\left(\frac{\partial g_{{}_P}}{\partial \bm{F}}\right)}\right|_{t_{n+1}}\,\Delta \bm{F}\label{lwf_e}
\end{eqnarray}

\begin{eqnarray}
    \Delta g_{{}_S} =& \left.\displaystyle{\left(\frac{\partial g_{{}_S}}{\partial c^0_{{}_P}}\right)}\right|_{t_{n+1}}\,\Delta c^0_{{}_P}\nonumber\\
    &+ \left.\displaystyle{\left(\frac{\partial g_{{}_S}}{\partial c^0_{{}_T}}\right)}\right|_{t_{n+1}}\,\Delta c^0_{{}_T}\nonumber\\
    &+ \left.\displaystyle{\left(\frac{\partial g_{{}_S}}{\partial c^0_{{}_E}}\right)}\right|_{t_{n+1}}\,\Delta c^0_{{}_E}\nonumber\\
    &+ \left.\displaystyle{\left(\frac{\partial g_{{}_S}}{\partial \rho^0_{{}_S}}\right)}\right|_{t_{n+1}}\,\Delta \rho^0_{{}_S}\nonumber\\
    &+ \left.\displaystyle{\left(\frac{\partial g_{{}_P}}{\partial \bm{F}}\right)}\right|_{t_{n+1}}\,\Delta \bm{F}\label{lwf_S}
\end{eqnarray}

\begin{eqnarray}
    \Delta g_{{}_u} =& \left.\displaystyle{\left(\frac{\partial g_{{}_u}}{\partial c^0_{{}_E}}\right)}\right|_{t_{n+1}}\,\Delta c^0_{{}_E}\nonumber\\
    &+ \left.\displaystyle{\left(\frac{\partial g_{{}_u}}{\partial \rho^0_{{}_S}}\right)}\right|_{t_{n+1}}\,\Delta \rho^0_{{}_S}\nonumber\\
    &+ \left.\displaystyle{\left(\frac{\partial g_{{}_u}}{\partial \bm{F}}\right)}\right|_{t_{n+1}}\,\Delta \bm{F}\label{lwf_u}
\end{eqnarray}
The discretized weak form is constucted as follows:
\begin{eqnarray}
    g^h_{{}_P} = &=& \displaystyle{\bigcup^{n_{e}}_{i = 1}} \delta (\bm{c}^0_{{}_P})_e^T \bm{R}^e_{{}_P}\nonumber\\
    g^h_{{}_T} = &=& \displaystyle{\bigcup^{n_{e}}_{i = 1}} \delta (\bm{c}^0_{{}_T})_e^T \bm{R}^e_{{}_T}\nonumber\\
    g^h_{{}_E} = &=& \displaystyle{\bigcup^{n_{e}}_{i = 1}} \delta (\bm{c}^0_{{}_E})_e^T \bm{R}^e_{{}_E}\nonumber\\
    g^h_{{}_S} = &=& \displaystyle{\bigcup^{n_{e}}_{i = 1}} \delta (\bm{\rho}^0_{{}_S})_e^T \bm{R}^e_{{}_S}\nonumber\\
    g^h_{{}_u} = &=& \displaystyle{\bigcup^{n_{e}}_{i = 1}} \delta \bm{U}_e^T \bm{R}^e_{{}_u}
\end{eqnarray}
The discretized and linearized weak forms hence read
\begin{eqnarray}
    \Delta g^h_{{}_P} &=& \displaystyle{\bigcup^{n_{e}}_{i = 1}} \delta (\bm{c}^0_{{}_P})_e^T \left[\bm{K}^e_{{}_{PP}}\,\Delta(\bm{c}^0_{{}_P})^e + \bm{K}^e_{{}_{PT}}\,\Delta(\bm{c}^0_{{}_T})^e 
     + \bm{K}^e_{{}_{PS}}\,\Delta(\bm{\rho}^0_{{}_S})^e + \bm{K}^e_{{}_{Pu}}\,\Delta \bm{U}^e \right]\nonumber\\
     \Delta g^h_{{}_T} &=& \displaystyle{\bigcup^{n_{e}}_{i = 1}} \delta (\bm{c}^0_{{}_T})_e^T \left[ \bm{K}^e_{{}_{TT}}\,\Delta(\bm{c}^0_{{}_T})^e 
     + \bm{K}^e_{{}_{TS}}\,\Delta(\bm{\rho}^0_{{}_S})^e + \bm{K}^e_{{}_{Tu}}\,\Delta \bm{U}^e \right]\nonumber\\
     \Delta g^h_{{}_E} &=& \displaystyle{\bigcup^{n_{e}}_{i = 1}} \delta (\bm{c}^0_{{}_E})_e^T \left[ \bm{K}^e_{{}_{TP}}\,\Delta(\bm{c}^0_{{}_P})^e + \bm{K}^e_{{}_{EE}}\,\Delta(\bm{c}^0_{{}_E})^e
     + \bm{K}^e_{{}_{ES}}\,\Delta(\bm{\rho}^0_{{}_S})^e + \bm{K}^e_{{}_{Eu}}\,\Delta \bm{U}^e \right]\nonumber\\
     \Delta g^h_{{}_S} &=& \displaystyle{\bigcup^{n_{e}}_{i = 1}} \delta (\bm{\rho}^0_{{}_S})_e^T \left[ \bm{K}^e_{{}_{SP}}\,\Delta(\bm{c}^0_{{}_P})^e + \bm{K}^e_{{}_{ST}}\,\Delta(\bm{c}^0_{{}_T})^e + \bm{K}^e_{{}_{SE}}\,\Delta(\bm{c}^0_{{}_E})^e
     + \bm{K}^e_{{}_{SS}}\,\Delta(\bm{\rho}^0_{{}_S})^e + \bm{K}^e_{{}_{Su}}\,\Delta \bm{U}^e \right]\nonumber\\
     \Delta g^h_{{}_u} &=& \displaystyle{\bigcup^{n_{e}}_{i = 1}} \delta \bm{U}_e^T \left[  \bm{K}^e_{{}_{uE}}\,\Delta(\bm{c}^0_{{}_E})^e
     + \bm{K}^e_{{}_{uS}}\,\Delta(\bm{\rho}^0_{{}_S})^e + \bm{K}^e_{{}_{uu}}\,\Delta \bm{U}^e \right]\nonumber\\
\end{eqnarray}
The vectors $\bm{R}^e_{{}_{(\bullet)}}$ and the matrices $\bm{K}^e_{{}_{(\bullet)(\bullet)}}$ are constructed using the shape function vectors $\bm{N}^L$ and the shape function derivative matrices $\bm{B}$ and $\bm{B}_{{}_U}$ as defined in Eqs. \ref{interp} - \ref{grad_u}. All the derivatives that are necessary to be calculated for the discretized and linearized weak forms are obtained using algorithmic differentiation via the software package \textit{AceGen} \citep{korelc2002,korelc2009}.

\subsection{Parameter sensitivity study for patient-specific parameters}
\label{appendix:pst}
The following figures depict the sensitivity of the model to the parameters that can be tuned patient-specifically. The volume change due to growth at point P, seen in Fig \ref{fig_cube_problem}(a), is used here for the comparative study. The rest of the parameters remain the same as in Table \ref{model_params} except for those specified in the respective captions. 

\begin{figure}[htb!] 
    \centering
    \includegraphics[width=0.5\linewidth]{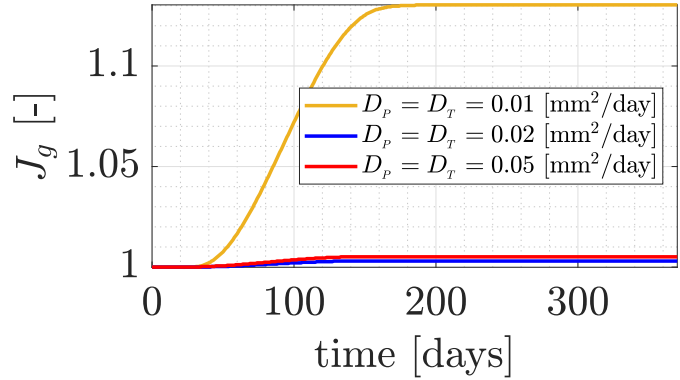}
    \caption{\textbf{Variable diffusivities:} $\chi_{{}_C} = 1.0 \times 10^{12}$ [mm$^5$/mol/day], $\chi_{{}_C} = 0$ [mm$^5$/mol/day], $\eta_{{}_S} = 0$ [mm$^3$/cell/day]} 
\end{figure}
\vspace{0.4in}  
\begin{figure}[htb!] 
\centering
  \begin{subfigure}[b]{0.5\linewidth}
    \centering
    \includegraphics[width=\linewidth]{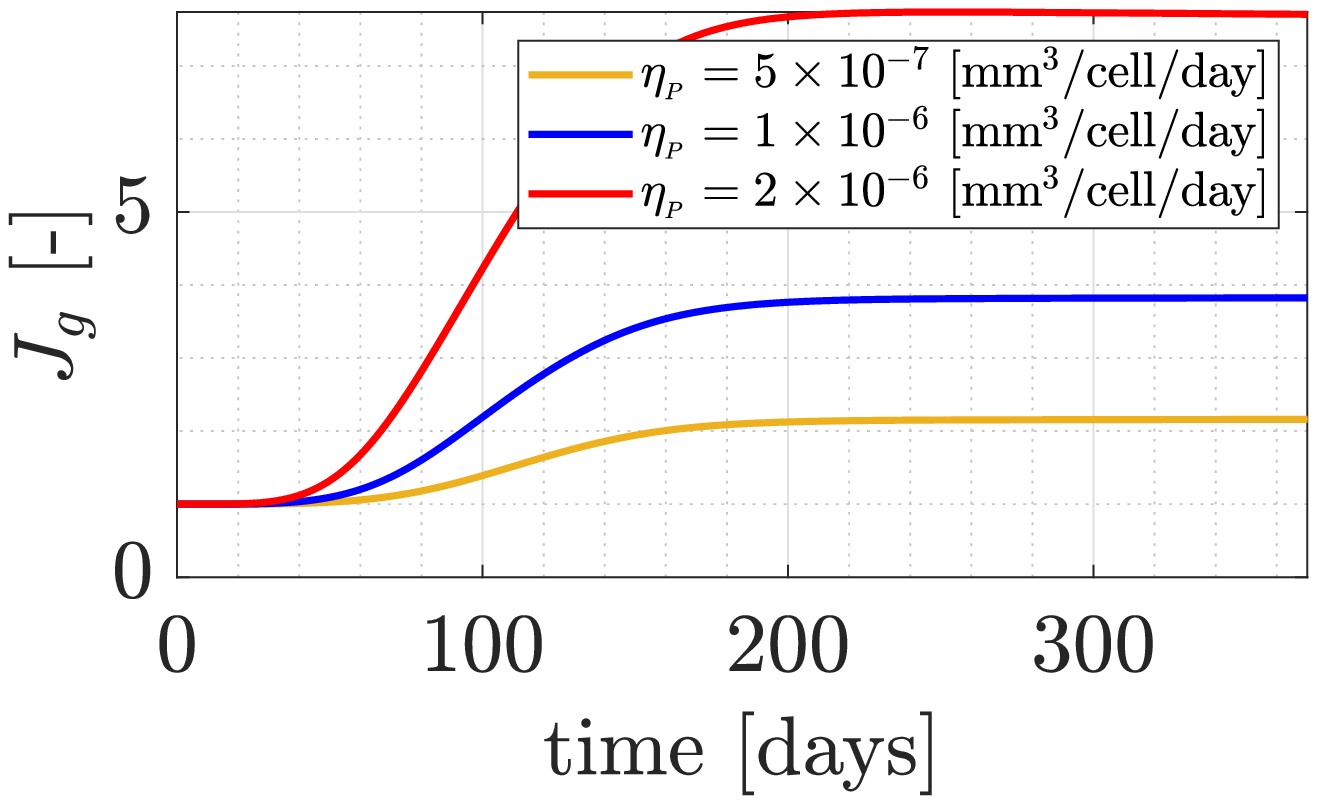} 
  \end{subfigure}
  \begin{subfigure}[b]{0.5\linewidth}
    \centering
    \includegraphics[width=\linewidth]{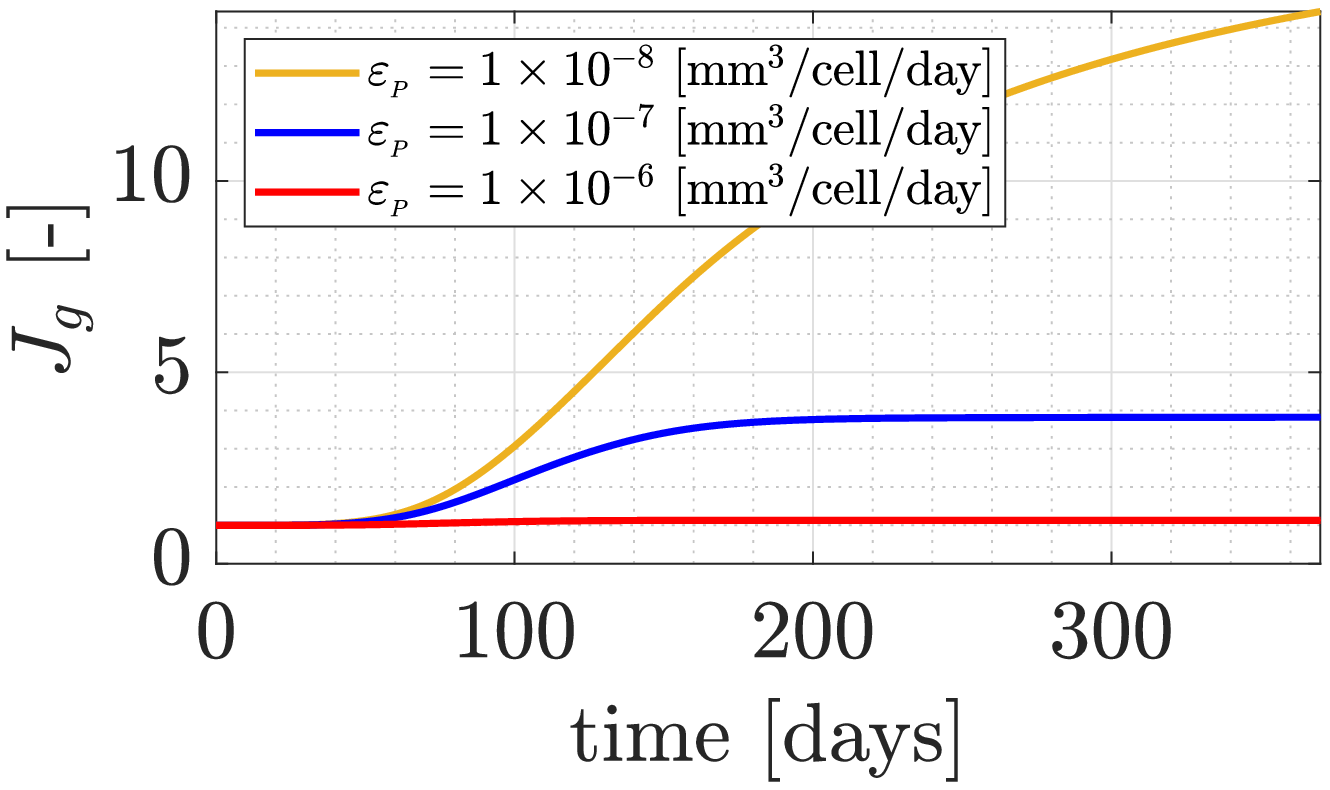} 
  \end{subfigure} 
  \caption{\textbf{Variable PDGF secretion and receptor internalization coeefficients}}
\end{figure}

\pagebreak

\begin{figure}[htb!] 
\centering  
  \begin{subfigure}[b]{0.5\linewidth}
    \centering
    \includegraphics[width=\linewidth]{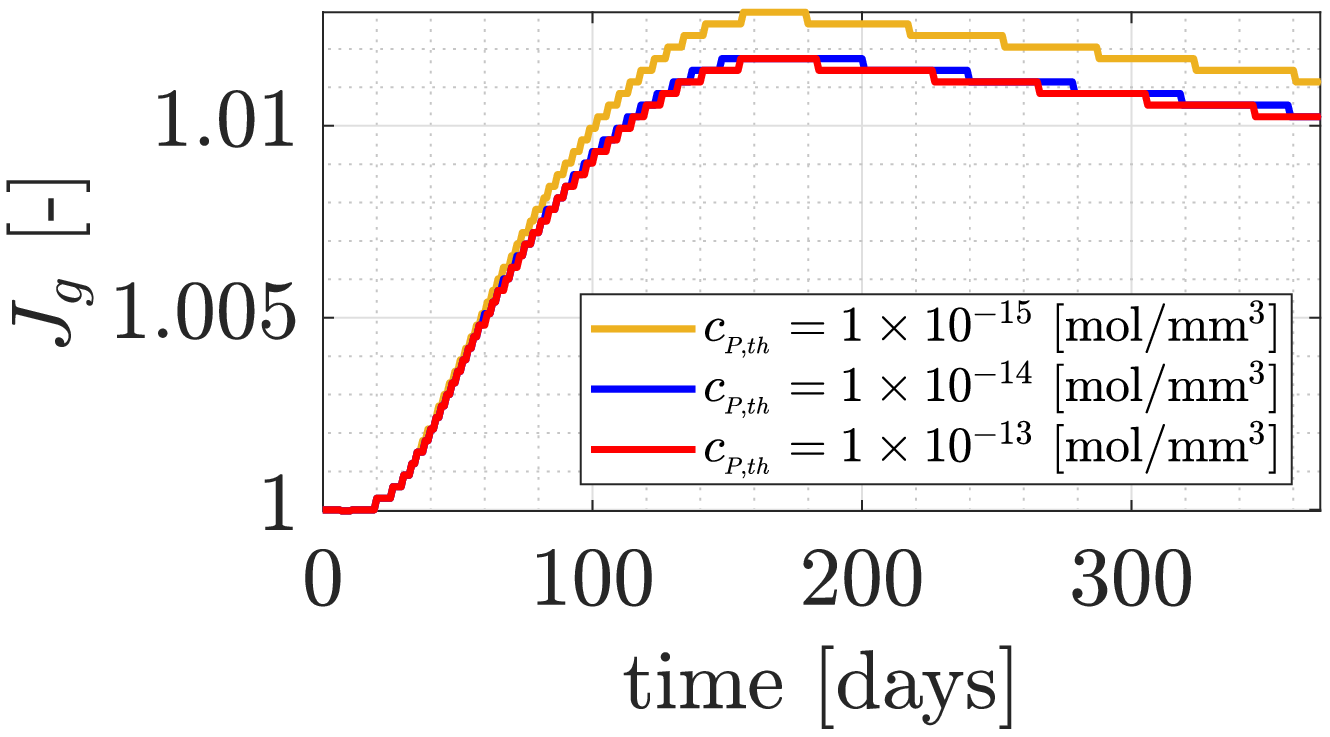} 
  \end{subfigure}
  \begin{subfigure}[b]{0.5\linewidth}
    \centering
    \includegraphics[width=\linewidth]{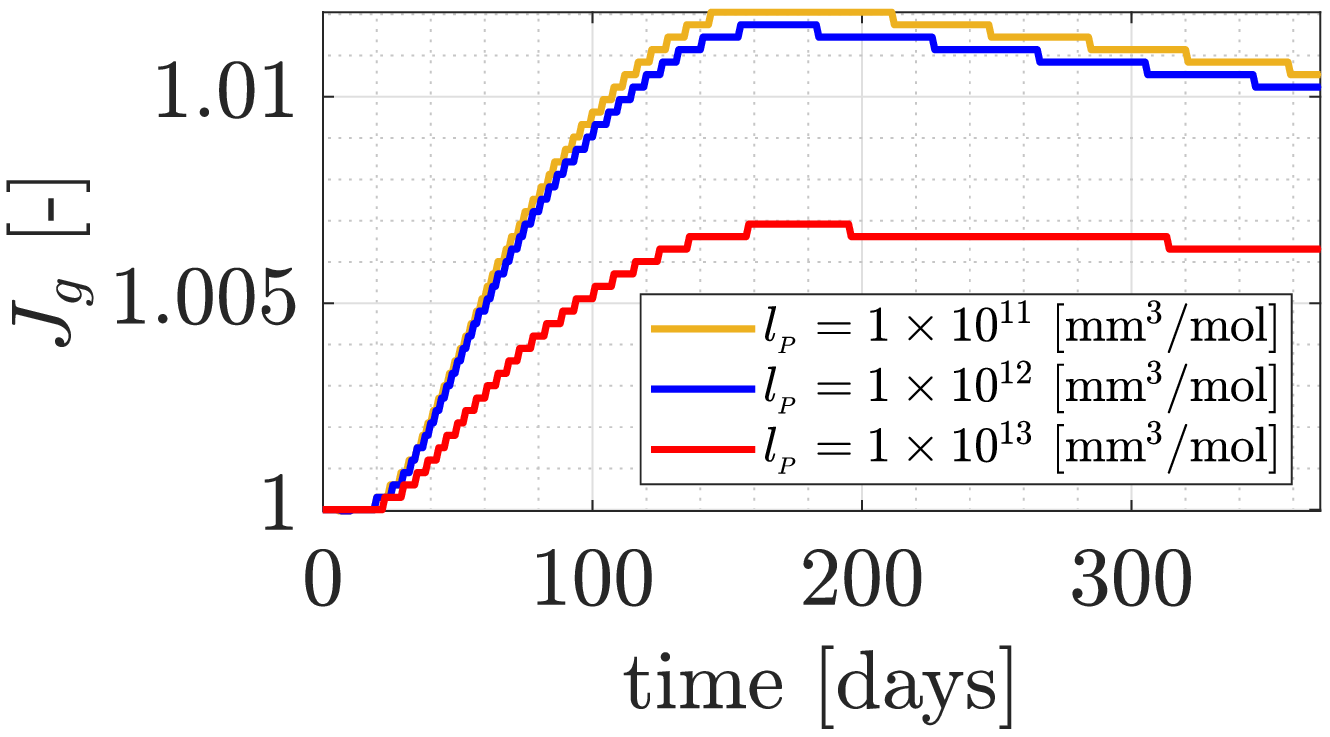} 
  \end{subfigure} 
  \caption{\textbf{Variable PDGF threshold and steepness coefficient:} $D_{{}_P} = D_{{}_T} = 0.01$ [mm$^2$/day], $\chi_{{}_H} = 1.0 \times 10^{5}$ [mm$^5$/mol/day], $\chi_{{}_C} = 0$ [mm$^5$/mol/day], $\eta_{{}_S} = 0$ [mm$^3$/cell/day]}
\end{figure}

\vspace{0.4in}

\begin{figure}[htb!] 
    \centering
    \includegraphics[width=0.5\linewidth]{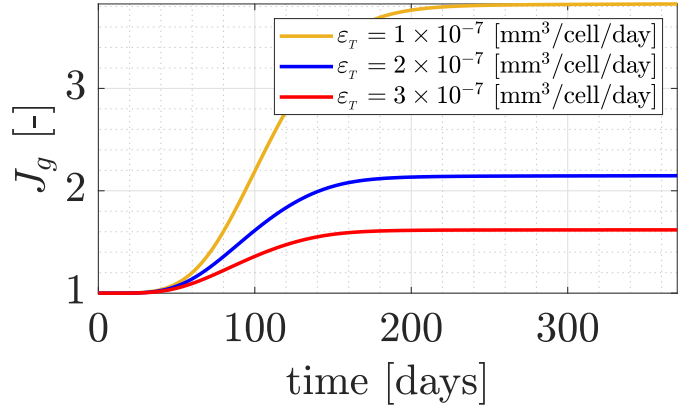}
    \caption{\textbf{Variable TGF$-\beta$ receptor internalization coefficient}} 
\end{figure}
  
  \vspace{0.4in}
\begin{figure}[htb!] 
\centering
  \begin{subfigure}[b]{0.5\linewidth}
    \centering
    \includegraphics[width=0.95\linewidth]{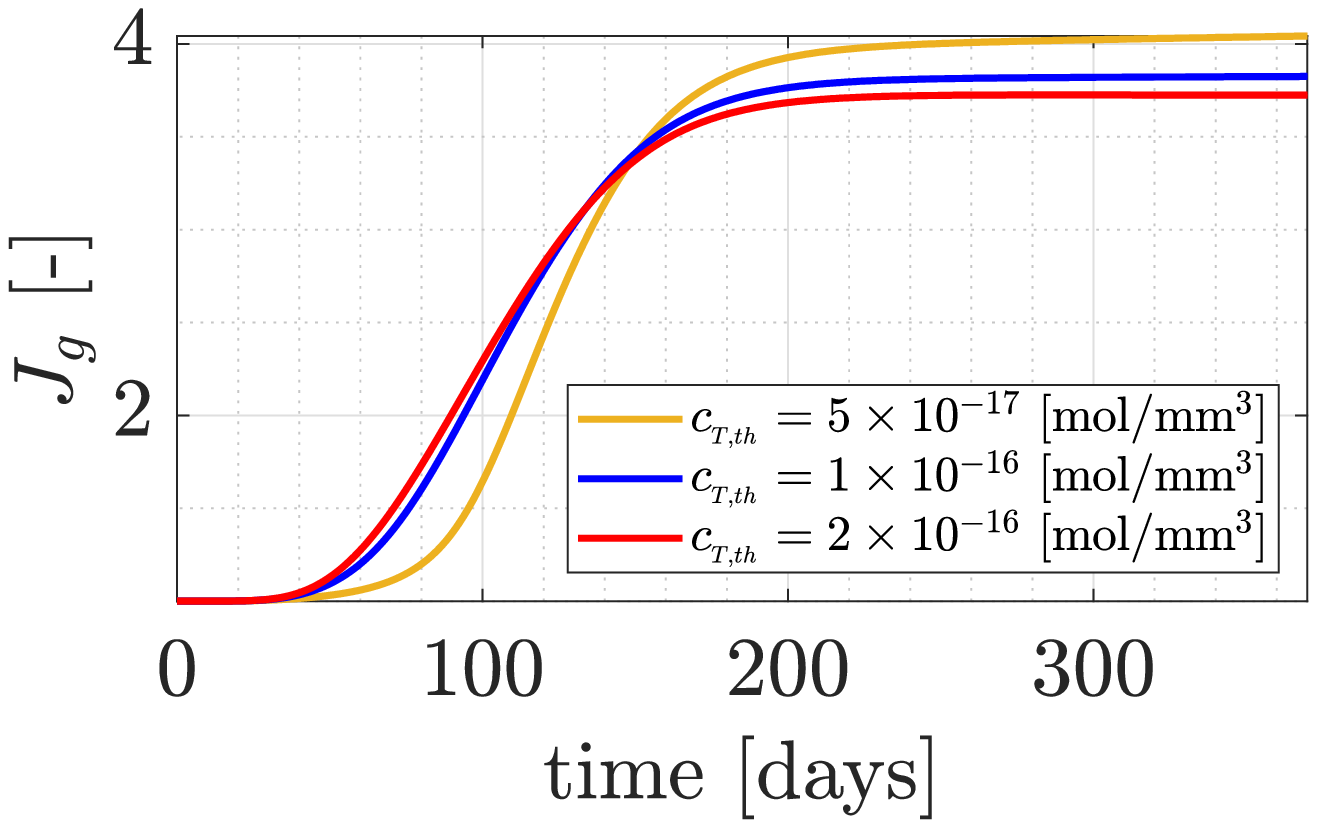} 
  \end{subfigure}
  \begin{subfigure}[b]{0.5\linewidth}
    \centering
    \includegraphics[width=0.95\linewidth]{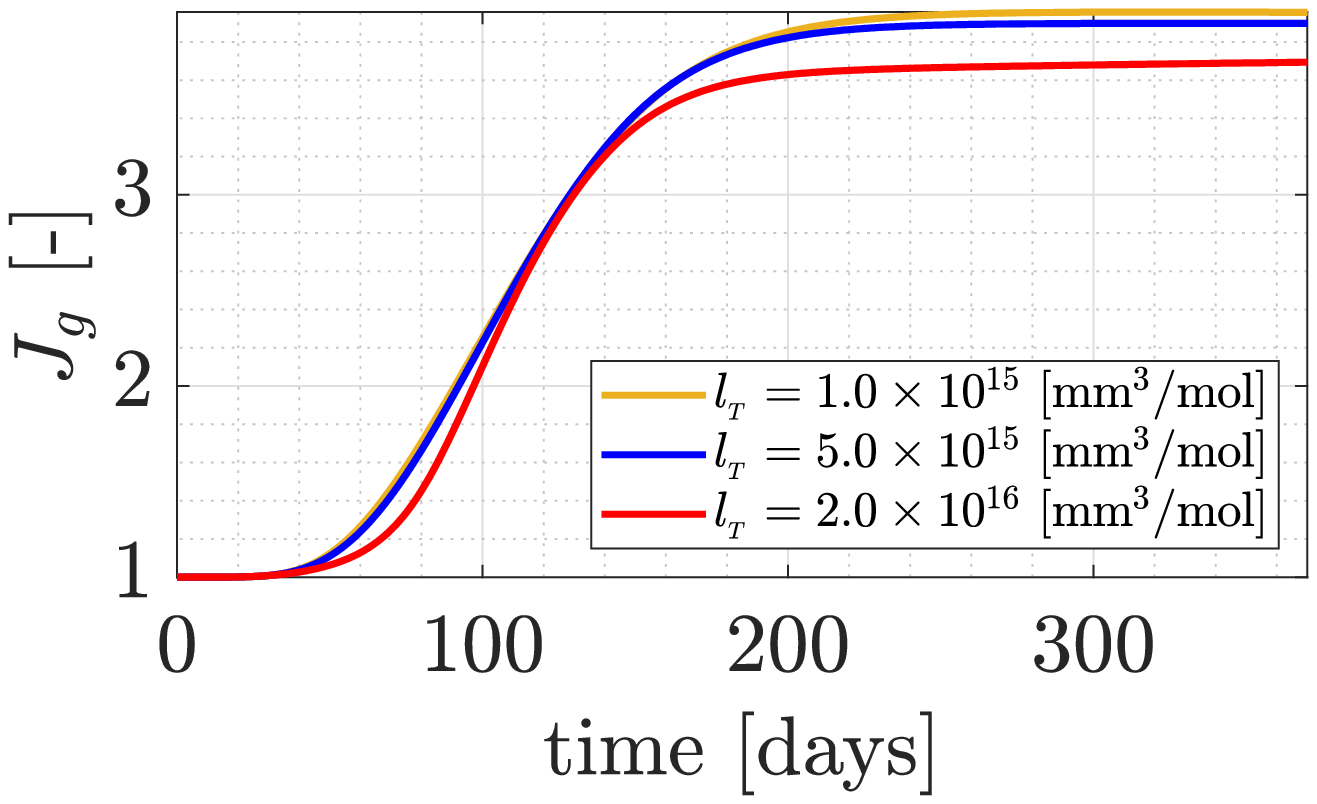} 
  \end{subfigure} 
  \caption{\textbf{Variable TGF$-\beta$ threshold and steepness coefficient}}
\end{figure}

\pagebreak

\begin{figure}[htb!] 
\centering  
  \begin{subfigure}[b]{0.5\linewidth}
    \centering
    \includegraphics[width=\linewidth]{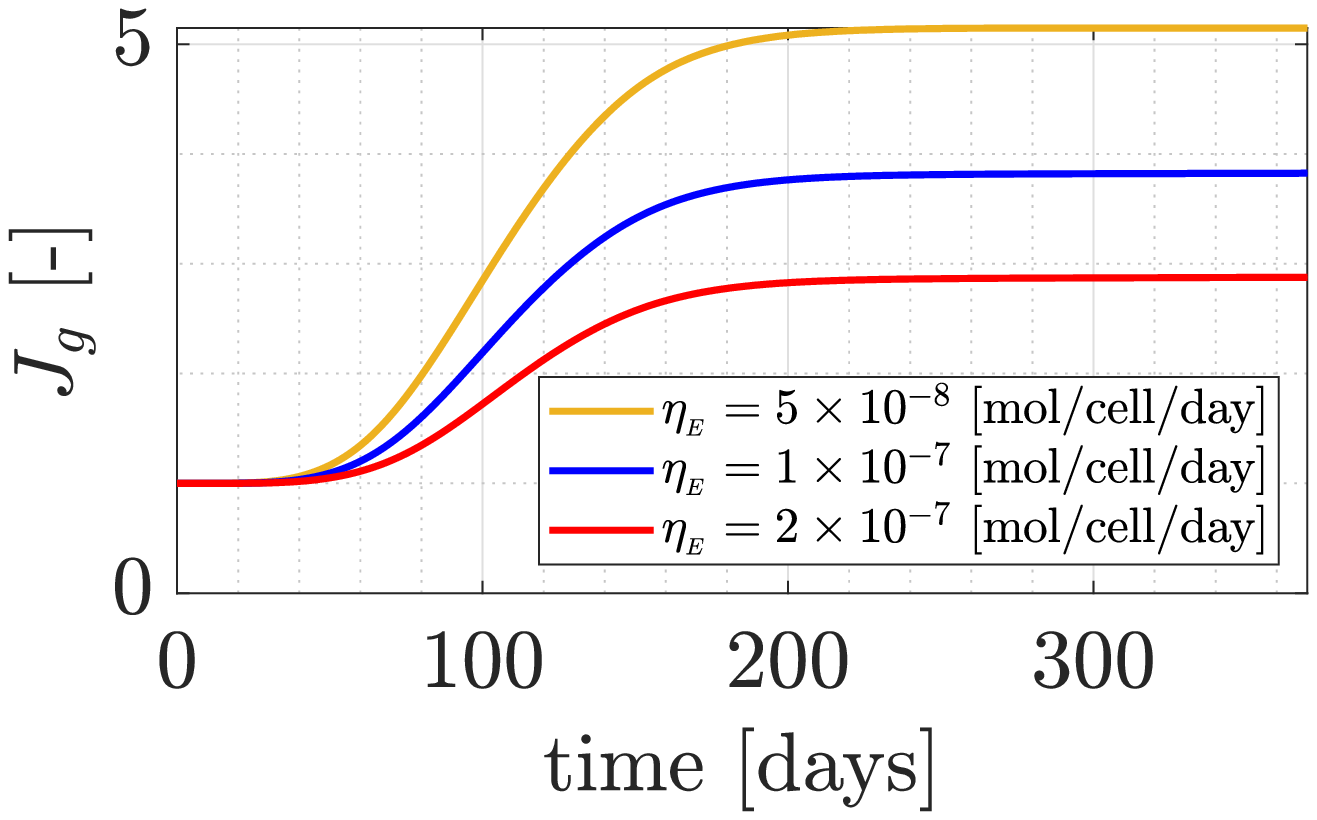} 
  \end{subfigure}
  \begin{subfigure}[b]{0.5\linewidth}
    \centering
    \includegraphics[width=\linewidth]{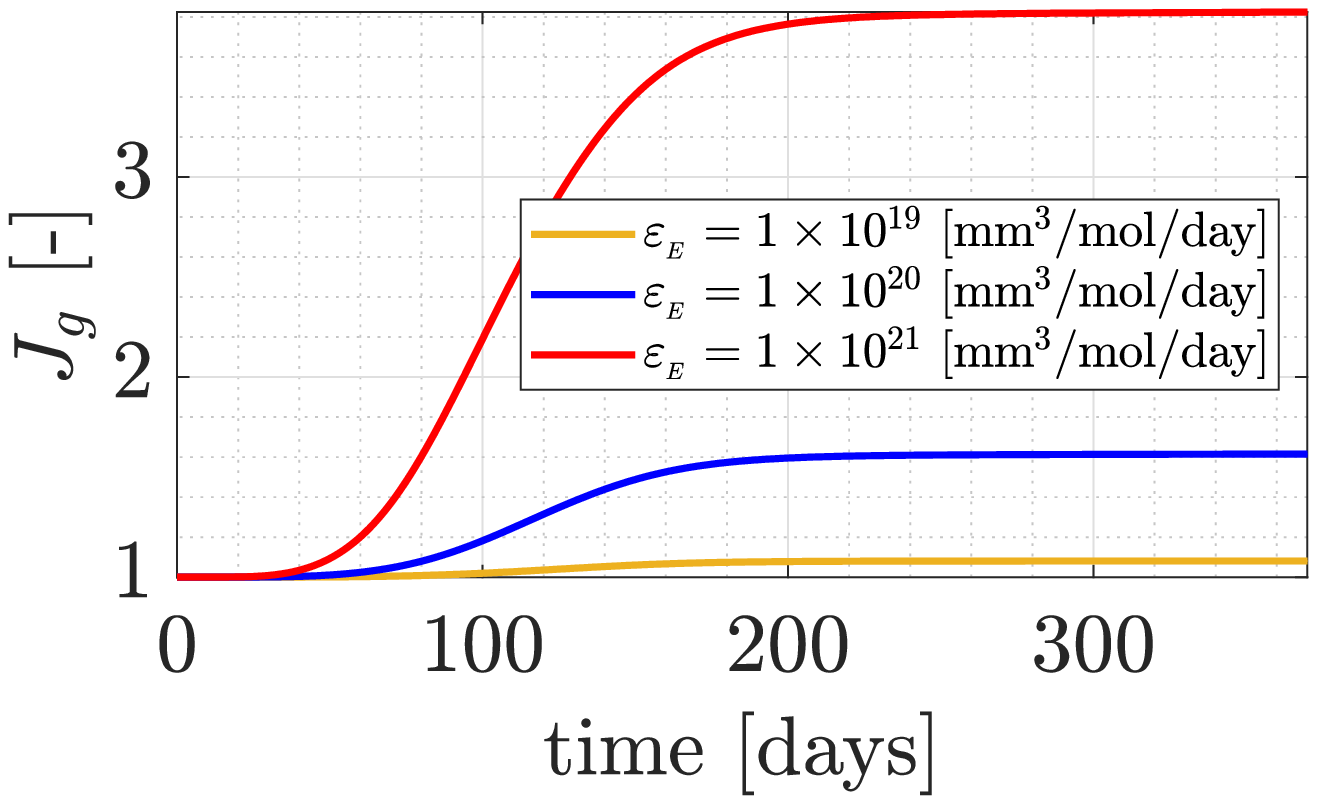} 
  \end{subfigure} 
  \caption{\textbf{Variable ECM secretion and degradation coefficients}}
\end{figure}

\vspace{0.4in}
\begin{figure}[htb!] 
    \centering
    \includegraphics[width=0.5\linewidth]{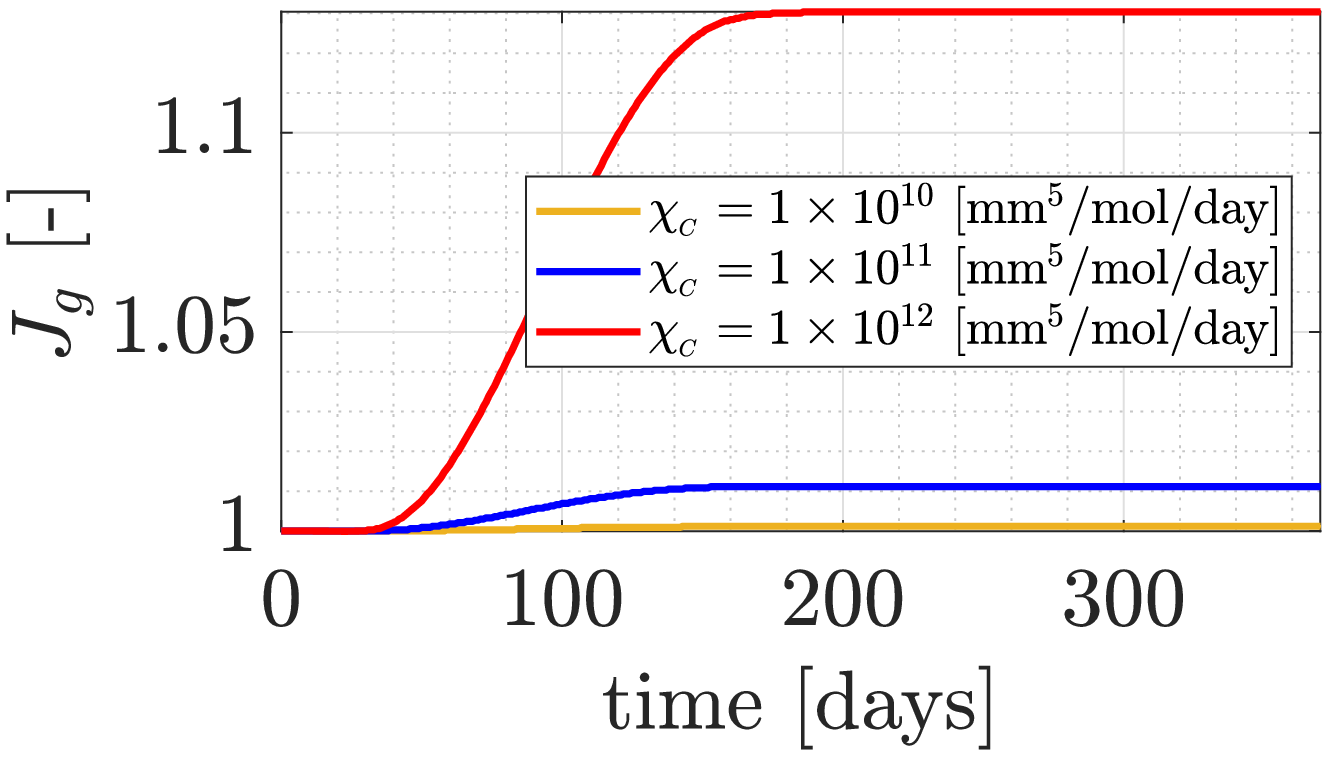}
    \caption{\textbf{Variable chemotactic sensitivity:} $D_{{}_P} = D_{{}_T} = 0.01$ [mm$^2$/day], $\chi_{{}_H} = 0$ [mm$^5$/mol/day], $\eta_{{}_S} = 0$ [mm$^3$/cell/day]} 
\end{figure}

\vspace{0.4in}
\begin{figure}[htb!] 
    \centering
    \includegraphics[width=0.5\linewidth]{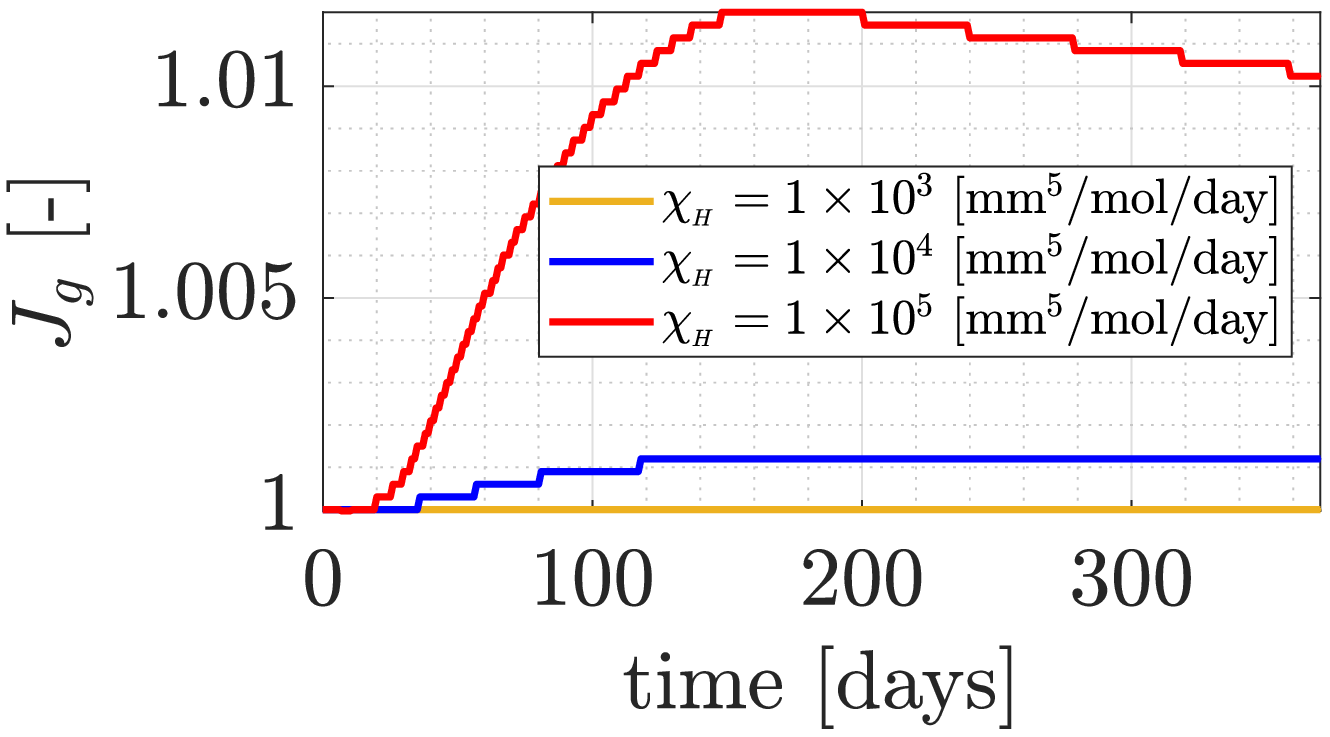}
    \caption{\textbf{Variable haptotactic sensitivity:} $D_{{}_P} = D_{{}_T} = 0.01$ [mm$^2$/day], $\chi_{{}_C} = 0$ [mm$^5$/mol/day], $\eta_{{}_S} = 0$ [mm$^3$/cell/day]} 
\end{figure}

\begin{figure}[htb!] 
    \centering
    \includegraphics[width=0.5\linewidth]{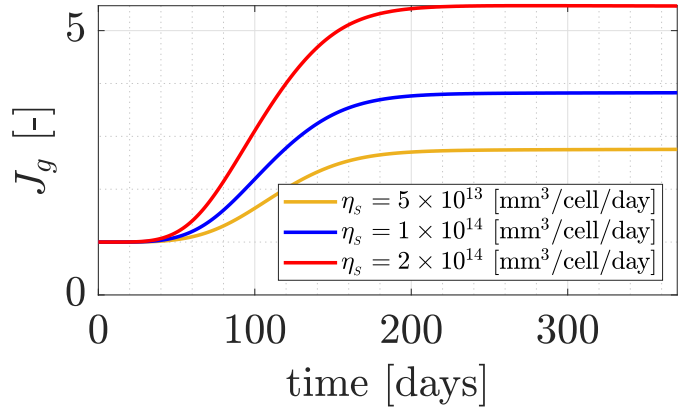}
    \caption{\textbf{Variable SMC proliferation coefficient}} 
\end{figure}
\pagebreak
\section{Declarations}
\subsection{Funding}
This work has been funded through the financial support of German Research Foundation (DFG) for the projects ``Drug-eluting coronary stents in stenosed arteries: medical investigation and computational modelling" (project number 395712048: RE 1057/44-1, RE 1057/44-2), and ``Modelling of Structure and Fluid-Structure Interaction during Tissue Maturation in Biohybrid Heart Valves", a subproject of ``PAK961 - Modeling of the structure and fluid-structure interaction of biohybrid heart valves on tissue maturation" (project number 403471716: RE 1057/45-1, RE 1057/45-2).

\subsection{Conflict of interest}
The authors certify that they have no affiliations with or involvement in any organization or entity with any financial interest (such as honoraria; educational grants; participation in speakers’ bureaus; membership, employment, consultancies, stock ownership, or other equity interest; and expert testimony or patent-licensing arrangements), or non-financial interest (such as personal or professional relationships, affiliations, knowledge or beliefs) in the subject matter or materials discussed in this manuscript.

\subsection{Availability of data} 
The data generated through the course of this work is stored redundantly and will be made available on demand.

\subsection{Code availability}
The custom written routines will be made available on demand. The software package FEAP is a proprietary software and can therefore not be made available.

\subsection{Contributions from the authors}
K. Manjunatha determined the state of the art by reviewing relevant literature, created the user subroutines for implementation in FEAP, interpreted the results and wrote this article. K. Manjunatha and S. Reese developed the theoretical framework contained within this work. M. Behr and F. Vogt participated in project discussions, provided conceptual advice, gave feedback on the interpretation of results, read the article and provided valuable suggestions for improvement. All the authors approve the publication of this manuscript. 

\bibliography{isrpaper}

\end{document}